\documentclass[12pt,preprint]{aastex}               	

\newcommand{\km}{${\rm km\,s}^{-1}$}
\newcommand{\fuse}{{\em FUSE}}
\let\la=\lesssim

\newcommand{\hi}{H$\;${\small\rm I}\relax}
\newcommand{\di}{D$\;${\small\rm I}\relax}
\newcommand{\nni}{N$\;${\small\rm I}\relax}
\newcommand{\nai}{Na$\;${\small\rm I}\relax}
\newcommand{\ari}{Ar$\;${\small\rm I}\relax}

\newcommand{\cii}{C$\;${\small\rm II}\relax}
\newcommand{\ciii}{C$\;${\small\rm III}\relax}
\newcommand{\civ}{C$\;${\small\rm IV}\relax}

\newcommand{\nv}{N$\;${\small\rm V}\relax}
\newcommand{\oi}{O$\;${\small\rm I}\relax}

\newcommand{\ov}{O$\;${\small\rm V}\relax}
\newcommand{\ovi}{O$\;${\small\rm VI}\relax}
\newcommand{\ovii}{O$\;${\small\rm VII}\relax}
\newcommand{\oviii}{O$\;${\small\rm VIII}\relax}

\newcommand{\siii}{Si$\;${\small\rm II}\relax}

\newcommand{\siiv}{Si$\;${\small\rm IV}\relax}

\newcommand{\mgii}{Mg$\;${\small\rm II}\relax}

\newcommand{\feii}{Fe$\;${\small\rm II}\relax}
\newcommand{\feiii}{Fe$\;${\small\rm III}\relax}
\newcommand{\piv}{P$\;${\small\rm IV}\relax}
\newcommand{\pv}{P$\;${\small\rm V}\relax}

\slugcomment{Version of \today}
\shortauthors{Savage \& Lehner}
\shorttitle{Properties of \ovi\ Absorption in the LISM}
\begin{document}

\title{Properties of \ovi\ Absorption in the Local Interstellar Medium}

\author{Blair \ D.\ Savage and
	Nicolas \ Lehner
}
	
\affil{Department of Astronomy, University of Wisconsin, 475 North Charter Street, Madison, WI 53706}

\begin{abstract}
The measurement of absorption over short distances in the local 
interstellar medium (LISM) provides an excellent opportunity to study 
the properties of \ovi\ absorption and its kinematic relationship to 
absorption by tracers of cool and warm gas. We report on the properties 
of LISM \ovi\ absorption observed with 20 \km\  resolution {\em Far 
Ultraviolet Spectroscopic Explorer (FUSE)}\  observations of 39 white 
dwarfs (WDs) ranging in distance from 37 to 230 pc with a median 
distance of 109 pc.  LISM \ovi\ is detected with $\ge 2\sigma$ significance along 
24 of 39 lines of sight. The column densities range from $\log N($\ovi$) = 12.38$ 
to 13.60  with a median value of 13.10. The line of sight 
volume density, $n($\ovi$) = N($\ovi$)/d$,  exhibits a large dispersion 
ranging from (0.68 to $13.0)\times 10^{-8}$  cm$^{-3}$ with an average 
value  $n($\ovi$) = 3.6\times 10^{-8}$ cm$^{-3}$  twice larger than 
found for more distant sight lines in the Galactic disk.   The Doppler parameter, $b$, 
for the $\ge 4\sigma$ \ovi\ detections range from $15.0 \pm 3.5$ to $36.2 \pm 7.3$ \km,  
with average and dispersion of $23.0 \pm 5.6$ \km.  The narrowest profiles are consistent 
with thermal Doppler broadening of \ovi\ near its temperature of peak 
abundance, $2.8\times 10^5$ K. $\log N($\ovi) is observed to correlate with  $b$. 
The broader profiles are tracing a combination of \ovi\ with $T > 2.8\times 10^5$  K, 
multiple component \ovi\ absorption either in interfaces or hot bubbles,   
and/or  the evaporative outflow of conductive interface gas. Comparison of the average 
velocities of \ovi\ and \cii\ 
absorption reveals 10 cases where the \ovi\ absorption is closely aligned 
with the \cii\ absorption as expected if the \ovi\ is formed in a condensing 
interface between the cool and warm absorption traced by \cii\ (and also \oi\ and \ciii) 
and a  hot exterior gas.  The comparison also reveals 13 cases where \ovi\ 
absorption is displaced to positive velocity by 7 to 29 \km\ from 
the average velocity of \cii.  The positive velocity \ovi\ is mostly found 
in the north Galactic hemisphere.    The positive 
velocity \ovi\ appears to be tracing the evaporative flow of \ovi\ from a 
young interface between warm gas and a hot exterior medium. However, it is 
possible the positive velocity \ovi\ is instead tracing cooling hot Local Bubble  (LB) gas.   The 
properties of the \ovi\ absorption in the LISM are broadly consistent with the 
expectations of the theory of conductive interfaces caught in the old condensing phase 
and possibly in the young evaporative phase of their evolution. 

\end{abstract}


\section{Introduction}
An important source of information about hot gas in interstellar space
is the lithium-like \ovi\ ion with doublet transitions in the
far-ultraviolet at 1031.93 and 1037.62 \AA.  The conversion of \ov\ into
\ovi\ requires 113.9 eV.   Since normal hot stars rarely have large
radiative fluxes at energies exceeding the He$^+$ absorption edge at 54.4
eV, the \ovi\ found in the interstellar medium is most likely created
by electron collisions in hot gas rather than by photoionization in
warm gas.   For gas in collisional ionization equilibrium, \ovi\ peaks
in abundance at $T = 2.8\times 10^5$ K with an ionization fraction, 
$n($\ovi$)/n({\rm O}) = 0.22$ (Sutherland \& Dopita 1993).  Gas with
temperatures in the range from T $\sim 10^5-5\times 10^5$ K cools very rapidly and
is often referred to as ``transition temperature gas."  In fact,
emission by \ovi\ is the primary coolant of transition temperature gas. 
\ovi\ found in the interstellar medium (ISM),  therefore, likely traces regions where hot gas
is cooling through transition temperatures in cooling bubbles or in interface regions
where cool and warm gas  ($10^2$--$10^4$ K) comes into contact with hot
($10^6$ K) gas resulting in layers containing transition temperature gas.    

	The first studies of \ovi\ absorption in interstellar space with the
{\em Copernicus}\ satellite strongly suggested the existence of a hot phase
to the interstellar gas (Jenkins \& Meloy 1974: York 1974).  At the
same time, Williamson et al. (1974) recognized that the diffuse soft
X-rays observed by sounding rocket instruments were tracing X-ray
emission from hot gas in the ISM.   These ultraviolet and X-ray
observations confirmed the theoretical prediction of Spitzer (1956) of
the probable existence of hot gas in the ISM.  The hot gas was
proposed to provide the pressure confinement of neutral clouds found
in the low halo.  A subsequent survey of \ovi\ absorption toward 72
stars with the {\em Copernicus}\ satellite by Jenkins (1978a, b) established
the widespread existence of a hot phase of the ISM.   A recent analysis of
the physical implications of the Copernicus \ovi\ absorption line data
is found in Shelton \& Cox (1994). 

	Except for brief observing programs with the {\em Hopkins Ultraviolet
Telescope} (Davidsen 1993) and with the {\em Orbiting and Retrievable Far
and Extreme Ultraviolet Spectrometers} (Hurwitz et al. 1998; Widmann et
al. 1998), observational studies of \ovi\ in the ISM were not possible
from 1980 to 1999.  However, insights about transition temperature gas
were obtained by ultraviolet observations of \siiv, \civ, and \nv\ with 
the {\em International Ultraviolet Explorer}\ ({\em IUE}) satellite and the {\em Hubble Space
Telescope}\ ({\em HST}).  For reviews of this work and hot gas in the ISM see
Savage (1995) and Spitzer (1990).   The possible role of \ovi\ in the
three-phase ISM is reviewed by Cox (2005) while Indebetouw 
\& Shull (2004) consider various \ovi\ production processes possibly operating in the Galactic halo. 

	With the launch of the {\em Far Ultraviolet Spectroscopic Explorer}\ ({\em FUSE})
satellite in 1999  (Moos et al. 2000), we entered a new era in our
ability to study \ovi\ in hot gas in the ISM and IGM.  With a
resolution of 20 \km, a wavelength coverage from 912 to 1187 \AA, and
a high radiometric efficiency, {\em FUSE}\ has provided unique information
about the distribution and properties of interstellar and
intergalactic \ovi.   {\em FUSE}\ measurements now include observations of \ovi\ 
absorption in many astrophysical sites including: the LISM
(Oegerle et al. 2005),  the Galactic disk (Bowen et al. 2005), the low
halo (Zsarg\'o et al. 2003), the distant halo (Savage et al. 2000,
2003),  high velocity clouds (Sembach et al. 2000, 2003), the
Magellanic Bridge (Lehner 2002), the Large and Small Magellanic Clouds
(Howk et al. 2002;  Hoopes et al. 2002),  the outflow from starburst
galaxies (Heckman et al. 2002), and the intergalactic medium at low
redshift (Oegerle et al. 2000;  Tripp et al. 2000; Savage et al. 2002; 
Richter et al. 2004; Sembach et al. 2004; Danforth \& Shull 2005).  The
{\em FUSE}\ studies have also revealed the presence of \ovi\ emission from gas
in shocked interstellar regions in the Milky Way and LMC (Sankrit \&
Blair 2002; Sankrit, et al. 2004), the Milky Way halo (Shelton et al.
2001, Shelton 2002) and the halos of nearby starburst galaxies (Otte
et al. 2003). 

	The studies of \ovi\ in the different environments all require an
understanding of the physical processes controlling the production and
destruction of \ovi\ in hot plasmas.    While the studies of \ovi\ in
the more distant sites are of great interest, the interpretations of
the measurements are hampered by the large amount of line of sight
averaging that occurs when studying absorption phenomena over kpc to
hundreds of kpc distances.  By studying \ovi\ absorptions over the
smallest distance scales, we have the opportunity to investigate \ovi\
production in a small number of cool/warm-hot gas interface zones rather than
in the superposition of many such zones.  Over short distances it is
possible to test the ionic column density, line width, and kinematical
predictions of the models of conductive interfaces in
different stages of evolution and to search for other types of \ovi\ absorption sites.  

	 The local interstellar medium (LISM) provides an excellent
laboratory to probe \ovi\ production processes since the line of sight
distances to continuum sources for absorption line spectroscopy are
small ($\sim$40 to 200 pc). The LISM is considered to be comprised of the clouds and hot gas
within the boundary of the LB which has an asymmetric extension
ranging from $\sim$60 to 250 pc.   Extensive \nai\ absorption line studies
(Sfeir et al. 1999, Lallement et al. 2003) to map the extent of the LB reveal that the bubble
has a sharp boundary determined by a large gradient in the amount of
neutral gas.   The neutral outer boundary to the local bubble has been
traced by \nai\ (D2) absorption with equivalent widths exceeding 20 m\AA\ 
which corresponds to $\log N($\hi$) \sim 19.3$.  Models for the structure of
various clouds situated in the LB are discussed by Linsky et al.
(2000) and Redfield \& Linsky (2000).  In particular Fig.~1 of Linsky
et al. (2000) shows the directions where the line of sight to stars in
the LB pass through the Local interstellar cloud (LIC), the G cloud,
the north Galactic pole cloud, and the south Galactic pole cloud.  The
distinct cloud identifications suggested by Linsky et al. (2000) are
based on the kinematical properties of UV absorption lines seen in the
spectra of stars sampling the various cloud directions. 

	The {\em Copernicus}\ satellite probed \ovi\
absorption toward several relatively nearby stars in the LISM (York
1974; Jenkins 1978a). However, {\em FUSE}\ has completely opened up the range of
possible studies.  Because of its high radiometric efficiency,
measures of \ovi\ absorption toward relatively close-by WDs
with {\em FUSE}\ can be achieved in integration times ranging from 10 to 50
ksec.   Oegerle et al. (2005) have reported a survey of \ovi\
absorption in the LISM toward 25 white dwarfs ranging in distance from
37 to 224 pc.   They reported \ovi\ detections with $\ge 2 \sigma$ significance
toward 14 of the 25 stars with \ovi\ column densities ranging from
somewhat less than $10^{13}$ cm$^{-2}$ to $1.7\times 10^{13}$ cm$^{-2}$.   Oegerle et al. (2005)
found an  average \ovi\ volume density in the LISM  of $2.4\times 10^{-8}$ cm$^{-3}$,  
which is slightly larger than the medium value of  $1.7\times 10^{-8}$ cm$^{-3}$
measured over kpc distances in the Galactic disk by {\em FUSE}\ (Bowen et al.
2005).   Oegerle et al. (2005) attribute the \ovi\ absorption they
observe in the LISM to the evaporative interface regions between warm
clouds and hot diffuse gas in the Local Bubble (LB).  The size and
shape of the LB are roughly estimated from the intensity of the
diffuse soft X-ray background (Snowden et al. 1998) and the extent of
the local region of very weak absorption in the \nai\ D lines at
optical wavelengths (Sfeir et al. 1999, Lallement et al. 2003). 

	Since the LISM \ovi\ observations discussed by Oegerle et al. (2005)
were obtained and analyzed, there have been many additional
observations obtained by {\em FUSE}\ of WDs in the LISM.  The new
observations provide information about \ovi\ absorption along
additional lines of sight and provide improved signal-to-noise (S/N)
spectra for many of the lines of sight studied by Oegerle et al.  In
addition, the {\em FUSE}\ data reduction and handling techniques have
improved since the version 1.8.7 of the {\em FUSE}\ data reduction pipeline
used by Oegerle et al.   With these improvements in S/N and data
extraction and calibration, it is now possible to not only determine
the column densities of \ovi\ toward the WDs but in the higher
S/N cases to obtain important information about \ovi\ absorption line 
velocities and  line widths.   This allows an evaluation of how the \ovi\ 
absorption profiles relate  to the absorption  produced by cooler
gas along the line of sight to each white dwarf.  Since conductive
interface models make specific predictions about the expected \ovi\
line widths and velocity offsets from the cooler gas producing the
interface  (B\"ohringer \& Hartquist 1987; Slavin 1989; Borkowski, Balbus
\& Fristrom 1990), it is important to fully quantify the observed
character of the \ovi\ absorption in order to test the interface model
predictions.

This paper is organized as follows:  In \S2 we discuss the sample of
WDs observed by {\em FUSE}\ and the LISM structures they probe.  In
\S3.1 the {\em FUSE}\ observations and data handling techniques are discussed. 
In \S3.2 and \S3.3 problems associated with stellar and circumstellar
contamination are discussed.  In \S3.4 we consider the effects of the
WD gravitational redshift.  In \S3.5 the techniques for determining
column densities, line velocities, line widths and their errors are
discussed. Comments about absorption along individual lines of sight
are found in the Appendix.  The observed properties of the LISM \ovi\
absorption are presented in \S4.  The processes responsible for the presence of \ovi\ in the 
LISM are considered in \S5 with the emphasis being on
observational tests of the theory of conductive interfaces.  An
extension of the results found for the LISM to more distant lines of
sight is discussed in \S6.   The results are summarized in \S7.
The appendix gives detailed comments about results for selected WDs from the 
sample.

\section{The Sample}
We have studied \ovi\ absorption in the {\em FUSE}\ spectra of the 46 
WDs listed in Table~1.  The sample includes most WD
measurements available through the publicly accessible {\em FUSE}\ data
archive up to 1 December 2004 but excludes observations where stellar
line blending was previously known to severely compromise the
interstellar \ovi\ measurement (see \S3.2) or where the signal-to-noise
level is not adequate for the LISM \ovi\ program.  Our sample has
considerable overlap with the sample studied by Oegerle et al. (2005),
although we use the most recent {\em FUSE}\ extraction and calibration
software and when possible have obtained measurements with higher
signal-to-noise by incorporating {\em FUSE}\ observations obtained after  the
2003 cut off date of the Oegerle et al. data processing.  

The various entries in Table~1 include the WD 1950.0 coordinate name
and other names, the Galactic latitude, the Galactic longitude, the
distance in pc, the estimated effective temperature of the WD,  the WD
type,  the stellar heliocentric velocity, $v_\star$,  and  the  heliocentric
velocity of  low ionization LISM absorption, $v_{\rm LISM}$.  The sources for
these quantities are given in the footnotes to the table.    The
distances with listed, errors are from Hipparcos parallax measurements
except for WD\,1314+293 (HZ 43) where we adopt the distance and error
from Van Altena et al.  (1995).  When no distance error is listed the
distance is from photometry and the uncertainty is typically  $\pm 20$--30\%. 
The absolute velocities from the literature listed in Table~1 are
based on measurements from the {\em IUE}\ satellite (reference 1) or the {\em HST}\
(references 3, 4, and 5) and are more reliable than absolute velocities
from {\em FUSE}\ given in parenthesis  (this work, reference 2).  If no
stellar metal lines are detected in the {\em FUSE}\ observations, ``NM" for no
metals is listed in the column for the stellar heliocentric velocity
(see \S3.2). 

The distribution of the observed WDs on the sky for the 39 stars for
which the LISM \ovi\ absorption could be measured or upper limits
determined is shown in Fig.~1 where the symbol sizes are inversely
proportional to the distance to each object and the symbol grey scale
shading denotes the measured column density of \ovi\  (see \S4.1). 
Seven of the 46 stars are not included in Fig.~1 because stellar
contamination of the LISM absorption was judged to be likely (see
\S3.2).   The directions where LISM \ovi\ is detected with $\ge 2\sigma$
significance are displayed as circles with non-detections displayed as
triangles.   The 39 WDs range in distance from 37 to 230 pc with a
median distance of 109 pc.

For several of the brighter WDs in the sample, high resolution spectra
($\lambda/\Delta \lambda \sim 100,000$ and $\sim 40,000$) from the {\em HST}\ 
obtained with either the Goddard High Resolution Spectrograph (GHRS) or the Space Telescope
Imaging Spectrograph (STIS) exist which provide information about the
number of discrete LISM absorbing components along the line of sight
(see Redfield \& Linsky 2002, 2004).  Also, in several cases very
detailed analysis of the LISM absorption toward selected WDs has been
performed with {\em FUSE}\ measurements in order to determine D/H. In some of
these cases the analysis has revealed the existence of multiple
absorbing components.  The component information is important for
studying the possible occurrence of \ovi\ in the cloud boundary
interface regions.  Comments about the WDs with supporting high
resolution {\em HST}\ measurements or other observations revealing component
structure are given in the Appendix as part of the detailed comments
about the \ovi\ absorption toward many of the survey objects.   

The picture that emerges from the detailed studies of LISM absorption
is that although the lines of sight to the WDs in the LISM extend over
relatively small distances, multiple structures are often encountered.  For example, 
along many lines of sight the LISM absorption will be a superposition of absorption in 
several different structures including the LIC or the G cloud and other more distant
clouds. However, in other directions the line of sight structure is very simple.

\section{Observations and Measurements}

\subsection{The {\em FUSE}\ Spectroscopic Data Handling}

Descriptions of the {\em FUSE}\ instrument  design and  inflight performance
are found in Moos et al. (2000) and Sahnow et al. (2000).  The
instrument consists of four channels.  Two are optimized for short
wavelengths (SiC\,1 and SiC\,2; 905-1100 \AA)  and two for longer
wavelengths (LiF\,1 and LiF\,2; 990-1185 \AA).  The four channels place
spectra on two delay line UV detectors and the resulting {\em FUSE}\ data set
consists of eight $\sim$90 \AA\ wide spectra identified according to the
segments:  LiF\,1A, LiF\,2A, LiF\,1B, LiF\,2B, SiC\,1A, SiC\,1B, SiC\,2A, SiC\,2B. ).  The
LiF\,1A, LiF\,2B, SiC\,1A, and SiC\,2B segment spectra cover the wavelength
region of the \ovi\ $\lambda$$\lambda$1031, 1037 doublet.  However, we have
chosen not to use the SiC\,2B data because of the strong fixed pattern
noise and relatively low resolution of observations near \ovi\ when
using this segment.  

Observations in the SiC\,1A and LiF\,2B segments have  $\sim$30\% and $\sim$60\%  of
the sensitivity of the LiF\,1A segment, respectively.  The \ovi\ ISM
absorption toward WDs is relatively weak.  We have combined the
measurements from the three different segments to obtain as high a
signal-to-noise (S/N) as possible in the final spectrum even though
the spectral resolution is degraded from 20 to approximately 23 \km\ when combining spectra. 
In a number of cases early in the {\em FUSE}\ mission, SiC measurements were
not obtained because of misalignment of the various channels.

The data set we have processed is listed in Table~2 where we give WD
name, the {\em FUSE}\ observation ID, the {\em FUSE}\ aperture, the segments that
have been combined into the final co-added spectrum, and the
integration times for the segments.  The observations were mostly
obtained in the LWRS (30\arcsec) aperture with a few observations using the
MDRS (4\arcsec) aperture and 1 line of sight (WD\,1634--573) 
obtained through the HIRS ($1\farcs25$).  Some of the more recent observations obtained
with program ID P204 were designed to place the WD at different
locations in the LWRS aperture to reduce the effects of detector fixed
pattern noise.  

The current version of the calibration pipeline software (version
2.0.5 and higher) was employed to extract and calibrate the
observations.  The details of the processing steps are described by
Lehner et al. (2003).  The extracted spectra for the separate
exposures and segments were aligned in wavelength by a
cross-correlation technique and co-added.  The cross-correlation
provides an excellent alignment because of the presence of strong \cii\ 
and \oi\  ISM lines near the weak \ovi\ lines.  The combined
observations were binned to 4 pixels which corresponds to $\sim$0.025 \AA\ or
$\sim$7.5 \km.  This provides $\sim$3 samples per 20--23 \km\ resolution
element. 

When possible, the zero point in the final wavelength scale was
established by forcing the ISM lines of \cii\ $\lambda$1036.34 to have the LISM
heliocentric velocities found from the more accurate LISM velocities 
from the literature  listed in Table~1 (references 1, 3, 4 or 5). 
However, when an accurate LISM heliocentric velocity was not
available, we simply adopted the heliocentric wavelengths and
velocities provided by the standard {\em FUSE}\ pipeline processing of the
LiF\,1A channel observations.  Those velocities are also listed in Table~1
but inside parenthesis.   The absolute velocities will therefore be
more accurate for those observations incorporating the velocity
corrections to the literature.   However, since we are primarily
interested in the velocity similarities and differences between the \ovi\ 
absorption and tracers of the neutral gas (\oi\ and \cii), this
approach is quite acceptable because the velocities of \ovi\ relative
to \oi\ and \cii\ are reliable in all cases.  Independent studies of \ovi\
absorption toward extragalactic objects by Wakker (private
communication) have shown that when using Version 2.05 or later of the
data reduction software, the two \ovi\ $\lambda$$\lambda$1031.93, 1037.62 lines are
aligned in velocity to better than $\sim$5 \km\ when using the standard
extractions.  Since the \ovi\ $\lambda$1037.62 line lies between the ISM \oi\ 
$\lambda$1039.23 and \cii\ $\lambda$1036.34 lines, it appears that the standard
reduction process produces a relative velocity calibration of the 
\ovi, \oi, and \cii\ absorption lines that is accurate to $\sim$ 5 \km. 
This is a substantial improvement over the 10 \km\ velocity error
between the two \ovi\ lines present in the earlier version 1.8.7 of the
calibration pipeline (see Fig.~2 in Wakker et al. 2003).  

Since the path lengths through the LISM are through relatively low
density gas, we do not need to worry about possible contamination of
the \ovi\ $\lambda$1031.93 measurements from the H$_2$  (6-0) P(3) and R(4)  lines
at  1031.19 and 1032.36 \AA, respectively.  Similarly the \ovi\ $\lambda$1037.62
absorption is not contaminated by the H2 (5-0) R(1) or P(1) lines at
1037.15 and 1038.62 \AA, respectively.

\subsection{The Effects of  WD Gravitational Redshifts}

 For the WDs listed in Table~1 with detected stellar metal line
absorption, it is possible to compare the stellar velocity to the
velocity of neutral and weakly ionized gas in the LISM.  If the nearby
WDs of our sample and the gas of the LISM roughly share a common
pattern of motions we would expect the values of the stellar velocity,
$v_\star$,  to exceed that for gas in the LISM by an amount equal to the WD
gravitational  redshift.  Gravitational redshifts of WDs have mostly
been studied by observing the radial velocities of  WDs  and
non-degenerate stars in wide binary systems.   In a recent study, 
Silvestri et al. (2001) determined gravitational redshifts for 41 DA
WDs by this method and found a distribution of redshifts extending
from 8.8 to 58.3 \km\ for 36 WDs and from 103.6 to 132 \km\ for 5
WDs. We find the  average velocity and velocity dispersion for the
lower redshift Silvestri et al. sample are $33.2 \pm 11.0$ \km.   With
such a large average gravitational redshift, the effect should be
easily seen in a plot of  the stellar velocity versus the LISM
velocity for a large sample of WDs near to the Sun.  In Fig.~2a we
compare the values of $v_\star$ and $v_{\rm LISM}$ listed in Table~1 and in 
Fig.~2b display a histogram of number versus $v_\star - v_{\rm LISM}$.  Although the
absolute values of  $v_\star$ and $v_{\rm LISM}$ are not well determined from our {\em FUSE}\ 
observations, the relative velocity is accurate to $\sim$ 5 \km\ when the
WD stellar reference lines are the  \piv\ absorptions which are near
the LISM \cii\ $\lambda$1036.34 absorption we use to trace neutral and weakly
ionized LISM gas.  We see from  Fig.~2  that $v_\star$  exceeds  $v_{\rm LISM}$  in
all cases except one  and the accompanying histogram of  number versus
$v_\star - v_{\rm LISM}$  can be interpreted as an approximate  histogram of
gravitational redshifts modified somewhat by the different kinematics
of gas and stars in the LISM near to the Sun.   The average and
dispersion of $v_\star - v_{\rm LISM}$ we obtain for the 18 stars is $35.2 \pm  20.4$ \km\  
which is similar to the average obtained from the Silvestri et al.
(2001) sample.  It is interesting that the average is similar since the
Silvestri et al. (2001) sample is for relatively cool WDs with  median 
temperatures  $\sim$10,000 K while our sample is for relatively hot WDs
with median  temperatures $\sim$40,000 K.   A study of this effect could be
done more carefully with a larger sample of WDs.   However, the
measurements are clearly revealing the presence of the WD
gravitational redshifts with a simple reference to the rest frame of
the LISM.   The important result for our investigation of \ovi\ in the
LISM is the fact that the WD stellar contamination present in our data
set will generally be $\sim$20 to 50 \km\ redward of the LISM absorption
because of the effects of the gravitational redshift.

\subsection{Stellar Contamination}

Stellar blending of the LISM \ovi\ absorption in the spectra of WDs is
a serious problem (Oegerle et al.  2005).   The contaminating lines
are stellar \ovi\ $\lambda$$\lambda$1031.93 and 1037.62 and in some situations stellar
\feiii\ $\lambda$1032.12 which lies 49 \km\ redward of the stellar \ovi\ $\lambda$1031.93 line.  

For WDs with $T_{\rm eff} <  40,000$ K, the  NLTE model atmosphere calculations
of  Oegerle et al. (2005; see their Fig.~5)  suggest that the stellar
\ovi\ absorption should be very weak provided the oxygen abundance is
less than $\log ({\rm O/H}) = -5.0$.   Above 40,000 K,  the stellar \ovi\ will be
weak or strong depending on the abundance of oxygen in the
photosphere.  Contamination from \feiii\  near \ovi\ can occur in the
cooler WDs.  When stellar \ovi\ or \feiii\ is present, stellar lines of
other elements and ions will also generally be observable in the
wavelength region covered by the {\em FUSE}\ spectra.   The Oegerle et al.
model atmosphere calculations for particular WDs with stellar
comtamination in the lines of \ovi\ and \feiii\ shown in their Fig.~6
illustrate some of the absorption features accompanying stellar \ovi\ 
and/ or stellar \feiii.  In the higher temperature WDs stellar \ovi\ is
accompanied by stellar \piv\ $\lambda$$\lambda$1030.51, 1033.11.  In the four hot WD
cases shown by Oegerle et al., the \piv\ $\lambda$1031.51 line has a strength
relative to \ovi\ $\lambda$1031.93 ranging from $\sim 3 \times$ stronger to $\sim 2 \times$ weaker
although there are WDs with extremely strong stellar \ovi\ with no
detectable \piv\ (see Fig.~3 for WD\,2156--545 in Oegerle et al. 2005 and
Fig.~2 for WD\,0131--163 in Lehner et al. 2003).  However, WD\,0131--163
also shows strong stellar \siiv\ $\lambda$1066.63 and \pv\ $\lambda$1117.98.   This
suggests that a first evaluation of the possibility of stellar \ovi\ 
contamination can be based on the presence or absence of absorption by
\piv\ $\lambda$$\lambda$1030.51, 1033.11, \pv\ $\lambda$1117.98, and \siiv\ $\lambda$1066.63.  For \siiv\ 
$\lambda$1066.63,  allowance must be made for ISM contamination from \ari\ 
$\lambda$1066.66 which can be evaluated with reference to the blend free  ISM
Ar I $\lambda$1048.22 absorption.   The possible stellar contamination from \feiii\ 
$\lambda$1032.12 can be evaluated with reference to other stellar \feiii\ 
lines including \feiii\ $\lambda$$\lambda$1030.92, 1033.30. 

For all the stars in our program listed in Table~1 we produced
velocity plots showing a combination of potential stellar and
potential ISM absorption lines including \piv\ $\lambda$$\lambda$1030.51, 1033.11, \pv\ 
$\lambda$1117.98, \feiii\ $\lambda$$\lambda$1030.92, 1032.12, 1033.30, \siiv\ $\lambda$1066.63,  
\ari\ $\lambda$$\lambda$1048.22, 1066.66, \feii\ $\lambda$1144.94,  and \ovi\ $\lambda$1031.93.  
The ISM \feii\ $\lambda$1144.94 observation allows us to calibrate the velocity of the \pv\ 
$\lambda$1117.98 stellar absorption with the detector segments containing the
\ovi\ absorption.   An evaluation of the velocity plots for  the
presence of  stellar absorption lines allowed us to identify the
possibility of stellar contamination of the LISM \ovi\ $\lambda$1031.93
absorption and the velocity of the contamination.  Additional
information about possible stellar line blending can be inferred from
the WD studies of Holberg et al. (1998) and Bannister et al.  (2003)
of spectra in  the wavelength range from $\sim$1150 to $\sim$1850 \AA\ obtained by
the  {\em IUE} satellite and/or the spectrographs flown on the {\em HST}\ including
the  GHRS and STIS. 

In Table~1 we list values of the WD stellar velocity, $v_\star$, when known
either from the literature (Table~1 reference 1, 3, 4, and 5) or from
the {\em FUSE}\ observations of one or more of the stellar lines of  \piv, \pv\, 
\siiv, and \feiii\ listed above (Table~1  reference~2).  Velocities
based on the {\em FUSE}\ observations alone are given in parenthesis since
the absolute velocity is not well determined but the relative
velocities are accurate.   In a large number of cases denoted with
``NM" in Table~1, no stellar metal lines have been seen with {\em FUSE}\ or
{\em IUE}\ or {\em HST}.  

Based on our evaluation of the stellar/interstellar absorption line
velocity plots and other sources of information about stellar
contamination we have rejected from our LISM study the following 7 WDs
from Table~1 because of likely or possible contamination of the LISM \ovi\
$\lambda$1031.92 absorption by stellar \ovi\ absorption.  We give the
reasons for the rejection in the following brief discussions of each
object. 

WD\,0027--636  ($T_{\rm eff} = 63,724$ K).  The \piv\ and \feiii\ lines are not
detected. However, the \siiv\ and \pv\ absorption is very strong and
occurs near 30.2 \km\ which is  close to the velocity of the
observed \ovi\ absorption at $23.2 \pm 4.4$  \km.   A substantial amount
of stellar \ovi\ contamination is likely given the high stellar
temperature. 

WD\,0050--332  (GD 659) ($T_{\rm eff} = 36,000$ K).   We detect strong \pv\ but do
not detect \piv\ or \siiv.  Bannister et al. (2003) note that GD 659
has photospheric  absorption at $34.3  \pm 0.17$ \km\  and possibly weaker
circumstellar absorption at $-2.97 \pm 3.00$ \km.   The photospheric
absorption is seen in \siiv, C IV and N V in STIS and {\em IUE}\ spectra and
in \pv\ in the {\em FUSE}\ observations. This photospheric absorption is well
aligned with the \ovi\ we detect at $39.8 \pm 3.1$ \km\ and we conclude
stellar \ovi\ contamination is possible even though $T_{\rm eff}  <  40,000$ K. 
This is a conservative conclusion since O V $\lambda$1371.30 is evidently not
seen in the {\em IUE}\ spectrum discussed by Holberg et al. (1998). 

WD\,1942+499 ($T_{\rm eff} = 34,400$ K).  Strong \siiv\ $\lambda$1066.63 absorption is
detected near $-5$ \km\ which is close to the \ovi\ absorption velocity
of $-0.8 \pm 5.1$ \km.  Stellar \ovi\ contamination is possible even
though $T_{\rm eff} <  40,000$ K. 

WD\,2111+498 ($T_{\rm eff} = 37,360$ K). Very strong stellar \piv, \siiv, and \feiii\
are found at 28.9 \km\ which is near the velocity of  the
observed \ovi\ absorption at $21.1 \pm 8.4$ \km. The \feiii\ $\lambda$1032.12
feature near the stellar or interstellar \ovi\  is also evident in the
spectrum.  Since the \piv\ and \siiv\ absorption is so strong, the $W_\lambda =
9.5 \pm 3.1$ m\AA\   \ovi\ $\lambda$1031.93 feature detected is probably contaminated
by stellar \ovi\ even though $T_{\rm eff} <  40,000$ K. This is a conservative
conclusion since \ov\ $\lambda$1371.30  is evidently not seen in the {\em IUE}\ 
spectrum studied by Holberg et al. (1998).

WD\,2000--561 ($T_{\rm eff} = 47,299$ K).  Very strong stellar \piv\ and \siiv\ are
found at $-15.4$ \km\  which is close to the $-19.6 \pm 4.1$ \km\ velocity
of  the observed \ovi\ absorption.  Given the large value of $T_{\rm eff}$,
stellar \ovi\ contamination is likely.  

WD\,2146--433  ($T_{\rm eff} = 67,912 K$). Very strong stellar \piv\ and \siiv\ are
found at 27.0 \km\ which is near  $13.5 \pm 6.4$ \km,  the observed \ovi\ 
velocity. Given the high value of $T_{\rm eff} $, stellar \ovi\ contamination
appears likely.  

WD\,2321--549 ($T_{\rm eff} =45,860$ K). Very strong stellar \piv\ and \siiv\ are
found at 9.9 \km.   Approximately 50\% of the observed \ovi\ 
absorption occurs near this velocity and is likely stellar in origin
because of the large value of $T_{\rm eff} $.  Rather than trying to separate
the stellar and interstellar \ovi\ absorption, the object was rejected
from inclusion in the final LISM sample. 

We also rejected from consideration and inclusion in Table~1 the other
WDs  that Oergele et al. (2005) have previously shown to be
contaminated.  These include WD\,0501+527, WD\,2156--546, WD\,2211--495,
and WD\,2331--475. 

In several cases, we identify strong stellar features in  the {\em FUSE}\ WD
spectrum but have retained the star for our LISM \ovi\ study.   The 4
stars include: 

WD\,0455--282 ($T_{\rm eff} =57,200$ K).  Strong stellar \piv, \siiv, \civ\ and \nv\
absorption is seen at 69.6 \km\ which is well displaced from the
LISM \ovi, \cii, and \oi\ absorption.  \ov\ $\lambda$1371.30 is evidently not
seen in the {\em IUE}\ spectrum studied by Holberg et al. (1998) although
stellar \ovi\ near 80 \km\ is evident in the {\em FUSE}\ spectrum shown in
Fig.~3. 

WD\,0603--483 ($T_{\rm eff}=35,322$ K).   Stellar \pv\ is detected near 41 \km\ 
but is not aligned with the \ovi\ absorption at $25.6 \pm 7.1$ \km.  

WD\,1634--573 (HD\,149499B) ($T_{\rm eff} = 49,500$ K).  We detect what could be \piv\
$\lambda$1033.11 near  $-15$ \km\ but do not see the corresponding \piv\
$\lambda$1030.51 absorption.   Holberg et al. (1998) report stellar \siiv\ and
\civ\ at $v_\star = 0.61 \pm 2.15$ \km\ and LISM absorption at $v_{\rm LISM} = -19.6$ \km.  
The \ovi\ absorption we see toward this star is well aligned with the
LISM absorption of \cii\ and \oi.   Although some stellar \ovi\ 
contamination may be present, it does not appear to be affecting our
measurement. 

WD\,1950--432 ($T_{\rm eff}= 41,339$ K). Stellar \pv\ is detected at 40 \km\ 
but  it is not aligned in velocity with the \ovi\ absorption at
$-4.9 \pm 7.3$ \km. 

There are a number of stars that potentially have stellar
contamination of \ovi\ as judged from the stellar/ISM velocity plots
but for which no \ovi\ (stellar or interstellar) has been detected. 
Since these objects provide good upper limits to the interstellar
value of $\log N($\ovi),  they have been retained for our LISM study. WDs
in this category include:

WD\,0416+402 ($T_{\rm eff}= 35,227$ K).  Strong stellar \feiii\ $\lambda$1030.92 and 
\siiv\ $\lambda$1066.63  are found near 79 \km.  However, stellar or ISM \ovi\ 
is not detected. 

WD\,0802+413 ($T_{\rm eff} = 45,394$ K).   Photospheric \pv\ $\lambda$1117.98 is strong
at 59 \km.  However, stellar or ISM \ovi\ is not detected.  

We believe it is unlikely that the 39 WDs retained for our LISM study
are seriously affected by stellar \ovi\ blending.  However, it would
take a very extensive stellar spectroscopic investigation, well beyond
the scope of our ISM study, to actually prove in all cases that
stellar blending can be ignored.

\subsection{Circumstellar Contamination}

Non-photospheric circumstellar absorption has been found in the
spectra of 5 of 44 DA WDs  observed by  Holberg et al. (1998).  An
additional four examples have been identified by  Bannister et al.
(2003)  among a group of 23 WDs.  The features have shown up as highly
ionized ions (\siiv, \civ and \nv) seen in {\em IUE} spectra (Holberg et al.
1998) and/or STIS spectra (Bannister et al. 2003). A number of
origins for these features include photoionization of the local
interstellar environment of the WD (Dupree \& Raymond 1983), matter in
the WD gravitational potential well, mass loss in a WD wind, matter in
a surrounding ancient planetary nebula, and matter from a companion
star.   One object on our list, WD\,0050--332 (GD 659) with possible
circumstellar contamination, has already been discussed \S3.2 and has
been rejected from our LISM study because of the possibility of strong
stellar contamination. 

WD\,0455--282 with $T_{\rm eff} = 57,200$ K is another star in our sample
possibly having circumstellar contamination problems.   Holberg et al.
(1998) note that \siiv\ and \civ\ absorption is seen in {\em IUE} spectra at 
$v= 16.21  \pm 2.66$ \km\  while the photospheric velocity is $69.60 \pm 1.97$ 
\km.  Photospheric \ovi\ is detected in this WD but at a velocity well
removed from the LISM absorption.  While circumstellar contamination
is potentially a problem, our observations show the \ovi\ absorption
near 16 \km\ provides only a small ($\sim$10--15 \%) contribution to the
total \ovi\ absorption.  

We are not aware that any of the other stars in our sample have
been identified as possibly having circumstellar contamination
problems.  While the frequency of occurrence of circumstellar
contamination appears to be relatively low, we need to be aware that
circumstellar  contamination may adversely affect some of the WDs in
our LISM study.

\subsection{Measurements of  \ovi\ Column Densities, Velocities, and Line
Widths}

Continuum normalized absorption line profiles for the lines of \ovi\ 
$\lambda$1031.93, \cii\ $\lambda$1036.34, and \oi\ $\lambda$1039.23 displayed on a heliocentric
velocity scale are shown in Fig.~3 for the 24 WDs for which the LISM
\ovi\ $\lambda$1031.93 line has been detected with $\ge 2\sigma$ significance along with
profiles for 4 WDs where no \ovi\ has been detected.  Absorption lines
produced by contaminating stellar features are identified on several
of the panels.  In Fig.~4 we display the line profile plots for the
7  WDs for which we believe the LISM \ovi\ absorption is likely
contaminated by stellar \ovi\ as discussed in \S3.2. 

\ovi\ absorption line measurements are listed in Table~3 for all the
objects listed in Table~1.  The \ovi\ $\lambda$1031.93 absorption line
equivalent width, $W_\lambda$, is listed in column 9 with the heliocentric
velocity range of the integration listed in column 11. When we believe
stellar contamination is likely affecting the measurements as
discussed in \S3.2, the equivalent width of the \ovi\ $\lambda$1031.93 line is
given in parenthesis.

The continuum placement procedure for measuring the absorption
utilized Legendre polynomials and the algorithm described by Sembach \&
Savage (1992) which provides an estimate of the continuum uncertainty. 
The listed  $\pm 1\sigma$ errors for the equivalent widths in Table~3 are from a
quadrature addition of the statistical error and the continuum
placement error.  The fixed pattern noise errors for these combined
observations from multiple segments are estimated to be $\sim$2--3 m\AA.  When
an equivalent width and error is listed in Table~3, the detection
significance is  $\ge 2 \sigma$. When LISM \ovi\ is not detected with a
significance of 2$\sigma$ or greater we list the 2$\sigma$ upper limit to the value
of $W_\lambda$.  The actual significance of the detection, $W_\lambda/\sigma$, is given in
column 10. 

Two methods have been used to measure the \ovi\ absorption velocities,
line widths, and column densities listed in Table~3.  For method 1,
the entries for $\bar{v}_f$, $b_f$, $\log N_f$, in columns 2, 3, and 4 are based on
Voigt profile fits of single absorption components with velocity, $\bar{v}_f$, 
Doppler parameter, $b_f$, and logarithmic \ovi\ column density, $\log N_f$. 
The fit results and the errors were obtained using the Voigt component
fitting software of Fitzpatrick \& Spitzer  (1997).  The fit results 
listed are based on a simultaneous fit to both members of the \ovi\
$\lambda$$\lambda$1031.93, 1037.62 doublet even though the $\lambda$1037.62 line is only
clearly detected with $\ge 2 \sigma$ significance in 8 of the 30 WDs (see Table~4). 
The {\em FUSE}\ instrumental function was assumed to be a Gaussian with
FWHM\,$= 20$ \km.  The errors on $b_f$  are large for the lower
significance detections of the $\lambda$1031.93 line with $2\sigma < W_\lambda < 4\sigma$.  In
those cases the values of $b_f$ are not listed in Table~3.
Two examples of the simultaneous component fit process are shown in
Fig.~5. These examples are for the well-observed stars WD\,1254+223
and WD\,1314+293. 

For method 2, the entries for $\bar{v}_a$, $b_a$, $\log N_a$, in columns 5, 6, and 7
in Table~3  are based on the apparent optical depth (AOD) method of
Savage \& Sembach (1991) applied to the $\lambda$1031.93 absorption line.  The
\ovi\ absorption profiles were converted into apparent column densities
per unit velocity $N_a(v) = 3.768\times 10^{14} \ln[F_c/F_{\rm obs}(v)]/(f\lambda)$, where $F_c$ is
the continuum flux,  $F_{\rm obs}(v)$ is the observed flux as a function of
velocity,  $f$ is the oscillator strength of the absorption and $\lambda$ is in
\AA.  The values of $\bar{v}_a$, $b_a$, $\log N_a$, listed in columns 5, 6, and 7 are
obtained from 
$\bar{v}_a= \int v N_a(v)dv / N_a $,	$b_a = [2 \int  (v-\bar{v}_a)^2 N_a(v)dv / N_a]^{1/2}$, and 
$N_a= \int N_a(v)dv$. 

Again for the lower significance detections with $2\sigma < W_\lambda < 4\sigma$, the
errors on $b_a$ are so large the values of $b_a$ are not listed.  In
those cases where \ovi\ $\lambda$1031.93 absorption was not detected  with a
significance $\ge 2 \sigma$, we report the 2$\sigma$ upper limit to the column density 
in the column labeled $\log N_a$ in Table~3 by  assuming the line is on the
linear part of the curve of growth. This is a valid assumption given the breadth and weakness of the \ovi\ absorption.  The $f$-values for the \ovi\ 
$\lambda$$\lambda$1031.93, 1037.62 lines of 0.133 and 0.0658 used in our derivations
of the \ovi\ column density are taken from Morton (2003).

For the 8 WDs for which the \ovi\ $\lambda$1037.62 line was also detected with
a significance $\ge 2 \sigma$ the equivalent widths and ADO values of 
$\bar{v}_a$, $b_a$, $\log N_a$  are listed  in Table~4.  When both \ovi\ absorption lines are
detected the best estimates of the \ovi\ absorption properties are from
the simultaneous Voigt profile fits listed in Table~3 or from an
average of the two reported values of $N_a$. 	

The \ovi\ absorption lines are broad, relatively weak and are unlikely
to contain unresolved saturated structures.   Since \ovi\ peaks in
abundance at $T = 2.8 \times 10^5$ K, the thermal Doppler width of a single
component containing \ovi\ is expected to be $b = 17.1$   \km\ 
corresponding to FWHM\,$= 28.5$ \km.  With a resolution of 20--23 \km, 
{\em FUSE}\ is expected to nearly fully resolve the \ovi\ absorption
implying the observed absorption profiles are probably a good
representation of the true \ovi\ absorption.  In the 8 cases where both
components of the \ovi\ doublet have been detected (see Table~3 and 4)
a comparison of the two values of $\log N_a$ based on the weak and strong
line of the doublet reveals good agreement.  This confirms that line
saturation problems are not a difficulty when deriving values of
$\log N($\ovi).   If saturation were a problem we would expect the value
of $N_a$ from the strong line to be smaller than the value from the weak
line. 

	In Fig.~6 we compare the results from the two methods we have used
to derive the \ovi\ absorption parameters for the WDs not affected by
stellar contamination.  In Fig.~6a the AOD and profile fit column
densities are compared.   The solid line shows the 1:1 relationship.  
The correspondence is very good.  However, we prefer the AOD results
since they are not dependent on assumptions about the line shape or
component structure.  In Fig.~6b we compare the AOD results from
Table~3 with the earlier measures of $\log N($\ovi) based on the
equivalent widths measured by Oegerle et al. (2005).  The agreement is
good, although in a number of cases our measurements have a higher
statistical significance since for some of the WDs our results are
based on combining more recent measurements with the older
observations studied by Oegerle et al.  Therefore, we were able to
convert several of the earlier upper limits into detections.  In
Fig.~6c we compare the AOD and profile fit values of the average \ovi\ 
velocity.  The correspondence between the two ways of measuring
$\bar{v}($\ovi) is good.   In Fig.~6d we compare the AOD and profile fit
results for $b($\ovi) for the more reliable cases, where the \ovi\ 
detection significance for the strong line of the doublet exceeds
4$\sigma$.  The two methods yield similar results
although the AOD method does not require assumptions about line shape
or component structure.   We conclude that both the profile
fit method and the AOD method yield good measures of $\log N($\ovi) and
$\bar{v}($\ovi).   The two methods provide reliable estimates of $b($\ovi)
provided $W_\lambda \ge 4\sigma $. 

	An important aspect of our investigation involves studying the
kinematical relationships between the absorption by LISM \ovi\ and
other tracers of gas in the LISM. We therefore compare the absorption
velocities of \ovi\ with those of \oi\ and \cii\ in Table~5 for the 24
WDs listed in Table~3 not affected by stellar contamination and with
$W_\lambda\ge 2 \sigma$. For all three ions we list the average absorption
velocity for the ion determined from the AOD method.  Table~5
therefore lists for each WD with detected \ovi\ absorption,  $\bar{v}_a($\ovi),
$\bar{v}_a($\cii), $\bar{v}_a($\oi), and $\bar{v}_a($\ovi$) -\bar{v}_a($\cii).  Table~1 identifies those cases where the absolute velocities are referenced to the more
reliable {\em IUE}\ or {\em HST}\ measurements (Table~1 references 1, 3, 4, and 5). 
However, velocity differences among \ovi, \oi, and \cii\ within the
{\em FUSE}\ observations listed in Table~5 are reliable in most cases since
the relative velocity calibration error is $\sim 5$ \km.

\section{Properties of \ovi\ Absorption}

\subsection{\ovi\ Column Densities}

Results are reported for \ovi\ LISM absorption unaffected by stellar
contamination toward 39 WDs in Table~3.   The \ovi\ $\lambda$1031.93 line is
detected with $\ge 2\sigma$ significance toward 24 of the 39 WDs.  If we
increase the required significance of a detection to $\ge 3\sigma$, these
numbers change to 17 of 39 WDs.   In 12 cases where the detection
significance $\ge 4\sigma$ it is possible to obtain high quality information
about the line width and shape.  For 8 of these higher significance
detections there is also reliable information provided by the \ovi\ 
$\lambda$1037.62 absorption line.  The \ovi\ column densities for the 24  $\ge 2\sigma$
detections  range from $\log N_a($\ovi$) = 12.38$ (WD\,1211+332) to 13.60 
(WD\,0809--728) with a median of 13.10.  In the following discussions we
will make use of the AOD value of $\log N($\ovi) for the stronger $\lambda$1031.93
absorption since no assumptions are required about profile shape or
component structure when determining a column density via the AOD
method. 

The distribution of \ovi\ absorption on the sky is illustrated with the
symbols and grey scale coding of Fig.~1.
The irregular distribution of \ovi\ on the sky is well illustrated in the figure.  The
most striking column density contrast is in the direction of the north Galactic pole
with  $b >80\degr$  where there are two WDs  including  WD\,1254+223 and WD\,1314+293 with
similar distances (67 pc) and similar column densities  ($\log N($\ovi$) = 13.10$ and 12.94).
The results for these two stars are in  stark contrast to the result for the third north
Galactic polar WD\,1211+332 with $d = 115$ pc and $\log N($\ovi$) = 12.38$.  The range in average line of sight \ovi\ volume density among these three stars is a factor of six.

In Fig.~7 $\log N_a($\ovi) is plotted against $\log d$ (pc) with the distance
to each WD taken from Table~1.  The filled circles are the $\ge 2\sigma$ 
detections and the open circles with the attached arrow are the 2$\sigma$
upper limits.   The open diamonds  (without or with limit arrows) show
the values of $\log N($\ovi) or  the $\sim$2$\sigma$ limits  from the {\em Copernicus}\
Satellite survey of \ovi\  (Jenkins 1978a) toward hot stars with $d <
1000$ pc.  There is a general trend of increasing $N_a($\ovi) with
distance, although the dispersion in the measurements indicates a
considerable amount of irregularity in the \ovi\ absorption in the LISM
and beyond.  The solid line in Fig.~7 shows the relation between
$\log N_a($\ovi) and $\log d$ corresponding to the average value of   $n_a($\ovi$)  
= 3.6\times 10^{-8}$  cm$^{-3}$  we obtain for  the {\em FUSE}\ WD sample (see below).  The
dashed line shows the median value, $1.7\times 10^{-8}$ cm$^{-3}$, obtained when the
{\em FUSE}\ measurements of \ovi\ in the Galactic disk toward 150 stars with
distances up to $\sim$8 kpc (Bowen et al. 2005) are combined with the
{\em Copernicus}\ measurements of Jenkins (1978a) of stars with distances up to $\sim$2
kpc. 

\subsection{\ovi\ Average Density}

Although the actual volume density distribution of \ovi\ in the ISM is unknown, it is useful to examine the behavior of the \ovi\ volume density, $n_a($\ovi$) = N_a($\ovi$)/d$, averaged along each WD line of sight. The volume density along different lines of sight and its dispersion can be studied as a function of position in the Galaxy and the values can be compared to theoretical expectations for different \ovi\ production processes.

The value for $n_a($\ovi) or the
density limit for each WD is listed in column 8 of Table~3 for 39 WDs.
The limit to $n_a($\ovi) for WD\,0113+002 assumes a WD distance of 100 pc. 
The values of $n($\ovi) for the 24 $\ge 2\sigma$ detections range from 
$(0.68-13.0)\times 10^{-8}$ cm$^{-3}$.
The distribution on the sky of the average line of sight physical
density is displayed in Fig.~8 with the same symbol coding that was
used in Fig.~1.  There is considerable irregularity to the
distribution of the values of $n_a($\ovi)  on the sky although  the
highest average density line of sight cluster in the general
direction toward  $l \sim270\degr$ and $b \sim -15 \degr$.  

Medians, averages, and dispersions for $n_a($\ovi)   derived from several
subsets of the 39 WD sample are listed in Table~6. The first entries
in Table~6 are the values for the 39 stars treating the 15  2$\sigma$ upper limits as detections.   The second entries are the values for the 24 stars where \ovi\ is detected with $\ge 2\sigma$ significance. 

Since 15 of 39 measurements are upper limits to  $n_a($\ovi), we need to
seek a better method to determine the average value of  $n_a($\ovi) from
our data sample.  Feigelson \& Nelson (1985) and Isobe et al. (1986)
discuss the Kaplan-Meier product-limit estimator method, which is a
nonparametric univariate survival analysis technique that can be used
to estimate the mean of a set of data points containing upper limits. 
The technique can be applied to variables with any underlying
distribution so long as the observations are a random sampling of the
distribution and the detections and limits are part of the same distribution.  
The Kaplan-Meier analysis technique applied to the 24
detections and 15  2$\sigma$ upper limits  yields  $n_a($\ovi$) = (3.6 \pm 0.5)\times 10^{-8}$ cm$^{-3}$, where the error is the Kaplan-Meier error on the average value rather than the dispersion about the average.   

The average density we estimate for \ovi\ is considerably larger than the value $2.4\times 10^{-8}$cm$^{-3}$  found by Oegerle et al. (2005) along the sight line to 25 LISM WDs.   \ovi\ was detected with greater than 2$\sigma$ significance toward  13 of the 25 stars in the Oegerle et al. sample.   The derived value of $n($\ovi) is sensitive to how the detections and  limits are included in the averaging process.  Oegerle et al. averaged all the data values for the 25 WDs in their sample treating non-detections as detections and even treating negative non-detections as negative densities in the averaging. Many of the non-detections have values of $n($\ovi) close to zero which introduces a large downward bias in the estimate of the average value of $n($\ovi).   The Kaplan-Meier analysis technique we have adopted for dealing with the upper limits is a much better approach for estimating the average value of \ovi. To test if there is a fundamental difference between the Oergerle et al. measurements and our measurements, we converted the Oegerle et al. column densities into a set of 13 detections and 12 2$\sigma$ upper limits.  After adopting our distances from Table 1, we determined \ovi\ volume densities and 2$\sigma$ upper limits.  We then used the Kaplan-Meier analysis technique to estimate the average volume density from the Oegerle et al. sample and found $n($\ovi$)= (3.1 \pm 0.4)\times 10^{-8}$ cm$^{-3}$ which is very similar to the average density we have obtained from our measurements. The estimate of the average value of the LISM \ovi\ volume density from the Oegerle et al. data set agrees with our value when the upper limits are properly included in the averaging process.

If we restrict our estimate of the average value of $n($\ovi) to the 17 WDs with $d < 100$
pc (including 9 $\ge 2\sigma$ detections and 8 upper limits) we obtain a sample dominated by WDs
within the LB.  If we again use the Kaplan-Meier analysis technique to evaluate the
effects of the upper limits on the average, we obtain $n($\ovi$) = (4.0\pm 0.8) \times 10^{-8}$  cm$^{-3}$ 
which is similar  to the value of $3.6\pm 0.5$  cm$^{-3}$ obtained for the entire sample of 39 WDs.

The average value of $n($\ovi$)= 3.6\times 10^{-8}$ cm$^{-3}$ we obtain  for \ovi\ in 
the LISM toward 39 WDs is  2.1  times larger than the  median value  $n($\ovi$)  =
1.7\times 10^{-8}$ cm$^{-3}$ obtained  for the much more distant sample of hot stars
surveyed by {\em FUSE}\ by Bowen et al. (2005). There is a
modest excess of \ovi\ in the LISM compared to the
more distant interstellar regions.

\subsection{$N($\ovi) vs $N($\oi)  in the LISM}

It is of interest to compare the observed column density  of \ovi\ to
that for other tracers of LISM gas along the line of sight to each WD. 
Currently, the largest body of measurements for other important
species toward our WD sample exists for \oi\ from Lehner et al. (2003).   
In Table~7 we list the WD,  distance from Table~1, $\log N_a($\ovi)
from Table~3,  and values of $\log N($\oi)  which are mostly from
Lehner et al. (2003).   \oi\ is an important tracer of gas in the LISM
because the measured value of $\log N($\oi),  is useful in determining if
the line of sight passes through the dense  boundary wall of the LB. 
For O/H\,$= (3.45 \pm 0.19)\times 10^{-4}$  which is the value measured in the LISM
(Oliveira et al. 2005),  $\log N($\oi$) = 15.8$ implies $\log N($\hi$) \sim 19.3$
corresponding to the bubble wall definition of Sfeir et al. (1999).
Those lines of sight where $\log N($\oi$) <  15.6$ probably do not extend
through the boundary of the LB wall. For $\log N($\oi$) > 15.8 $
the LB wall is probably penetrated. For
the intermediate cases with  $15.6 < \log N($\oi$) <  15.8$, the line of
sight may or may not pass through the LB wall.   If the wall does
occur in this column density range, one might expect to observe a
change in the behavior of \ovi\ if  the \ovi\  absorption occurs in  or
near the  boundary of the LB.

Fig.~9 shows $\log N($\ovi) versus $\log N($\oi) for the  18 WDs for which
both \ovi\ and \oi\ column density measurements exist (ignoring 
WD\,0455--282).   There are only 4 WDs with $\log N($\oi$) <  15.6$. \ovi\ is detected toward 3 of these 4 WDs and the median  value  of $\log N_a($\ovi) is 12.94.  The 14 WDs with $\log N($\oi$) > 15.6$ show a large spread in $\log N($\ovi) with a median value of 12.95.  Three of the WDs that
should lie near or beyond the bubble wall show the smallest values of
$\log N($\ovi) in the sample. These include WD\,1211+332 with $\log N($\ovi$) =
12.38\pm\,^{0.17}_{0.28}$, WD\,1631+781 with $\log N($\ovi$) = 12.52\pm\,^{0.12}_{0.17}$, 
and WD\,2309+105 with $\log N($\ovi$) <  12.53$.   There are 3 
WDs out of a total of 8 beyond the bubble wall displaying  an excess of \ovi\ compared to the closer WDs.  These are WD\,0715--703 with $\log N($\ovi$) = 13.23 \pm 0.07$,  
WD\,1017--138 with $\log N($\ovi$) = 13.29 \pm 0.11$, and WD\,1528+487 with 
$\log N($\ovi$)= 13.27 \pm 0.06$.   The  0.2 to 0.3 dex excess in $\log N($\ovi) toward
these 3 objects could be due to the bubble wall. The excess corresponds to a bubble wall contribution to $\log N($\ovi) of $\sim$ 13.0.  Note that the objects beyond the bubble wall with high and with low values of $\log N($\ovi) are widely distributed on the sky. The presence of \ovi\ in the LB wall is suggested by the excess \ovi\ recorded along 3 of the 8 lines of sight through the wall.  The absence of a strong LB wall signature in \ovi\ for the 5 lines of sight could be due to the magnetic suppression of conduction in the warm-hot gas interface  or  to the absence of hot gas at the LB wall.

\subsection{\ovi\ Line Widths } 

When LISM \ovi\ is detected with $\ge 4\sigma$ significance, it is possible to
obtain reliable measures of the line width, $b$ (\km), either from
the AOD or profile fit methods (see Fig.~6d).   The AOD values of $b$
are slightly larger than the profile fit values, probably partly due
to the fact that the profile fit values of $b$ allow for the
instrumental blurring while the AOD values of $b$ are simply based on a
direct integration of the observed profile.   The observed profile fit
line widths for the 11 WDs with $W_\lambda/\sigma>  4$  tracing LISM gas  range
from $15.0 \pm 3.5$ \km\ for WD\,1800+685  to $36.2 \pm 7.3$ \km\ for WD\,1528+487.  
The median and the  average and dispersion of  $b_f$ are 20.5 
\km\ and $23.0 \pm 5.6$  \km\  respectively.   In CIE \ovi\ peaks in
abundance at $2.8\times 10^5$ K (Sutherland \& Dopita 1993).  At this
temperature the thermal Doppler line width of  \ovi\ is  $b = 17.1$ \km\
which is close to the minimum value of $b_f$ observed.   A line with
$b_f$ as large as  36.2 \km\ implies $T = 1.3\times 10^6$ K if the broadening is
dominated by thermal Doppler broadening.  Very little \ovi\ exists at
temperatures this large under conditions of CIE (Sutherland \& Dopita
1993). Therefore, the broader \ovi\ profiles could be tracing multiple
component \ovi\ absorption or the kinematic flow of the \ovi\ in a
single absorbing structure. 

In several of the highest S/N cases the observed line profiles are
strongly suggestive of the blending of multiple absorption components. 
Examples include WD\,0455--282 and WD\,1528+487 (see Fig.~3).   However,
the S/N in most of the observations is not adequate to warrant fitting
multiple absorption components to the observed line profiles of the
weak \ovi\ absorption. 

The \ovi\ column density and Doppler parameter for LISM lines of sight
are correlated.  In Fig.~10a $\log N_a($\ovi) is plotted against $b_a$ for
the 11 $\ge 4\sigma$ LISM detections listed in Table~3. The
histogram of the values of $b_a$  for the 11 $\ge 4\sigma$ LISM detections is shown
in Fig.~10b.   The correlation of $\log N_a($\ovi) and $b_a$ has previously
been shown to extend from the LISM to the Galactic  disk and halo and 
beyond (Heckman et al. 2002; Savage et al. 2003).  The trend observed
in the LISM could be the result of the superposition of contributions
to $\log N_a($\ovi) and $b_a$ from multiple interfaces displaced in velocity. 
This explanation is supported by the fact that those lines of sight with very broad \ovi\ profiles also have very strong and broad multicomponent \cii\ 
absorption.  Examples include WD\,0455--282, WD\,0830--535, and  WD\, 1528+487 (see Fig.~3).  However, several lines of sight 
with particularly simple and narrow \cii\ absorption profiles including 
WD\,1254+223 and WD\,1314+293 exhibit \ovi\ absorption that may be affected 
by the outflow from an evaporative interface (see \S5.1.2). The \ovi\ profile 
breadths for gas in the LISM evidently are influenced by a combination of 
several effects in addition to the thermal Doppler broadening that certainly 
affects each component of the \ovi\ absorption profiles.  Heckman et al. (2002) have proposed that  
the correlation is similar to that predicted for radiatively cooling gas if 
it is assumed that the cooling flow velocity and line width are strongly correlated. 
While radiative cooling could be  applicable  for the higher column density systems included in the Heckman et al. study, the discussion above shows that such a simple cooling gas  flow process is not a valid explanation for the origin of the correlation observed for lines of sight only sampling gas in the  LISM.
 
\subsection{ \ovi\ Absorption Velocities }

	The kinematical relationships between the absorption by \ovi\ and
other tracers of interstellar gas are important for evaluating the
origin(s) of the \ovi.  If the \ovi\ is produced in the interface
region between the cool and warm ISM and the hot ISM,  one would
expect the \ovi\  to absorb at velocities near the absorption tracing
the cool and warm gas. If there are outflows or inflows in the \ovi\ 
interface zone, modest ($\sim$ 10 to 20 \km)  velocity shifts might
occur.   If the some of the \ovi\ in the LISM is produced in bubbles of
hot gas with $T \sim 2-5\times 10^5$ K there should be very little associated
absorption from tracers of cool or warm gas and the \ovi\ absorption
might be found at velocities where there is no cool and no warm gas.    

In Fig.~11a   $\bar{v}_a($\oi) is compared to  $\bar{v}_a($\cii\ ) for the 24 WDs
where  LISM \ovi\ is detected with $\ge 2\sigma$ significance.  The velocities
are from Table~5 and were calculated by the AOD method and refer to
the dominant \cii\ and \oi\ absorbing components.  The solid diagonal
line assumes $\bar{v}_a($\cii$) = \bar{v}_a($\oi).   The correspondence between $\bar{v}_a($\cii)
and  $\bar{v}_a($\oi) is excellent.  In the LISM \oi\ and \cii\ generally have
very similar absorption velocities.  The histogram of $\bar{v}_a($\oi$) - \bar{v}_a($\cii) 
is shown in Fig.~11c. 

There are several objects for which \cii\ can be seen in weak extra
components not detected in \oi.  These include: WD\,0455--282 which has
a \cii\ component at $-42$ \km; WD\,1528+487 with a \cii\ component at
$-57.5$ \km; WD\,0603--483 with a \cii\ component at $-40$ \km.   There
are other cases where the actual shape of the \cii\ and \oi\ absorption
differ slightly in the wings of the line.  However, even in those
cases the measured values of $\bar{v}_a($\cii) and $\bar{v}_a($\oi) are very similar. 
The extra \cii\ components without associated \oi\ absorption is likely
the result of \cii\ being a more sensitive tracer of neutral and
ionized gas than \oi\ which only traces neutral gas.  

In Fig.~11b $\bar{v}_a($\ovi) is compared to  $\bar{v}_a($\cii)  and the histogram of
$\bar{v}_a($\ovi$) - \bar{v}_a($\cii) is shown in Fig.~11d.	The solid line in Fig.~11b 
is for $\bar{v}_a($\ovi) $= \bar{v}_a($\cii).   \ovi\ and \cii\ are correlated in
velocity but  not as well as for \oi\ versus \cii.   

There are 10 WDs where $\bar{v}_a($\ovi$) = \bar{v}_a($\cii) to within  $\pm 6$ \km.  These
include:  WD\,0004+330, WD\,0715--703, WD\,0809--728, WD\,0830--535, 
WD\,1603+432, WD\,1631+781, WD\,1634--573, WD\,1800+685, WD\,1950--432, and 
WD\,2004--605.

There are 5 WDs where $\bar{v}_a($\ovi) exceeds  $\bar{v}_a($\cii) by +7 to +14 \km. 
These include: WD\,0603--483, WD\,1234+481, WD\,1528+487, WD\,1648+407, and
WD\,2124--224.

There are 8 WDs where $\bar{v}_a($\ovi) exceeds  $\bar{v}_a($\cii) by +20 to +29  \km. 
These include: WD\,0937+505, WD\,1017--138, WD\,1100+716,  
WD\,1211+332,  WD\,1254+223, WD\,1314+293, WD\,1636+351,  and WD\,2116+736. 

This summary of the \ovi\ and \cii\ velocity difference has ignored the
most discrepant case of WD\,0455--282 with $\bar{v}_a($\ovi$) - \bar{v}_a($\cii$) =
-37.6 \pm 4.6$.  Most of the \ovi\ absorption toward WD\,0455--282 appears to
be associated with the unusual high negative velocity \cii\ absorption
seen at $-42$ \km\  in the multicomponent \cii\ profile (see Fig.~3). 
Note that with respect to the $-42$ \km\ \cii\ component, the \ovi\ 
observed in this direction has $\bar{v}_a($\ovi$) -\bar{v}_a($\cii$) = +18.4 \pm 4.6$ \km.  

The positive velocity offset between the \ovi\ and \cii\
absorption for the 13 WDs  could be the result of an evaporative flow
in an interface containing \ovi\ or from multi-component  \ovi\
absorption where some of the \ovi\ components do not have associated
absorption in  \cii\ or \oi.  The highest quality \ovi\ line profiles
for stars exhibiting large velocity offsets between \ovi\ and \cii\ are
for WD\,1254+223 and WD\,1314+293 (see Figs.~3 and 5).   For both lines
of sight, some of the \ovi\ absorption extends over the range of the \cii\ 
and \oi\ absorption.  However, the large velocity shift between $\bar{v}_a($\ovi)
and   $\bar{v}_a($\cii) is caused by strong \ovi\ absorption also occurring
where there is no \cii\ absorption.  For example for WD\,1254+223 the \ovi\ 
absorption from 15 to 60 \km\ occurs in a velocity range where \oi\ 
and \cii\ are not detected.  The same is true for WD\,1314+293 for the
10 to 50 \km\ velocity range.  While the profiles for the other WDs
with large values of $\bar{v}_a($\ovi$) - \bar{v}_a($\cii) are not as well defined, 
many of  them imply there is substantial \ovi\ absorption occurring at
velocities where there is no evidence for \cii\ or \oi\ absorption. 

It is interesting that the \ovi\ absorption displaced in velocity from
the  \cii\ or  \oi\ absorption always occurs on the positive velocity
side of the \cii\ and \oi\ absorption.  In order to see if this positive
velocity \ovi\ absorption is possibly associated with a more highly
ionized tracer of gas in the LISM we show in Fig.~4 for WD\,1314+393
LISM absorption in the \ciii\ $\lambda$977.00 line.  The \ciii\ follows the \cii\ 
absorption and shows no evidence for component structure at the larger
positive velocities where the \ovi\ is detected.  Therefore a
substantial fraction of the \ovi\ absorption toward WD\,1314+394 occurs
at velocities where there is no associated   \oi, \cii, and \ciii. 

In Fig.~12 we show the directions in the sky where $\bar{v}_a($\ovi$) - \bar{v}_a($\cii)
has small, medium,  and large values.  In that figure the circles
denote the different WDs with size scaled inversely according to the
WD distance.  The grey scale is coded according to the observed values
of $\bar{v}_a($\ovi$) - \bar{v}_a($\cii), with white  to black indicating small values 
to large positive values.  The WDs with the larger values of $\bar{v}_a($\ovi$)
- \bar{v}_a($\cii) are all found north of the Galactic plane.   The three
objects closest to the north Galactic pole (WD\,1211+332, WD\,1254+223, 
and WD\,1314+293) all have $\bar{v}_a($\ovi$) - \bar{v}_a($\cii$) > 20$ \km.  In the
case of the closely aligned pair of stars in the direction of the
north Galactic pole (WD\,1254+223 and WD\,1314+293),  the observations
are of very high quality and approximately the same \ovi\ to \cii\ 
velocity  shift is measured toward each WD. 

Since we have found 8 cases where the \ovi\ LISM absorption
is redshifted by 20 to 29  \km\ with respect to the cooler LISM gas traced by \oi, \cii,
and \ciii\ it is reasonable to ask could the redshifted \ovi\ absorption
be  from contaminating stellar absorption redshifted by the
gravitational redshift of the WDs.   The required redshift is $\sim$25 \km\
and the average gravitational redshift estimated for the WDs
showing stellar absorption lines discussed in \S3.3 was $35.2 \pm 20.4$ \km. 
In the case of the two best observed stars in our sample showing
the shifted \ovi\ (WD\,1254+223 and WD\,1314+293),   there is no evidence
for any stellar metal line absorption in the spectra from {\em FUSE},  {\em IUE}
or STIS.  Furthermore WD\,1254+223 and WD\,1314+293 with effective
temperatures of 38,686 and 50,560 K, respectively would need to
produce nearly identical stellar \ovi\ blend contributions for the two
\ovi\ profiles displayed in Fig.~3 to be so similar.  While WD\,1314+293 
is hot enough to have stellar \ovi, WD\,1254+223 appears to be
too cool to produce stellar \ovi.  We believe the most likely
explanation for the similar \ovi\ profiles toward WD\,1254+223 and 
WD\,1314+293 is that the absorption occurs in the LISM and  the gas
properties are fairly similar along the  line of sight to  these two
stars which are only separated by $\sim$10 degrees on the sky.  We conclude
that the redshifted \ovi\ absorption compared to \cii\ and \oi\ is a
common characteristic of the absorption by \ovi\ in the LISM.

\section{The Origin of  \ovi\ in the LISM }

\subsection{LISM \ovi\ in Warm-Hot Gas  Interfaces}
\subsubsection{Theory of Evaporating and Condensing Interfaces}

A leading theory for the origin of  \ovi\ in the LISM is that it forms
in the interface region between warm ($T\sim5000$--8000 K)  and hot  
($T\sim 10^6$ K) gas.  In the interface, detectable amounts of transition
temperature gas with $T \sim (1-6)\times 10^5$ K  should exist with the gas column
density depending on  a number of parameters including the 
temperature of the hot gas, the viewing angle through the interface,
the age of the interface, the  presence or absence of a magnetic
field,  and the  orientation  of the field with respect to the
interface.  Non-equilibrium ionization processes are important in the
interfaces and must be considered when calculating expected column
densities of particular ions such as \ovi.   In addition to
predictions of the column densities  of the highly ionized atoms,  the
various interface models also make  predictions for expected line
widths from thermal broadening and for  possible velocity offsets
between the warm gas forming the interface and the transition
temperature gas in the interface.   These velocity offsets arise from
flows associated with the evaporation of the warm gas into the hot
medium or  the condensation of hot gas into the warm medium.  The
interface models, therefore, make specific predictions for $\log N($\ovi),
$b($\ovi), and $\Delta v$, the velocity offset between the \ovi\ absorption and
some tracer of the warm gas such as \oi\ or \cii. 

	A short  summary of the model predictions from the theoretical papers
of  B\"ohringer \& Hartquist (1987), Borkowski et al. (1990), and Slavin
(1989) follows:

{\em The evaporative phase}.   In the initial evaporating phase of an
interface the evolution is dominated by conductive heating because of
the steep temperature gradient in the interface.  During this phase
which can last for up to $\sim 2 \times 10^6$ years,  the column density of \ovi\ for
the magnetic field perpendicular to the interface becomes detectable
at $\log N($\ovi) $\sim$ 12.5 after  $10^5$ years and increases to $\sim$ 13.0 after
$3\times 10^5$ years (see Fig.~6 in Borkowsky et al. 1990).  The amount of \ovi\ 
in the interface is reduced by up to $\sim$1 dex for field orientations
nearly parallel to the interface for the Borkowsky et al. model
assumptions.  Although very few results are reported in the
literature, it is during this evaporative phase that substantial
evaporative outflows of \ovi\ are possible.  For example, B\"ohringer \& 
Hartquist (1987) report \ovi\ outflow velocities as large as 18 to 25
\km\ in their models D, A, and C.   These same models have values of
$\log N($\ovi) ranging from 12.85 to 13.15.  In the evaporative stage the
mean temperature of \ovi\ in these models ranges from  $\sim (7-8)\times 10^5$ K  
which is higher than the CIE temperature of peak \ovi\ abundance
because of the time lag in the ionization of the evaporating gas.  \ovi\ 
with $T = 7.5\times 10^5$ K would have a thermal Doppler parameter $b = 28$ \km.  
The Borkowsky et al. models also predict velocity shifts and
enhanced thermal line widths in the evaporative stage of the interface
evolution, although the  sizes of the effects are smaller than those
determined by B\"ohringer \&  Hartquist (1987).    In the evaporative
phase, the \ovi\ lines produced in an interface are expected to be
shifted in velocity and to be  broad.  The \ovi\ column density in the
evaporative phase could be as large as $\log N($\ovi) $\sim$ 13.0 for a path
perpendicular to the interface.  For oblique paths through the
interface the column density will be several times larger.  The value
of $\log N($\ovi) could be suppressed by up to $\sim$1 dex if the magnetic
field limits the evaporation.  The overall column density range  is
therefore expected to  be from $\log N($\ovi) $\sim$ 12.0  to 13.3 per
interface. 

{\em The condensation phase}.   At later times in the evolution of the
interface, radiative cooling dominates the evolution and the gas flow
is from the hot phase into the cold phase.  This condensation phase
begins after  $\sim 2\times 10^6$  years for the Borkowski et al. (1990) ISM
parameters.  During the condensation phase the expected \ovi\ column
density stabilizes at $\log N($\ovi$) \sim 13.0$ for the magnetic field
perpendicular to the interface. Again the expected \ovi\ column density
is substantially suppressed by up to $\sim$1 dex for fields more nearly
parallel to the interface.  The condensing flow velocities are only
several \km, therefore, the internal motions in the front are
negligible compared to thermal Doppler broadening.  In the condensing
front the average temperature of \ovi\ is smaller than in the
evaporation front.  The Borkowski et al. models suggest values of $T =
(3-4)\times 10^5$ K for hot exterior gas temperatures of $(0.75 -1.0)\times 10^6$ K.  
At $T = 3.5\times 10^5$ K, $b($\ovi$) = 19$ \km.  Therefore, in the condensing
phase the \ovi\ absorption should be well aligned with the warm gas
absorption and the \ovi\ line widths should be relatively narrow.  The
values of the \ovi\ column density are expected to be similar to the
values found in the late stages of evolution of  the evaporating front
models discussed above. The  overall column density range  is
therefore expected to  be from $\log N($\ovi$) \sim 12.0$  to 13.3 per
interface depending on the amount of quenching from the magnetic
field. 

The Borkowski et al. (1990) interface theory described above assumed 
an interstellar oxygen abundance (O/H) $=8.13\times 10^{-4}$ from 
Grevesse (1984) which is 2.36 times larger than the value, 
(O/H) $=3.45\times 10^{-4}$, measured for warm gas in the LISM 
(Oliveira et al. 2005). Lowering the oxygen abundance will lower 
the predicted values of  $N($\ovi) for gas in condensing or evaporating 
interfaces.  However, as discussed by Savage et al. (2003) and Fox et al. (2004), 
the change in  $N($\ovi) does not scale linearly with the change in (O/H) because 
with a reduction in (O/H) the cooling rate of the interface gas is reduced, which 
allows \ovi\ to survive longer in the transition temperature gas of the interface. 
According to equation 8 in  Fox et al. (2004) a factor of 2.36 reduction in (O/H) 
will cause the predicted value of  $N($\ovi) to be reduced by only a factor of 1.13. 
The assumed oxygen abundance has a relatively small effect on the predicted column 
density provided the cooling of the gas is dominated by oxygen. 

The ages of warm-hot gas interfaces in the LISM will depend on explosive events 
that create hot gas and the nature and path of the expanding flow of that hot 
gas following the explosion. Smith \& Cox (2001) have concluded that several 
supernova explosions are needed to create the LB with the last explosion 
occurring within 5 Myr. Ma\'{i}z-Apell\'aniz (2001) has considered possible sites of those
supernovae and has proposed that supernovae occuring in the lower Centauris Crux
subgroup of the Scorpius-Centauris OB association are likely responsible for the creation
of the LB. The recent convincing detection of excess $^{60}$Fe in sea 
sediments in a layer 2.8 Myr old (Knie et al. 2004) provides strong evidence for a 
recent local supernova with an estimated distance of 15 to 120 pc (Fields et al. 2005).  
Such an event could be responsible for the reheating of the LB and for producing conductive 
interfaces in the LISM in both the old ($>2$ Myr) condensing phase and the young ( $<2$ Myr) 
evaporative phase of their evolution. 

The principal discriminators between evaporating interfaces and
condensing interfaces are the \ovi\ line width and the \ovi\ velocity
shift with respect to the warm gas velocity.  However, actually
looking for these predicted effects in the evaluation of \ovi\ 
absorption in the LISM becomes increasingly difficult as the line of
sight goes through more than one interface. 

\subsubsection{Observations}

We examine if the measurements of the \ovi\ 
absorption in the LISM are consistent with the theoretical
expectations of the interface theories discussed above.

The \ovi\ column densities we measure for the 24 LISM WD detections
plotted in Fig.~7 range from $\log N($\ovi$) = 12.38$ to 13.60 with a
median of 13.10.  The 2$\sigma$ upper limits span approximately the same
column density range.  The detections are consistent with a typical
maximum value of $\log N($\ovi$) \sim13.0$ per interface and the fact that
multiple interfaces are probably encountered toward the more distant
stars in the sample.  The detections with $\log N($\ovi) ranging from
12.4 to 12.7  are consistent with the expectation of smaller values of
the column density if the evaporation or condensation in the interface
is impeded by a magnetic field  or if an evaporation front is caught
very early  in  its evolution.   Thirteen of the 15 2$\sigma$ upper limits
with $\log N($\ovi)  ranging from $<12.84$ to $<13.48$ lack the
sensitivity to detect the typical value of the \ovi\ column density
expected to be produced by a single interface.   The two smallest
upper limits with $\log N($\ovi$) <  12.53$ and $<12.62$ for WD\,0549+158 and
WD\,2309+105 reveal that in some cases  there is very little \ovi\ associated with LISM warm gas absorbers. The absence of \ovi\ could be the result of an absence of the hot gas required to produce a warm-hot gas interface.  An alternate possibility is a warm-hot gas interface exists but  the evaporative or condensation flows are suppressed by a magnetic field.  The case of WD\,2309+195 is most impressive
in this context since the LISM in this direction has relatively strong
\oi\ and \cii\ absorption and virtually no \ovi\ absorption  as recorded
on a very flat and well defined continuum (see Fig.~3). 

In the LISM $N($\ovi) and $N($\oi) are not correlated (see Fig.~9).  A
poor correlation is expected between $N($\ovi) and $N($\oi) if interfaces
produce \ovi\ since the total column density of \ovi\ depends on the
number of interfaces along the line of sight and not on the total
amount of cool or warm gas in the structures producing the interfaces. 

The \ovi\ absorption velocities and line widths for some of the WDs are
most consistent with the expectations for \ovi\ absorption in
condensing interfaces where the \ovi\ velocity is expected to align
well with the \cii\ absorption.  These cases include the 10 WDs where
$\bar{v}_a($\ovi$) = \bar{v}_a($\cii) to within  $\pm 6$ \km.  High quality examples from
this group include WD\,0004+330, WD\,0715--703,  WD\,1634--573, and WD\,1800--432.  
In all four cases the \ovi\ absorption is well aligned with
the \cii\ absorption.  In these four cases the value of $b_f($\ovi) ranges
from  $15.0 \pm 3.5$ to $19.5 \pm 5.2$  \km\ with an average of 17.3 consistent
with the \ovi\ in gas with $T = 2.9\times 10^5$ K which is close to the
expected temperature  for \ovi\ in a condensing interface. 

There are 5 WDs where $\bar{v}_a($\ovi) exceeds $\bar{v}_a($\cii) by +7 to +14 \km
and 8 WDs where $\bar{v}_a($\ovi) exceeds $\bar{v}_a($\cii) by +20 to +29 \km\ (see
\S4.4).  The velocity shifts for some of these WDs could be the result
of looking through multiple interfaces with warm gas component
velocity offsets from one interface to the next and different values
of $\log N($\ovi) from one interface to the next.  However, the fact that
the velocity offsets are almost always in the positive sense suggests
that another possible explanation is that the paths to these WDs pass
outward from the warm medium of the local cloud into a hot medium and
we are seeing the \ovi\ in the evaporative outflow in the interface
region.   The two best observations illustrating this velocity offset
are for WD\,1254+223 and WD\,1314+293 (see the profiles in Fig.~3 and
5).  For both WDs the \ovi\ absorption overlaps the range defined by
the \cii\ absorption but also extends  $\sim$+40 \km\ beyond the center of
the \cii\ absorption so that the average velocity differences are
$24.2 \pm 3.1$ and $19.9 \pm 3.7$ \km, respectively.  The values of $b_f($\ovi)
for the two WDs are $20.5 \pm 3.6$ and $28.6 \pm 4.1$, respectively.  The velocity
shift and large value of $b_f$  for WD\,1314+293 are consistent with the
expectations for  the  evaporative flow described by the models of
B\"ohringer \&  Hartquist (1987).  The value $b_f = 20.5$ \km\ for WD\,1254+223 
implies $T = 4\times 10^5$ K which is more consistent with the
expected temperatures in the evaporative flow described by the
Borkowsky et al. models.  While the Borkowsky et al. models do not exhibit as large a
velocity offset as observed, it is evident from these simple
comparisons that changes in some of the basic ISM and model
assumptions might be implemented that would allow the models to span
the range of observables.  For example, an increase in the temperature of the hot exterior gas will cause the predicted evaporative flow velocity to increase. 

In the case of the evaporative interface explanation, the positive
value of the \ovi\ to \cii\ velocity offset implies we must be viewing
the interface from the warm medium into the hot medium.  Such a
viewing direction is possible since our viewing position is from
within the warm gas of the local cloud.  For look
directions toward the north Galactic pole and the Galactic center, 
the path length through the local cloud is small (Redfield \& Linsky 2000). 
In these directions very little warm gas from the local cloud would be sampled, 
although \ovi\ from an associated evaporating interface might still be sampled. 
Given the apparent complexity of the distribution of the warm gas in the LISM, 
we have not tried to associate the \oi\ and \cii\ warm gas tracers in our program 
with particular LISM cloud names as proposed for example by Redfield \& Linsky (2000) 
in their LISM model of the the gas distribution.  

If the evaporative interface explanation is correct for the origin of the
positive velocity offsets between \ovi\ and \cii, the evaporation must
be more prominent in the directions of  the northern Galactic
hemisphere as revealed in Fig.~11 where the larger values of 
$\bar{v}_a($\ovi$)-\bar{v}_a($\cii) are shown to occur in the north.   This could occur if
the interfaces in the northern Galactic hemisphere of the LISM are
younger than those in the southern Galactic hemisphere. 

We conclude that the LISM \ovi\ observations are consistent with an origin of the \ovi\ in conductive interfaces caught in the old condensing phase and in the young evaporative phase of their evolution.    However, other explanations are possible for that aspect of the \ovi\ absorption that appears to be tracing evaporative interface gas.

\subsection{Is there \ovi\ in the LISM that is not Associated with Warm-Hot Gas
Interfaces? }

Are there other possible explanations for the \ovi\ absorption we see
in the LISM, particularly that portion of the \ovi\ absorption that
occurs at velocities where there is no \cii\ or \oi\ absorption?   It
would be possible to produce the distribution of values of $\bar{v}_a($\ovi)
versus $\bar{v}_a($\cii)  shown in Fig.~11b if there existed in the LISM a
hot gaseous region or bubble with no associated \cii\ absorption in
addition to the \ovi\ associated with \cii.   A cooling bubble with
interior temperatures of $\ge 2\times 10^5$ K would contain \ovi\ but no detectable
\cii\ or \oi.  The hot gaseous region would need to exhibit \ovi\ 
absorption at velocities $\sim+20$ to +29 \km\ with respect to the bulk
of the warm gas in the LISM. This hot gaseous region needs to mostly
exist in the north Galactic hemisphere (see Fig.~12). Blending
between the absorption by the hot bubble gas and LISM conductive
interface gas would then explain the redshifted \ovi\ absorption
overlapping the \ovi\ absorption near the lower velocity of the warm
gas. A close inspection of the line profiles for the highest S/N
cases where the average velocity of \ovi\ is substantially shifted away
from the average velocity of \cii\ reveals the following:  For WD\,1254+223 
and WD\,1314+293 we see that a substantial portion ($\sim$70\%) of
the \ovi\ absorption occurs at velocities where \cii, \oi, and even \ciii\ 
are not detected.  This extra absorption corresponds to 
$\log N($\ovi$)\sim12.8-12.7$  out of a total of $\log N($\ovi$) = 13.1-12.9$.    
This extra \ovi\ absorption could be the outflowing evaporating interface gas as
discussed above in \S5.1 or it could be revealing the presence of a
second gaseous component that is only traced by \ovi. A possible
site is the cooler part of the more generally distributed hot gas of
the LB. It could, for example, be associated with hot gas related to
the local interstellar chimney extending into the low halo in the
northern Galactic hemisphere (Welsh et al. 1999). 

We can not rule out the possibility that the 20 to 29 \km\ positive
velocity shifts seen for a number of the \ovi\ profiles is from the
detection of a second spatially isolated hot phase of \ovi\ with no
associated warm gas absorption from \oi, \cii\ or \ciii.   

Additional observations of \ovi\ absorption toward WDs ranging in distance from 20 to 
100 pc in the general direction of the north Galactic
pole might provide information on the distance and extent of the
higher positive velocity \ovi\ absorption.  An extended bubble origin
for the \ovi\ would be favored if the positive velocity \ovi\ column
density were to increase with distance

\section{\ovi\ Absorption Beyond the LISM}

Understanding the properties of \ovi\ in nearby and relatively simple
warm-hot gas interfaces in the LISM is important for understanding the
origins of \ovi\ in more distant astrophysical sites including more
distant regions in the Galactic disk, gas in the Galactic halo,  gas
in high velocity clouds,  and gas in the IGM.  The observations in
this investigation reveal that whenever a cool/warm to hot gas
interface exists, \ovi\ will generally be detected with column
densities of $\log N($\ovi$) \sim 12.5$ to 13.0 provided the conductive heat
flow from the hot to the warm or cool medium is not suppressed by the
presence of a magnetic field.  In addition, the absorption by \ovi\ 
should be closely aligned with a tracer of the cool or warm gas if the
interface is an older condensing interface.  If the interface is in
the younger evaporative phase of its evolution offset velocities
between \ovi\ and the cool/warm gas might be as large as 15 to 30 \km.  

If conductive interfaces were the only explanation for \ovi\ found in
the Galactic disk, one might  expect  the dispersion in plots of
$\log N($\ovi) vs $\log d$ to decrease for larger values of  distance.  A
decrease might be expected since large values of $\log N($\ovi) would
require many interfaces each contributing $\log N($\ovi$)\sim13.0$ to the
total.  If 10 to 30 interfaces are required to produce $\log N($\ovi$) \sim
14.0$ to 14.5,   the dispersion in the large column density
distribution would be expected to be smaller than for the low column
density distribution because of the effects of the averaging over many
interfaces at large $\log N($\ovi).  However, such a decrease is not
evident in the {\em Copernicus}\ and {\em FUSE}\ observations of stars in the
Galactic disk with distances up to $\sim$8 kpc (Bowen et al. 2005).   The
fact that the dispersion does not decrease with distance possibly
implies that some other process could be responsible for the large
dispersion seen at large distances.  For example, perhaps the behavior
occurs because at large distances important contributions to
$\log N($\ovi) come from the presence of \ovi\ in the hot gas of cooling
supernova remnant bubbles as proposed by Slavin \& Cox (1993). 
Another simpler possibility for the effect is to assume that
most of the \ovi\ does occur in interfaces but that the distribution of
the hot gas required to produce the interfaces is irregularly
distributed over very large distance scales.  If the hot gas is
absent, there will be no interface even though numerous warm clouds
may exist. 

Moving into the Galactic halo, Savage et al. (2003) have shown that
the values of $\log N($\ovi) toward extragalactic objects associated with
the 2 to 5 kpc thick disk of the Milky Way reveal an irregular
distribution of  \ovi. With $\log N($\ovi$) = 14.36 \pm 0.18$ (standard deviation) 
through the halo to each side of the Galactic plane,  a typical line of sight would require $\sim$23 interfaces if the \ovi\ mostly occurs in conductive interfaces like those seen in the LISM.  With such a large number of interfaces, a reasonable way to
explain the irregularity of the distribution of \ovi\ on the sky would
be to have significant variations in the amount of the hot exterior
gas from one line of sight to the next. In addition, a population of supernova remnant 
bubbles (Slavin \& Cox 1993) and/or parcels of cooling Galactic 
fountain gas (Edgar \& Chevalier 1986) may also contribute to the irregularity. 

In the gaseous regions beyond the Galactic thick disk, high velocity \ovi\ 
has often been detected at velocities implying a direct association
with Galactic high velocity clouds (HVCs)  measured in \hi\ 21 cm
emission (Sembach et al. 2003).  For the carefully studied cases 
toward several HVCs  investigated by Fox et al.  (2004, 2005),  it
appears likely that the \ovi\ is produced in conductive or turbulent
interfaces between the cool/warm gas of the HVC and a hot exterior
medium assumed to be the highly extended $R > 70$ kpc Galactic corona as
proposed by Sembach et al. (2003).  The existence of such a hot medium
appears to be required to explain the detection of extremely strong \ovii\ 
and \oviii\ absorption lines at zero redshift (Fang et al. 2003;
Futamoto et al. 2004; McKernan et al. 2004; Savage et al. 2005; Yao \&
Wang 2005). However, Nicastro et al. (2002) have proposed that most of
the \ovii\ and \oviii\ absorption instead occurs in the Local Group
medium.   The close kinematic connection between \ovi\ and \cii\
observed in the conductive interfaces observed in the LISM gives
support to the conductive interface hypothesis for the origin of \ovi\ 
in some of the HVCs and appears to require the presence of hot gas in
the vicinity of the  \hi\ HVCs including the Magellanic Stream which is
likely at a distance  of $\sim$50 to 70 kpc from the Milky Way. 

Conductive interfaces could also exist in the intergalactic medium
(IGM).  If an IGM absorption line system is found implying a close
kinematical association of \ovi\ and lower ionization species such as \hi, 
\cii\ and \ciii, a conductive interface origin for the \ovi\ is a
distinct possibility provided the lower ionization gas does interface
with a hot (unseen) exterior gas.  A single interface is expected to
produce $\log N($\ovi$) \sim 13.0$ and the column density  should not depend 
sensitively on  metallicity as discussed by Fox et al. (2004).  If a
IGM absorption line system is found containing for example \hi, \cii,
and \ciii\ but no \ovi\ to a detection sensitivity of $<12.5-13.0$, it
would imply there is no conductive interface either due to magnetic
quenching of the conductive heat flow or because of the absence of the
exterior hot gas required to produce the interface.

\section{Summary}

We have investigated the properties of \ovi\ absorption in the LISM
observed in {\em FUSE}\ 20 \km\ resolution spectra of 39 WDs ranging in
distance from 37 to 230  pc with a median  distance 109 pc. The LISM 
\ovi\ absorption toward an additional 7 WDs in our sample appears to be
contaminated by stellar \ovi\ absorption.  Our investigation extends
the first {\em FUSE}\ \ovi\ LISM survey of Oegerle et al. (2005) by
approximately doubling the number of objects and substantially
increasing the S/N in the spectra for many cases.  In addition,
improvements in the data processing have provided more accurate
velocity calibrations.  With the larger sample, higher S/N and better
velocity calibrations, we are able to explore various properties of
the \ovi\ absorption including the absorption profile shapes and the
kinematic relationships between \ovi\ and other tracers of cool and
warm gas in the LISM such as \cii\ and \oi.   The investigation has
revealed the following:

1. LISM \ovi\ is detected with $\ge 2\sigma$ significance along 24 of 39 WD sight
lines and with $\ge 3\sigma$ significance along 17 of 39 sight lines.  For the 
$\ge 2\sigma$ significance detections, the observed \ovi\ column densities range
from $\log N($\ovi$) = 12.38$  to 13.60  with a median value of 13.10. 

2.  Values of the average line of sight volume density, $n($\ovi$) =
N($\ovi$)/d$, toward each WD exhibit a large dispersion ranging from
(0.88 to $13.0)\times 10^{-8}$  cm$^{-3}$ with a average value  and dispersion of
$(4.4 \pm 2.8)\times 10^{-8}$ cm$^{-3}$  obtained by treating the 15 2$\sigma$ upper limits as
detections.   A more reliable estimate of the average LISM \ovi\ 
density toward the 39 WDs  using the Kaplan-Meier survival analysis
technique to properly account for the 15 upper limits  yields $n($\ovi$)
= (3.6 \pm 0.5 )\times 10^{-8}$ cm$^{-3}$ which is 2.1 times larger than the average
value $1.7\times 10^{-8}$ cm$^{-3}$ found for more distant sight lines in the Galactic
disk by Bowen et al. (2005).  There is a modest excess of \ovi\ in the
LB compared to the more distant interstellar regions. 

3. Plots of $\log N($\ovi) versus $\log N($\oi) reveal an increase in  
 $\log N($\ovi) near the expected location of the LB wall at 
 $\log N($\oi$) \sim 15.8$ for only 3 of 8 WDs that lie beyond the wall. The increase is 
$\log N($\ovi$) \sim 13.0$.  The 5 WDs showing no signature of an enhancement in  \ovi\ from the LB wall may lie in directions where hot gas does not contact the wall or where magnetic fields suppress the conduction.

4.  For the 11 \ovi\ detections with $\ge 4\sigma$ significance, the \ovi\ line
widths can be reliably measured.  Gaussian profile fit values of the
Doppler parameter $b_f$ range from $15.0 \pm 3.5$ to $36.2 \pm 7.3$ \km, with
median value, average value and dispersion of 20.5 \km\ and $23.0 \pm 5.6$
\km, respectively.  The narrowest profiles are consistent with
thermal Doppler broadening of \ovi\ near its temperature of peak
abundance, $2.8\times 10^5$ K, where the Doppler parameter is 17.1 \km. The
broader profiles are probably tracing a combination of \ovi\ with $T
> 2.8\times 10^5$ K, multiple component \ovi\ absorption,   and/or  the
evaporative outflow of conductive interface gas.  

5.  $\log N_a($\ovi) and the Doppler parameter, $b_a$, are observed to be 
correlated for the sample of LISM lines of sight. The correlation 
is, in part, due to the detection of \ovi\ in multiple interfaces for many 
of the higher column density lines of sight which display strong multicomponent \cii\ absorption. 

6.   Comparisons of the average velocities and the profiles of \ovi\ 
and \cii\ absorption reveals 10  cases where the \ovi\ absorption is
closely aligned with the \cii\ absorption as expected if the \ovi\ is
formed in a condensing interface between the cool and warm absorption
traced by \cii\ (and also \oi\ and \ciii) and a hot exterior gas.  

7. There are 15 cases where no \ovi\ absorption is found at the
velocities tracing cool and warm LISM gas.  However, for 13 of the 15
cases the detection sensitivity is relatively large with $\log N($\ovi)
ranging from $<12.84$ to $<13.48$.  Low upper limits are obtained for
two WDs with $\log N($\ovi$) <  12.53$ and $<12.62$.  Low values of $N($\ovi)
can occur in an interface caught early in its evolution or if a
magnetic field inhibits the conduction of energy from the hot gas into
the cool gas.  Another explanation could be the absence of a hot
exterior gas required to produce an interface. 

8. The average velocity and profile comparison of \ovi\ and \cii\ 
reveals positive velocity \ovi\ absorption in the LISM toward 13 WDs that is
displaced +7 to  +29 \km\ from the velocity of  the tracers of the
warm and cool gas. This positive velocity \ovi\ absorbing LISM gas is mostly found in the north Galactic hemisphere.   The positive velocity \ovi\ appears to be tracing the evaporative flow of 
\ovi\ from the interface between warm gas and a hotter exterior gas. However, the positive velocity \ovi\ could also be explained as tracing cooling hot gas of the LB.

9.  The properties of \ovi\ absorption in the LISM are broadly consistent with the expectations of  the theory of conductive interfaces caught both in the old condensing phase and  possibly in the younger evaporative phase of their evolution.

\acknowledgments

We thank Jean Dupuis for allowing us to use his results prior to publication
and for enlightening discussions. We thank Don Cox, Andrew Fox, and 
Ken Sembach for their comments about a draft version of this manuscript.
An anonymous referee provided valuable suggestions for improving the manuscript.  This work is based on archival observations obtained by the
NASA-CNES-CSA {\em FUSE}\ mission operated by Johns Hopkins University.   
This research has made use of the NASA
Astrophysics Data System Abstract Service and the SIMBAD database,
operated at CDS, Strasbourg, France.
Financial support to B.D.S. has been provided by NASA
grant NNGO4GK12G.   B.D.S also acknowledges support from the
University of Wisconsin Graduate School.

\begin{center}
{\bf Appendix}
\end{center}

\begin{center}
{\em Comments on LISM Absorption Toward Individual Objects}
\end{center}

	In this appendix we provide brief comments about the \ovi\ absorption toward most WDs with detected LISM \ovi\ listed in Table 3.  We also discuss the measurements  toward several WDs with no detected \ovi.  The comments refer to the \ovi, \cii\ and \oi\ absorption profiles shown in Fig.~3 and the measured parameters given in Tables 3, 4, and 5. In selected cases we include comments about the velocity structures along the line of sight as determined from higher resolution data from STIS.  

	{\em WD\,0004+330}.--  The \ovi\ line is well detected (4.8$\sigma$) and is closely
aligned in velocity with the \cii\ and \oi\ absorption. The strong and weak lines of the \ovi\ doublet give very similar values of $\log N_a($\ovi$)$  ($12.79 \pm0.09$ versus  $12.75 \pm0.12$).   

	{\em WD\,0455--282}.--  The \cii\ absorption contains a strong component at 14
\km\ and a weak component at $-42$ \km. The weak component is not
seen in \oi.  The strong and well detected \ovi\ extends from $-60$ to 
$\sim+20$ \km.   Much of the \ovi\ absorption appears to be associated
with the weak \cii\ absorption although the positive velocity wing of \ovi\ 
absorption could be associated with the strong \cii\ and \oi\ 
absorption.   The feature on the \ovi\ panel between 65 and 95 \km\ 
could be stellar \ovi\ but is $<2\sigma$ in significance.

	{\em WD\,0603--483}.-- The complex multicomponent \cii\ absorption spans the
velocity range from $-65$ to +65 \km\ with the strongest components
near +15 and $-40$ \km.  \oi\ is not detected in the $-40$ \km\ 
component.  The 2.2$\sigma$ feature identified as \ovi\ has $\bar{v}_a = 25.6 \pm 7.1$ \km\ 
and appears to be associated with the strongest part of the \cii\ 
absorption. 

	{\em WD\,0715--703}.--  The strong and well detected (4.9$\sigma$) \ovi\ absorption is
closely aligned with the \cii\ and \oi\ absorption.  The strong and weak lines of the \ovi\ doublet give similar values of $\log N_a($\ovi$)$  ($13.23 \pm0.07$ versus  $13.30  \pm0.15$).   

	{\em WD\,0809--728}.--   Airglow emission affects the \oi\ profile for $v = -70 $
to $-30$ \km. The absorption identified as \ovi\ extending from $-60$ to
50 \km\ has 4.8$\sigma$ significance and extends well beyond the narrower
velocity range from -$35$ to 30 \km\ of the \cii\ absorption.  The
apparent absorption in the \ovi\ panel from $-150$ to $-120$ \km\ is of
unknown origin.  

	{\em WD\,0830--535}.-- \cii\ and \oi\ exhibit principal absorption near 9 to 13
\km\ but also have a weak positive velocity wing of absorption
extending to 50 \km\ implying a second component of absorption near
30 \km. The \ovi\ absorption with 3.4$\sigma$ significance, which may have
two components, is contained within the velocity range of the \cii\ and
\oi\ absorption.  

	{\em WD\,0937+505}.-- The strong multicomponent \cii\ absorption with 
$\bar{v}_a = -5.1 \pm 1.5$ \km\ extends from $-40$ to +40 \km.  The \oi\ absorption is
contaminated by \oi\ airglow emission. The broad \ovi\ absorption
detected with 3.6$\sigma$ significance extends from $-25$ to +60 \km\ or $\sim$20
\km\ beyond the positive velocity limit of the \cii\ absorption. 

{\em WD\,1017--138}.-- The \ovi\ toward this object illustrates a low
significance (2.2$\sigma$) detection. It is possible there are two \ovi\ 
features, one aligned with the \cii\ and \oi\ absorption and the other
displaced by approximately +45 \km.  Approximately half of the \ovi\ 
absorption occurs at velocities $> 20$ \km\ where \cii\ is not
detected. 

{\em WD\,1100+716}.--  The \oi\ profile is affected by airglow emission. The \ovi\ 
absorption with 7.0$\sigma$ significance extends from $-30$ to +40 \km
while   \cii\ is not detected at velocities exceeding 10 \km.   The
\ovi\ absorption is probably multicomponent with weak and strong
absorption at $-25$ and +15 \km, respectively.

{\em WD\,1211+332}.-- The \ovi\ toward this object is  a low
significance (2.1$\sigma$) detection. The \ovi\ column density of 
$\log N_a($\ovi$) = 12.38$ is the smallest among the entire sample of WDs.  
The \ovi\ absorption is displaced +21.9 \km from the \cii\ absorption which is
very strong and broad. The velocity shift is similar to that seen toward three 
other WDs in the north Galactic polar direction including WD\,1234+481, WD\,1254+223 
and WD\,1314+293 (see below).

{\em WD\,1234+481}.-- The \ovi\ absorption with 5.2$\sigma$ significance is 
displaced $+14.2$ \km from the strong and broad \cii\ absorption. The velocity 
shift is similar to that seen toward three other WDs in the north Galactic polar 
direction including WD\,1211+332, WD\,1254+223 and WD\,1314+293.

{\em WD\,1254+223 and WD\,1314+293}.--  Both of these WDs are near the north
Galactic pole and they have  similar LISM  \ovi\ absorption profiles
(see Figs.~3 and 5).  In Fig.~5 the Voigt profile fits to each
component of the \ovi\ doublet are shown for both stars along with the
\cii\ profile.  The measurements are of high quality
permitting the detection of each member of the \ovi\ doublet.  The \ovi\ 
absorption toward each star overlaps the low velocity \cii\ and \oi\ 
absorption near 0 \km\ but shows a strong extension to large
positive velocities of approximately 50 \km.  The low velocity \ovi\ 
absorption is stronger toward WD\,1314+293.  Therefore, the overall \ovi\ 
to \cii\ velocity offset is larger for WD\,1254+223 (+24.4 \km)
than for WD\,1314+293 (19.9 \km).  More than half of the \ovi\ 
absorption occurs at velocities $> 10$  \km\ where \cii\ absorption is
not detected.  

The 67 pc line of sight to WD\,1254+223 in the direction $l = 317\fdg26$ and
$b = 84\fdg75$ appears to be kinematically very simple.   STIS absorption
line profiles obtained at a resolution of  $\lambda/\Delta \lambda= 114,000 $ of \di, 
\cii, \nni, \oi, and \siii\ are shown in Fig.~1e of Redfield \& Linksy
(2004).  Voigt profile fits to the measurements reveal only one
component to the absorption with the measured values of $v$ ranging from
$-4.91 \pm 0.12$ to $-5.37 \pm 0.21$ \km.  For \di\ the value of $b$ is $7.56 \pm 0.71$
\km\ while for the metals the range is from $2.43 \pm 0.52$ to $3.51 \pm 0.65$
\km.  With $\log N($\di$) = 13.13 \pm 0.03$ the implied value of $\log N($\hi$) =
17.91 \pm 0.05$ if we adopt the value D/H\,$= (1.66 \pm 0.14)\times 10^{-5}$ measured
toward WD\,1314+293 (see the following discussion).

The 68 pc line of sight to WD\,1314+293 in the direction $l = 54\fdg10$ and
$b = 84\fdg16$ lies  only 9.0 degrees away from the  direction to WD\,1254+223. 
STIS absorption line profiles obtained at a resolution of
$\lambda/\Delta \lambda= 114,000 $ of \di,  \cii, \nni, \oi, and \siii\ 
are shown in Fig.~1d of Redfield \& Linksy (2004). GHRS measurements at $\lambda/\Delta \lambda= 114,000 $ 
for \mgii\ and \feii\ are shown in Fig.~1 of Redfield \& Linsky (2002). 
The Voigt profile analysis of the STIS measurements for this line of
sight reveals a single absorbing component with values of $v$ ranging
from $-6.15 \pm 0.27$ to $-6.83 \pm 0.11$ \km.  For \di\ the value of $b$ is
$8.09 \pm 0.40$ \km\ while for the metals the range is from $2.8 \pm 0.37$ to
$3.59 \pm 0.29$ \km.  The GHRS observations of \mgii\ and \feii\  support
the single component result. A full analysis of the HZ\,43 STIS and
{\em FUSE}\ LISM data set has been reported by Kruk et al. (2002) who derive
$\log N($\hi$) = 17.93 \pm 0.03$, $\log N($\di$) = 13.15 \pm 0.03$, and $\log N($\oi$) =
14.49 \pm 0.04$.  Kruk et al. found it necessary to add a second broad low
column density \hi\ component with $\log N($\hi$) = 14.9\pm\,^{0.15}_{0.25}$ and
$v = -2$ \km\ in order to fit the Lyman series absorption.  The width
of this broad \hi\ component implies $T = 30,000 \pm 5,000$ K. The component
may be related to the hydrogen wall about the solar system  that may
arise from the accumulation of gas at the boundary of the heliosphere
(Linsky \& Wood 1996; Wood \& Linsky 1998).  Linsky et al. (2000)
attribute the LISM absorption along this line of sight mostly to a
cloud they call the north Galactic pole (NGP) cloud since the observed
cloud velocities are considered inconsistent with the velocities
expected for the LIC and G clouds.  If this NGP cloud interpretation
is correct, it would imply the LIC provides a negligible contribution
to the observed absorption toward this WD. A similar conclusion
would presumably also apply for the direction to WD\,1254+223. 

{\em WD\,1528+487}.-- \cii\ has strong and weak absorption at $-21.6 \pm 0.3$ and
$-57.5 \pm 1.0$ \km, respectively.  \oi\ is not detected in the weak \cii\ 
component. \ovi\ has absorption extending from $-100$ to +50 \km.   In
Table~3 we report measurements for the relatively isolated weak
feature with  $\bar{v}_a = -72.1 \pm 5.0$ \km\ and the stronger asymmetric feature
extending from $-50$ to +50 \km\ with $\bar{v}_a = -13.7 \pm 4.8$ \km.  The \ovi\ 
absorption wing extending from 5 to 50 \km\ occurs in a velocity
range where there is no \cii\ absorption.  The strong and weak lines of the \ovi\ doublet give similar values of $\log N_a($\ovi$)$ for the main absorption feature ($13.27 \pm0.06$ versus  $13.19 \pm0.10$).   

{\em WD\,1603+432}.-- The 2.3$\sigma$ feature identified as \ovi\ at $\bar{v}_a = -27.6 \pm 9.4$ \km\ 
is well aligned with \cii\ which has $\bar{v}_a = -26.2 \pm 1.2$ \km.  The
weak feature near -85 \km\  on the \ovi\ panel has less than 2$\sigma$
significance.

{\em WD\,1634--573}.-- A detailed analysis of the LISM absorption along the
$37 \pm 3$ kpc line of sight to WD\,1634-573 in the direction $l = 329\fdg88$ and
$b = -7\fdg02$  is presented in the D/H study of Wood et al. (2002).  The
line of sight exhibits absorption in the LIC and/or G clouds at  $v =
-19.6$ \km\ along with a weaker absorption component seen in Si\,{\sc ii}
$\lambda$1260.4 and the \hi\ Lyman lines  near  $v = -3$ \km.  The second
component which is mostly ionized (Lehner et al. 2003) appears to
contain $\sim 250 \times$ less \hi\ than the primary component which has $\log N($\hi$)
= 18.6 \pm 0.2$.  The \ovi\ absorption shown in Fig.~2 with $\log N_a($\ovi$) =
12.71 \pm 0.12$ is well aligned with the \oi\ and \cii\ absorption with 
$\bar{v}_a($\ovi$) -\bar{v}_a($\cii$) = -1.6 \pm 3.7$ \km. 

{\em WD\,1636+351}.-- The principal \cii\ absorption is at $\bar{v}_a = -33.4  \pm 0.6$ \km, 
with a possible weak \cii\ absorption at 25 \km. \oi\ is not
detected at 25 \km\ but the measurements are contaminated by
terrestrial \oi\ emission near that velocity. The \ovi\ absorption with
2.7$\sigma$ significance has $\bar{v}_a = -12.5 \pm 5.0$ \km.  Approximately half of
the \ovi\ absorption occurs for   $v > -10$ \km\  where there is no
corresponding \cii\ absorption.  The apparent \ovi\ absorption from $-115$
to $-70$ \km\ has $<2 \sigma $ significance.  

{\em WD\,1800+685}.-- The principal \cii\ absorption is at $\bar{v}_a= -15.9 \pm 0.5$ 
\km. There is a possible weak second \cii\ component at +48 \km\ which
is not seen in \oi.  The asymmetric \ovi\ absorption detected with 5.8$\sigma$
significance and $\bar{v}_a= -12.5 \pm 5.0$ \km\ is  within the velocity range
of  the \cii\ and \oi\ absorption.

{\em WD\,1844-223, WD\,1845+683m and WD\,2309+195}.--  All three of these WDs have strong LISM \oi\ and \cii\ absorption but no detectable \ovi\ absorption.  The smallest \ovi\ column density limit is for WD\,2309+195 with  $\log N_a($\ovi$) < 12.53$. The upper limit for WD\,2309+195 is 0.15 dex larger than the smallest detected value of $\log N_a($\ovi$) = 12.38$ toward WD\,1211+332.  WDs with no detectable \ovi\ are widely distributed over the sky (see Fig. 1).  

{\em WD\,1950-432}.--The \cii\ and \oi\ absorption is very strong and broad.  There is a possible weak \cii\ component at -90 \km. The \ovi\ absorption is well aligned in velocity with the broad \cii\ and \oi\ absorption. 

{\em WD\,2116+736}.-- The \cii\ and \oi\ absorption are strong and broad. The relatively weak \ovi\ absorption with $2.4 \sigma $ significance is shifted with respect to \cii\ and \oi\ to positive velocity by $21.5 \pm 7.9$ \km. 

{\em WD\,2124-224}.-- The \cii\ and \oi\ absorption are strong and broad. The peak of the \ovi\ absorption is well aligned with the centroid of the \cii\ and \oi\ absorption. However, possible weak and broad  \ovi\ absorption causes the average \ovi\ velocity to be shifted to positive velocity by $12.9 \pm 10.6$ \km.

\clearpage
\begin{deluxetable}{llcccccccc}
\tablecolumns{10}
\tablewidth{0pc} 
\tabletypesize{\scriptsize}
\tablecaption{Summary of the Observational Data \label{t1}} 
\tablehead{\colhead{WD name}    & \colhead{Other name}    &   \colhead{$l$}&   \colhead{$b$}&	\colhead{$d$} &   \colhead{$T_{\rm eff}$} &   \colhead{Type} & \colhead{$v_\star$}    & \colhead{$v_{\rm LISM}$}    &   \colhead{References}\\
\colhead{}    & \colhead{}    &   \colhead{(\degr)}&   \colhead{(\degr)} &  \colhead{(pc)}&  \colhead{(K)}&  \colhead{}&  \colhead{(\km)}&  \colhead{(\km)}&  \colhead{}}
\startdata
WD\,0004$+$330 & GD 2         	&	 112.48  &    $ -28.69 $ &   	  97		&  49360	& DA1	&  NM  	 	&    $ +0.1  $		&  a, 1    \\
WD\,0027$-$636 &MCT 0027-6341 	& 	306.98	 &    $-53.55$ &       238	     	& 63724 	& DA	& $(+30.2)$	& $  (+0.6)$		& b, 2       \\
WD\,0050$-$332 &GD 659        	&	 299.15  &    $ -84.12 $ &   	  58		&  36000	& DA1.5	& $+34.3$	&    $ +9.8 $		&  a, 3    \\
WD\,0113$+$002 &HS\,0111+0012	& 	134.85	 &    $-61.88$ &      \nodata	    	& 65000 	& DO	& NM 		& $(-18.1,+33.1)$	&  b,   2      \\
WD\,0147$+$674 &GD 421        	& 	128.58	 &    $+5.44 $ &       99	    	& 30210		& DA	& NM 		& $(-9.6)$      	&  b,	2      \\
WD\,0416$+$402 &        	& 	160.20	 &    $-6.95 $ &       228	    	& 35227		& DA	& $(+79.1)$	& $(-2.3,+19.6)$	&  a,   2       \\
WD\,0455$-$282 &MCT 0455-2812 	&	 229.29  &    $ -36.17 $ &   	  102		&  57200    	& DA1	& $+69.6$	&    $ +14.0 $		&  a, 1 	\\
WD\,0549$+$158 &GD 71         	&	 192.03  &    $ -5.34  $ &   	  49		&  32750    	& DA1.5	& NM   	 	&    $ +23.2 $		& a, 1    \\
WD\,0603$-$483 &        	& 	255.78	 &    $-27.36$ &       178	    	& 35332		& DA:	& $(+41.0)$	& $(-39.3,+15.3)$	&  a,   2       \\
WD\,0715$-$703 &   		&	 281.62  &    $ -23.50 $ &   	  94		&  43600    	& DA	& NM   	 	&    $ (-9.0)  $	& a, 2	  \\
WD\,0802$+$413 &        	& 	179.22	 &    $+30.94$ & 	230	   	& 45394		& DA	& $(+58.5)$     & $(+15.3,+70.7)$ 	&  b,   2	   \\
WD\,0809$-$728 &        	& 	285.82	 &    $-20.42$ & 	121	   	& 30585		& DA	& NM	      	& $ (-4.1)$       	& c,      2    \\
WD\,0830$-$535 &        	& 	270.11	 &    $-8.27 $ & 	 82	     	& 30500		& DA	& NM	      	& $( +9.2)$       	& c,      2       \\
WD\,0937$+$505 &        	& 	166.90	 &    $+47.12$ & 	 218	   	& 36200		& DA	& NM	      	& $(-5.1) $       	&  b,     2    \\
WD\,1017$-$138 &   		&	 255.74  &    $ +34.53 $ &   	  90		&  32000    	& DA	& NM        	&    $ (-7.5)  $	&  a, 2     \\
WD\,1041$+$580 & PG\,1041+580	& 	150.12	 &    $+52.17$ & 	 93	    	& 30800		& DA	& NM	      	& $(-10.1)$       	& a,     2	    \\
WD\,1100$+$716 &        	& 	134.48	 &    $+42.92$ & 	 141	   	& 43000		& DA	& NM	      	& $(-14.8)$       	& b,    2	 \\
WD\,1211$+$332 &HZ 21 		&	 175.03  &    $ +80.02 $ &   	  $115 \pm35 $  &  53000    	& DO2   & $(+14.8)$	& $ -18.0 $		& d, e, 2, 1  \\
WD\,1234$+$481 &   		&	 129.81  &    $ +69.01 $ &   	  129		&  56400    	& DA1   & NM	    	& $ -28.9 $		&a,    1    \\
WD\,1254$+$223 &GD 153    	&	 317.26  &    $ +84.75 $ &   	  67		&  38686    	& DA1.5 & NM	    	& $ -5.0  $		& a,   4    \\
WD\,1314$+$293 &HZ 43    	&	 54.10   &    $ +84.16 $ &   	  $68 \pm13$	&  50560    	& DA1   & NM	    	& $ -6.8  $		&f, g, 4    \\
WD\,1335$+$700 &        	& 	117.30	 &    $+46.80$ & 	 104	   	& 30289		& DA	& NM	      	& $(-24.5)$       	&b,      2	    \\
WD\,1440$+$751 & HS\,1440+7518 	& 	114.10 	 &    $+40.12$ & 	 98	    	& 42400		& DA:	& NM	      	& $(-18.1)$       	&a,     2	   \\
WD\,1528$+$487 &   		&	 78.87   &    $ +52.72 $ &   	  140		&  47600    	& DA1	& $(+48.6)$	&$(-85.0,-21.6) $	&a,  2	 \\
WD\,1603$+$432 & PG\,1603$+$432	& 	68.23  	 &    $+47.95$ & 	114	   	& 35075		& DA	& NM		& $(-26.2)$      	&b,     2   \\
WD\,1615$-$154 &EGGR 118    	&	 358.79  &    $ +24.18 $ &   	  55		&  29732    	& DA1.5	& NM   	 	&    $ -38.2 $		&f, h,  1         \\
WD\,1620$+$647 &               	& 	96.61  	 &    $+40.16$ & 	 174	   	& 30184		& DA	& NM		& $(-36.7)$       	&b,     2       \\
WD\,1631$+$781 &1ES 1631+78.1   &	 111.29  &    $ +33.58 $ &   	  67		&  44560    	&DA1+dMe& NM   	 	&    $ -11.8 $		&a,  1     \\
WD\,1634$-$573 &HD 149499 B     &	 329.88  &    $ -7.02  $ &    $37 \pm 3$	&  49500    	& DO+KOV& $+0.6$	&    $ -19.6 $		&h, i,  5    \\
WD\,1636$+$351 &   		&	 56.98   &    $ +41.40 $ &   	  109		&  37200    	& DA	& NM   	 	&    $ (-12.5)  $	&a,  2     \\
WD\,1648$+$407 &               	& 	64.64  	 &    $+39.60$ & 	200	    	& 38800		& DA:	& NM		& $(-27.2)$      	&a,     2    \\
WD\,1800$+$685 &   		&	 98.73   &    $ +29.78 $ &   	  159		&  46000    	& DA1   & NM		& $ -15.9 $		&a,  1    \\
WD\,1844$-$223 &   		&	 12.50   &    $ -9.25  $ &   	  62		&  31600    	& DA1   & NM	   	& $(-41.1)  $		&a,  2    \\
WD\,1845$+$683 &               	& 	98.84  	 &    $+25.65$ & 	125	  	& 37400		& DA	& NM		& $(-18.1)$       	&a,     2	    \\
WD\,1917$+$595 & HS\,1917+5954  & 	91.02  	 &    $+20.04$ & 	111	   	& 33000		& DA	& $(+12.2)$    	& $(-24.2)$       	& b,    2	    \\
WD\,1942$+$499 &               	& 	83.08  	 &    $+12.75$ & 	104	    	& 34400 	& DA:	& $(-8.0)$     	& $(-36.4)$       	&b,    2	\\
WD\,1950$-$432 &               	& 	356.49 	 &    $-28.95$ & 	 140	   	& 41339 	& DA	& $(+40:)$      & $(-7.5)$       	&b,    2	\\
WD\,2000$-$561 & MCT 2000-5611 	& 	341.78 	 &    $-32.25$ & 	 198	  	& 47229		& DA	& $(-15.4)$    	& $(-24.2)$      	&b,     2	  \\
WD\,2004$-$605 &   		&	 336.58  &    $ -32.86 $ &   	  58		&  44200    	& DA1	& NM        	&    $ -28.0 $		&a,  1    \\
WD\,2014$-$575 & RE J2018-572  	& 	340.20 	 &    $-34.25$ & 	 51	     	& 27700		& DA	& NM	      	& $(-34.4)$       	&c,     2	    \\
WD\,2111$+$498 & GD 394   	&	 91.37   &    $ +1.13  $ &   	  50		&  37360    	& DA1.5	& $+28.9$	&    $ -6.2  $		&a,  1    \\
WD\,2116$+$736 & RE J2116+735  	& 	109.39 	 &    $+16.93$ & 	 177	  	& 54680		& DA	& NM	      	&$ (-18.0)$       	&b,     2  \\
WD\,2124$-$224 &   		&	 26.81   &    $ -43.19 $ &   	  224 		&  49800    	& DA	& $+29.5$	&    $ -14.8 $		&j,  3    \\
WD\,2146$-$433 &BPS CS22951-0067& 	356.97 	 &    $-49.44$ & 	 362	    	& 67912		&\nodata& $(+27.0)$   	&$ (-7.7) $		&b,     2  	 \\
WD\,2309$+$105 &GD 246   	&	87.26	 &    $-45.11  $ &   	  79		&  58700    	& DA1	& $-12.9$	&    $-7.9   $		&a,  1    \\
WD\,2321$-$549 &  RE J2324-54  	& 	326.91 	 &    $-58.21$ & 	  192	  	& 45860		& DA:	& $(+9.9)$   	&$ (-11.1)$		& b,    2    
\enddata
\tablecomments{The distance and temperature of the WDs are from: (a) Vennes et al. 1997; 
(b) J. Dupuis et al. 2005, in prep.; (c) Vennes et al. 1997; (d) Perryman et al. 1997;   (e) Dreizler \& Werner 1996;
(f) Finley et al. 1997; (g) van Altenna et al. 1995; (h)  Napiwotzki et al. 1995; (i) Holberg et al. 1998; 
(j) Vennes et al. 1998. Distances with errors are from parallax measurements, others
are photometric distances.
The stellar ($v_\star$) and LISM ($v_{\rm LISM}$) heliocentric velocities are from: (1) {\em IUE}\ (Holberg et al. 1998), (2) {\em FUSE}\
(this work); (3) {\em HST}/STIS (Bannister et al. 2003); (4) {\em HST}/STIS  (Redfield \& Linksy 2004), (5) {\em HST}/GHRS (Wood et al. 2002). 
For {\em IUE}\ and {\em HST}\ the error on the absolute heliocentic velocity is $\la 3$ \km, for {\em FUSE}\ the error
on the absolute velocity is unknown and the values are therefore listed inside parentheses. 
``NM" stands for no metal detected in the {\em FUSE}\ bandpass. 
}
\end{deluxetable}

\begin{deluxetable}{ccclc}
\tablecolumns{5}
\tablewidth{0pc} 
\tabletypesize{\scriptsize}
\tablecaption{{\em FUSE}\ Data Summary \label{t1a}} 
\tablehead{\colhead{WD name}      &   \colhead{ID}&   \colhead{Aperture}&   \colhead{Segment}&   \colhead{Time$^a$}   \\
\colhead{}    & \colhead{}    &   \colhead{}&   \colhead{}&  \colhead{(ks)}}
\startdata
WD\,0004$+$330		&  P2041102	& MDRS & LiF\,1A+LiF\,2B 		& 80.0        \\
WD\,0027$-$636		&  Z9030201	& MDRS & LiF\,1A+LiF\,2B+SiC\,1A	& 14.6        \\
WD\,0050$-$332    	&  P2042001	& LWRS & LiF\,1A+LiF\,2B+SiC\,1A	& 8.5		\\
		    	&  M1010101	& LWRS & LiF\,1A+LiF\,2B		& 16.4		\\
WD\,0113$+$002 		&  A0130303	& LWRS & LiF\,1A+LiF\,2B+SiC\,1A	& 2.8         \\
WD\,0147$+$674 		&  Z9030301	& LWRS & LiF\,1A+LiF\,2B+SiC\,1A	& 4.5         \\
	 		&  Z9030302	& LWRS & LiF\,1A+LiF\,2B+SiC\,1A	& 9.8         \\
WD\,0416$+$402 		&  Z9030801	& LWRS & LiF\,1A+LiF\,2B		& 18.5        \\
WD\,0455$-$282    	&  P1041102	& MDRS & LiF\,1A+LiF\,2B		& 19.7		\\
		  	&  P1041103	& MDRS & LiF\,1A+LiF\,2B		& 10.1	  \\
		  	&  P1041104	& MDRS & LiF\,1A+LiF\,2B		& 17.7	  \\
WD\,0549$+$158    	&  P2041701	& LWRS & LiF\,1A+LiF\,2B+SiC\,1A  	& 13.9   \\
WD\,0603$-$483    	&  Z9030901	& LWRS & LiF\,1A+LiF\,2B+SiC\,1A  	& 14.2   \\
WD\,0715$-$703    	&  P2042101	& MDRS & LiF\,1A+SiC\,1A		& 10.2          \\
		    	&  M1050701	& LWRS & LiF\,1A+SiC\,1A		& 14.2          \\
WD\,0802$+$413    	&  Z9031101	& LWRS & LiF\,1A+LiF\,2B+SiC\,1A	& 9.4           \\
WD\,0809$-$728    	&  Z9031201	& LWRS & LiF\,1A			& 9.7           \\
		    	&  Z9031202	& LWRS & LiF\,1A+LiF\,2B		& 24.0          \\
WD\,0830$-$535    	&  Z9031301	& LWRS & LiF\,1A+LiF\,2B		& 14.9          \\
WD\,0937$+$505    	&  Z9031401	& LWRS & LiF\,1A+LiF\,2B+SiC\,1A	& 19.0          \\
WD\,1017$-$138    	&  P2041501	& LWRS & LiF\,1A+LiF\,2B+SiC\,1A 	& 18.9   \\
WD\,1041$+$580     	&  Z9031701	& LWRS & LiF\,1A+LiF\,2B+SiC\,1A 	& 21.2  \\
WD\,1100$+$716     	&  Z9031801	& LWRS & LiF\,1A+LiF\,2B+SiC\,1A 	& 34.1  \\
WD\,1211$+$332    	&  P2040801	& LWRS & LiF\,1A+LiF\,2B+SiC\,1A	& 12.4       \\
		    	&  P2040802	& LWRS & LiF\,1A			& 16.7       \\
		    	&  M1080201	& LWRS & LiF\,1A+LiF\,2B+SiC\,1A	& 4.9       \\
WD\,1234$+$481    	&  P2040901	& LWRS & LiF\,1A+LiF\,2B	  	& 12.2    \\
		    	&  M1052401	& LWRS & LiF\,1A+LiF\,2B	  	& 6.3    \\
		    	&  M1052402	& LWRS & LiF\,1A+LiF\,2B	  	& 12.7    \\
WD\,1254$+$223    	&  P2041801 	& LWRS & LiF\,1A+LiF\,2B+SiC\,1A	& 9.9   \\
    			&  M1010401 	& LWRS & LiF\,1A+LiF\,2B		& 6.3   \\
    			&  M1010402 	& LWRS & LiF\,1A+LiF\,2B		& 12.2 \\
   			&  M1010403 	& LWRS & LiF\,1A+LiF\,2B+SiC\,1A	& 8.1  \\
WD\,1314$+$293    	&  P1042302	& MDRS & LiF\,1A+LiF\,2B	        & 39.6      \\
WD\,1335$+$700    	&  Z9032001	& LWRS & LiF\,1A+LiF\,2B	        & 10.8      \\
WD\,1440$+$751    	&  Z9032201	& LWRS & LiF\,1A+LiF\,2B	        & 20.4      \\
WD\,1528$+$487    	&  P2040101	& LWRS & LiF\,1A+LiF\,2B+SiC\,1A 	& 20.3    \\
		    	&  D0580201	& MDRS & LiF\,1A		  	& 29.4     \\
WD\,1603$+$432    	&  Z9032401	& LWRS & LiF\,1A+LiF\,2B 		& 11.6       \\
WD\,1615$-$154    	&  P2041901	& MDRS & LiF\,1A+LiF\,2B 		& 14.0       \\
WD\,1620$+$647    	&  Z9032501	& LWRS & LiF\,1A	 		& 42.6       \\
WD\,1631$+$781    	&  P1042901	& MDRS & LiF\,1A+LiF\,2B	        & 22.2      \\
		    	&  P1042902	& MDRS & LiF\,1A+LiF\,2B	        & 30.2      \\
		    	&  M1052802	& LWRS & LiF\,1A+LiF\,2B+SiC\,1A	& 5.4      \\
		    	&  M1052803	& LWRS & LiF\,1A+LiF\,2B+SiC\,1A	& 5.3      \\
		    	&  M1052804	& LWRS & LiF\,1A+LiF\,2B	        & 16.3      \\
WD\,1634$-$573    	&  M1031107	& HIRS & LiF\,1A		        & 5.8	     \\
	 	   	&  M1031113 	& HIRS & LiF\,1A		        & 4.9	     \\
	 	   	&  M1031116 	& HIRS & LiF\,1A		        & 5.3	     \\
	 	   	&  M1031119 	& HIRS & LiF\,1A		        & 6.7	     \\
	 	   	&  M1031122	& HIRS & LiF\,1A		        & 5.3	     \\
	 	   	&  S5140203 	& HIRS & LiF\,1A		        & 9.5 	     \\
WD\,1636$+$351    	&  P2040201	& LWRS & LiF\,1A+LiF\,2B          	& 14.0   \\
WD\,1648$+$407    	&  Z9032601	& LWRS & LiF\,1A	          	& 10.8   \\
WD\,1800$+$685    	&  M1053001	& LWRS & LiF\,1A	 	    	& 13.2   \\
		    	&  M1053001	& LWRS & LiF\,1A	 	    	& 16.5   \\
		    	&  M1053002	& LWRS & LiF\,1A	 	    	& 12.4   \\
		    	&  M1053003	& LWRS & LiF\,1A+LiF\,2B		& 12.6   \\
		    	&  M1053005	& LWRS & LiF\,1A+LiF\,2B		& 16.6   \\
		    	&  M1053006	& LWRS & LiF\,1A+LiF\,2B		& 8.5   \\
WD\,1844$-$223    	&  P2040501	& LWRS & LiF\,1A+LiF\,2B+SiC\,1A  	& 13.0     \\
WD\,1845$+$683    	&  Z9032901	& LWRS & LiF\,1A+LiF\,2B+SiC\,1A  	& 36.3     \\
WD\,1917$+$595    	&  Z9033001	& LWRS & LiF\,1A+LiF\,2B	  	& 15.3     \\
WD\,1942$+$499    	&  Z9033101	& LWRS & LiF\,1A+LiF\,2B	  	& 28.2     \\
WD\,1950$-$432    	&  Z9033201	& LWRS & LiF\,1A+LiF\,2B+SiC\,1A	& 15.6     \\
WD\,2000$-$561    	&  Z9033301	& LWRS & LiF\,1A+LiF\,2B		& 12.4     \\
WD\,2004$-$605    	&  P2042203	& LWRS & LiF\,1A+LiF\,2B+SiC\,1A      & 9.6     \\
WD\,2014$-$575    	&  Z9033401	& LWRS & LiF\,1A+LiF\,2B	      	& 18.8    \\
WD\,2111$+$498    	&  P1043601	& LWRS & LiF\,1A+LiF\,2B+SiC\,1A	& 28.6   \\
WD\,2116$+$736    	&  Z9033801	& LWRS & LiF\,1A+LiF\,2B+SiC\,1A	& 16.8   \\
		    	&  Z9033802	& LWRS & LiF\,1A+LiF\,2B+SiC\,1A	& 38.2   \\
WD\,2124$-$224    	&  P2040601 	& MDRS & LiF\,1A+LiF\,2B	        & 19.0     \\
WD\,2146$-$433    	&  Z9033901 	& LWRS & LiF\,1A+LiF\,2B+SiC\,1A      & 17.4     \\
WD\,2309$+$105    	&  P2042401 	& MDRS & LiF\,1A+LiF\,2B+SiC\,1A	& 24.6     \\
WD\,2321$-$549    	&  Z9034201 	& LWRS & LiF\,1A+LiF\,2B+SiC\,1A	& 15.0     
\enddata
\tablecomments{$a$: Total exposure time in LiF\,1A. 
}
\end{deluxetable}

\begin{deluxetable}{lcccccccccc}
\tabcolsep=1pt
\tablecolumns{11}
\tablewidth{0pc} 
\tabletypesize{\scriptsize}
\tablecaption{Summary of the \ovi\ Measurements \label{t2}} 
\tablehead{\colhead{WD name}    & \colhead{$\bar{v}_f$}    &   \colhead{$b_f$}&   \colhead{$\log N_f$}   & \colhead{$\bar{v}_a$}    &	\colhead{$b_a$}&   \colhead{$\log N_a$}   &   \colhead{$ (N_a/d)$}&   \colhead{$W_\lambda^a$} & \colhead{$W_\lambda/\sigma$}&   \colhead{$[-v,+v]$}\\
\colhead{}    & \colhead{\km}    &   \colhead{(\km)}&   \colhead{(cm$^{-2}$)}& \colhead{\km}    &   \colhead{(\km)}&   \colhead{(cm$^{-2}$)}&   \colhead{($10^{-8}$ cm$^{-3}$)}&	\colhead{(m\AA)}&    \colhead{}& \colhead{(\km)}\\
\colhead{(1)}    & \colhead{(2)}    &   \colhead{(3)}&   \colhead{(4)}   &   \colhead{(5)}&   \colhead{(6)}&   \colhead{(7)}&	\colhead{(8)}&    \colhead{9}& \colhead{(10)}& \colhead{(11)}}
\startdata
WD\,0004$+$330  &$+3.2   \pm    1.9     $	&$ 	 17.6  \pm 4.1 	        $  &$  12.81  \pm   0.05 $	     &$-3.8   \pm    3.6     $       &$       21.3 \pm  4.1	    $  &$  12.79  \pm 0.09	     $&   2.06 &$7.6 \pm 1.6$ &   4.8 & $[-47,+38] $		  \\
WD\,0027$-$636 &  $   +5.8 \pm    1.9    $  	&$     23.6 \pm 3.4             $  &$	 13.55   \pm   0.04  $       &$+5.7    \pm    3.1    $       &$      23.2  \pm   4.4	    $  &$    13.55   \pm   0.04  $     &\nodata &$(39.3\pm 4.9)$ &   8.0 & $[-43,+56] $   \\
WD\,0050$-$332  &$+36.3   \pm    3.4    $  	&  	 \nodata  	           &$  12.52  \pm   0.22 $	     &$+37.2   \pm    3.1    $       &       \nodata  		       &$  12.70  \pm\,^{0.10}_{0.14}$& \nodata &$(6.0 \pm 1.7)$ & 3.5 & $[+14,+65]$   \\
WD\,0113$+$002 & \nodata		  	& \nodata		           & \nodata			     & \nodata  		     & \nodata  		       &$ < 13.06	      $        & $<3.72^b$ &$<14.4      $ &  $<2$& $[-68,+16] $        \\
WD\,0147$+$674 & \nodata		  	& \nodata		           & \nodata			     & \nodata  		     & \nodata  		       &$ < 13.01	      $        &$<3.38$&$<12.9      $ &  $<2$& $[-61,+41] $	    \\
WD\,0416$+$402 & \nodata		  	& \nodata		           & \nodata			     & \nodata  		     & \nodata  		       &$ < 13.35	      $        &$<3.31$&$<28.1      $ &  $<2$& $[-50,+50] $	    \\
WD\,0455$-$282  &$-26.0   \pm      3.8  $  	&$ 	 32.1   \pm	7.0     $  &$  13.40  \pm   0.06 $	     &$-23.6   \pm	4.6  $       &$       30.1   \pm    7.4     $  &$  13.42  \pm	0.07 $        &   8.36 &$30.5\pm 5.9$ &   5.2 & $[-74,+38] $		  \\
WD\,0549$+$158  &\nodata		  	& \nodata		           & \nodata			     & \nodata  		     & \nodata  		       &$ < 12.62	      $       & $<2.78$&$<5.3 $       &  $<2$& $[-45,+42] $	    \\
WD\,0603$-$483 &  $  +28.6 \pm    2.8    $  	&      \nodata                     &$	 13.33   \pm   0.19  $       &$+25.6   \pm    7.1    $       &\nodata			       &$13.22\pm\,^{0.15}_{0.22}$     &$2.01$&$18.2\pm 8.2$ &   2.2 & $[-20,+54] $   \\
WD\,0715$-$703  &$-7.0     \pm    2.5   $  	&$ 	 19.5   \pm	5.2     $  &$  13.19  \pm   0.06   $	     &$-6.7	\pm    3.8   $       &$       20.2   \pm     5.1    $  &$  13.23  \pm	0.07   $      &   5.86 &$19.8\pm 4.0$ &   4.9 & $[-41,+38] $		  \\
WD\,0802$+$413 & \nodata		  	& \nodata		           & \nodata			     & \nodata  		     & \nodata  		       &$ < 13.48	      $        &$<6.44$ &$<37.7      $ &  $<2$& $[-50,+50] $	    \\
WD\,0809$-$728 &  $   -4.0 \pm    4.2    $  	&$     26.2 \pm 8.7             $  &$	 13.50   \pm   0.08  $       &$-3.7    \pm    7.1    $       &$      33.2  \pm   8.9	    $  &$    13.60   \pm   0.09  $     &  10.6 &$44.4\pm 9.2$ &   4.8 & $[-66,+54] $   \\
WD\,0830$-$535 &  $   +3.0 \pm    6.4    $  	&     \nodata                      &$	 13.34   \pm   0.12  $       &$+10.5   \pm    7.7    $       &        \nodata  		       &$13.52\pm\,^{0.11}_{0.15}$     &  13.0 &$37.5\pm10.9$ &   3.4 & $[-39,+72] $   \\
WD\,0937$+$505 &  $  +30.1 \pm    8.1    $  	&     \nodata                      &$	 13.34   \pm   0.13  $       &$+23.1   \pm    6.4    $       &        \nodata  		       &$13.48\pm\,^{0.10}_{0.14}$     &  4.49 &$33.7\pm 9.3$ &   3.6 & $[-33,+74] $   \\
WD\,1017$-$138  &$+21.1                 $  	&     \nodata                      &$  12.94 \pm 0.25		$      &$+21.1   \pm	13.5   $     &        \nodata  		       &$  13.29   \pm  0.11 $        &   7.02 &$12.8\pm 5.8$ &   2.2 & $[-43,+67]$		  \\	  
WD\,1041$+$580 & \nodata		  	& \nodata		           & \nodata			     & \nodata  		     & \nodata  		       &$ < 12.99	      $        &$<3.44$&$<12.1      $ &  $<2$& $[-51,+27] $	    \\
WD\,1100$+$716 &  $  +11.7 \pm    2.0    $  	&$     20.6 \pm 4.0             $  &$	 13.32   \pm   0.05  $       &$+4.1    \pm    3.5    $       &$      27.2  \pm   3.9	    $  &$13.30 \pm 0.05 	 $     &  4.59 &$23.2\pm 3.3$ &   7.0 & $[-47,+46] $   \\
WD\,1211$+$332  &$+5.4   \pm    5.7    $  	& \nodata		           &$  12.30	\pm   0.21 $	      &$+4.0 \pm 5.5    $	     & \nodata  		       &$  12.38  \pm\,^{0.17}_{0.28} $&  0.68  &$3.0 \pm 1.4$ &   2.1 & $[-15,+21] $	 \\
WD\,1234$+$481  &$-17.4   \pm    3.6    $  	&$   20.1 \pm 8.6 	        $  &$  12.89   \pm   0.08 $	     &$-14.7\pm    3.8    $	  &$	17.9 \pm 5.6		     $ &$  12.91   \pm   0.08 $       &   3.16 &$9.8 \pm 1.9$ &   5.2 & $[-43,+26] $		  \\
WD\,1254$+$223  &$+22.9   \pm    1.8    $  	&$   20.5 \pm 3.6 	        $  &$  13.11   \pm   0.04 $	     &$+19.4\pm    3.1    $	  &$   23.1  \pm 4.1		     $ &$  13.10   \pm   0.05 $       &   4.21 &$15.0\pm 2.0$ &   7.5 & $[-26,+67] $		  \\
WD\,1314$+$293  &$+5.7   \pm     2.3    $  	&$       28.6   \pm    4.1      $  &$	      12.95   \pm    0.04 $  &$+13.1   \pm     3.7    $       &$       27.4   \pm    4.4     $  &$  12.94   \pm    0.06 $     &   4.15 &$10.6\pm 1.7$ &   6.2 & $[-34,+67] $\\
WD\,1335$+$700 & \nodata		  	& \nodata		           & \nodata			     & \nodata  		     & \nodata  		       &$ < 13.31	      $        &$<6.42$&$<25.6      $ &  $<2$& $[-60,+19] $	    \\
WD\,1440$+$751 & \nodata		  	& \nodata		           & \nodata			     & \nodata  		     & \nodata  		       &$ < 13.07	      $        &$<3.92$&$<14.8      $ &  $<2$& $[-71,+31] $	    \\
WD\,1528$+$487  &$-85.0   \pm     7.9 	$  	&     \nodata                      &$  12.40	\pm   0.25  $	     &$-72.1   \pm     5.0   $       &     \nodata                     &$  12.82  \pm\,^{0.09}_{0.13}$&   1.53 &$7.9 \pm 2.1$ &   3.8 & $[-106,-50]$			  \\
		&$-25.9   \pm     3.7 	$  	&$      36.2   \pm   7.3        $  &$  13.31	\pm   0.05  $	     &$-13.7   \pm     4.8   $       &$      31.7   \pm   7.4	    $  &$  13.27    \pm   0.06  $     &   4.31 &$21.9\pm 3.7$ &   5.9 & $[-50,+67 ]$			  \\
WD\,1603$+$432 &   $  -30.1 \pm    4.6    $  	&     \nodata                      &$	 12.93   \pm   0.18  $       &$-27.6   \pm    9.4    $       & \nodata  		       &$13.06\pm\,^{0.15}_{0.24}$     &  3.26 &$13.2\pm 5.7$ &   2.3 & $[-71,+11] $   \\
WD\,1615$-$154  &\nodata		  	& \nodata		           & \nodata			     & \nodata  		     & \nodata  		       &$ < 12.94	      $       & $<5.18$&$<10.8      $ &  $<2$& $[-90,+10] $	    \\
WD\,1620$+$647 & \nodata		  	& \nodata		           & \nodata			     & \nodata  		     & \nodata  		       &$ < 13.30	      $        &$<4.42$&$<25.2      $ &  $<2$& $[-110,+30] $	    \\
WD\,1631$+$781  &$-20.7   \pm      4.0 	$  	& \nodata		           &$	   12.51    \pm    0.11 $    &$-16.4   \pm	5.1  $       & \nodata		                &$  12.52 \pm\,^{0.12}_{0.17}$&   1.60 &$4.0 \pm 1.3$ &   3.1 & $[-51,+24] $\\
WD\,1634$-$573  &$-21.2\pm 2.1		$  	&$     17.2 \pm 4.7	        $  &$	   13.00    \pm    0.05 $    &$-21.2   \pm	3.7  $       &$      22.2 \pm 5.3   	     $  &$     13.04 \pm 0.06	     $&   9.6  &$13.0\pm 2.1$ &   6.2 & $[-46,+52] $\\
WD\,1636$+$351  &$ -16.4  \pm      3.8 	$  	& \nodata		           &$12.98   \pm    0.11   $	     &$ -12.5 \pm 5.0	     $       & \nodata		               &$12.95\pm\,^{0.12}_{0.20}$    &   2.65 &$10.4\pm 3.9$ &   2.7 & $[-47,+19] $	  \\
WD\,1648$+$407 &  $  -14.0 \pm    6.1    $  	& \nodata		           &$	 13.14   \pm   0.18  $       &$-15.0   \pm    6.5    $       & \nodata		               &$13.14\pm\,^{0.18}_{0.29}$     &  2.24 &$15.6\pm 7.9$ &   2.0 & $[-48,+20] $   \\
WD\,1800$+$685  &$ -10.5  \pm      1.6 	$  	&$    15.0 \pm 3.5              $  &$12.99   \pm    0.04   $	     &$ -14.3 \pm  2.9       $       &$19.9 \pm 3.3		    $  &$12.96\pm 0.07  	 $    &   1.86 &$11.0\pm 1.9$ &   5.8 & $[-53,+19] $	  \\
WD\,1844$-$223  &\nodata		  	& \nodata		           & \nodata			     & \nodata  		     & \nodata  		       &$ < 12.84	      $       & $<3.65$&$<  8.6     $ &  $<2$& $[-95,+5 ] $	    \\
WD\,1845$+$683 & \nodata		  	& \nodata		           & \nodata			     & \nodata  		     & \nodata  		       &$ < 12.95	      $        &$<2.33$&$<11.2      $ &  $<2$& $[-78,+38] $	    \\
WD\,1917$+$595 & \nodata		  	& \nodata		           & \nodata			     & \nodata  		     & \nodata  		       &$ < 13.10	      $        &$<3.71$&$<15.7      $ &  $<2$& $[-74,+27] $	    \\
WD\,1942$+$499 &  $  -2.7  \pm    2.1    $  	& \nodata		           &$	 13.12   \pm   0.13  $       &$-0.8    \pm    5.1    $       & \nodata  		       &$13.00\pm\,^{0.13}_{0.17}$     &\nodata&$(11.3\pm 3.8)$&   3.0 & $[-31,+33] $   \\
WD\,1950$-$432 &  $  -3.4  \pm    5.8    $  	&  \nodata		           &$	 13.19   \pm   0.10  $       &$-4.9    \pm    7.3    $       & \nodata		               &$13.30\pm\,^{0.10}_{0.13}$     &  4.62 &$23.0\pm 6.1$ &   3.8 & $[-73,+51] $   \\
WD\,2000$-$561 &  $  -19.7 \pm    2.6    $  	& $ 27.8 \pm 4.6                 $ &$	 13.69   \pm   0.05  $       &$-19.6   \pm    4.1    $       &$  25.1 \pm 5.8		    $  &$13.68 \pm 0.06 	 $     &\nodata &$(52.1\pm 8.1)$&   6.4 & $[-75,+31] $   \\
WD\,2004$-$605  &$-26.3   \pm    3.7 	$  	& \nodata		           &$  12.96  \pm    0.09 $	     &$-23.2   \pm    6.4    $       & \nodata		               &$  13.00  \pm	 0.10 $       &   5.59 &$12.0\pm 3.3$ &   3.6 & $[-64,+24] $	\\
WD\,2014$-$575 & \nodata		  	& \nodata		           & \nodata			     & \nodata  		     & \nodata  		       &$ < 13.19	      $        &$<9.93$&$<19.5      $ &  $<2$& $[-83,+24] $	    \\
WD\,2111$+$498  &$+30.8   \pm    10.9	$  	& \nodata		           &$  13.05  \pm    0.12 $	     &$+21.1   \pm    8.4    $       & \nodata		               &$  12.90  \pm	 0.13 $       & \nodata&$(9.5 \pm 3.1)$&   3.1 & $[-30,+60] $	\\
WD\,2116$+$736 &  $  -5.7  \pm    8.6    $  	& \nodata		           &$	 12.43 \pm 0.27      $       &$+3.5    \pm    7.9    $       & \nodata		               &$12.68\pm\,^{0.15}_{0.24}$     &  0.88 &$5.7 \pm 2.4$ &   2.4 & $[-36,+43] $	\\
WD\,2124$-$224  &$   -13.2\pm    3.4    $  	& \nodata		           &$	 12.88   \pm   0.11  $       &$-1.9    \pm    10.6   $       & \nodata		               &$    13.07   \pm   0.11  $    &   1.70 &$13.9\pm 4.4$ &   3.2 & $[-68,+53] $   \\
WD\,2146$-$433 &  $  +8.9  \pm    2.8    $  	& \nodata                          &$	 13.15   \pm   0.10  $       &$+13.5   \pm    6.4    $       & \nodata		               &$13.21\pm\,^{0.12}_{0.16}$     &\nodata &$(18.7\pm 6.1)$ &   3.1 & $[-19,+64] $   \\
WD\,2309$+$105  &\nodata 		       & \nodata			  & \nodata			    & \nodata			    & \nodata			      &$ < 12.53	     $        & $<1.40$&$<4.3	   $ &  $<2$& $[-60,+40] $	   \\
WD\,2321$-$549 &  $  -1.5  \pm    3.9    $  	& $ 33.9 \pm 7.2                 $ &$	 13.49   \pm   0.06  $       &$-4.5    \pm    4.2    $       &$  31.8 \pm 4.1		    $  &$13.47 \pm 0.06 	 $     & \nodata&$(52.1\pm 8.1)$&     6.4 & $[-75,+31] $  
\enddata
\tablecomments{A value preceded by ``$<$" is a 2$\sigma$ upper limit. 
WD\,1017$-$138: We fixed the velocity in the profile fit at 21.1 \km\ to be consistent with the AOD velocity.
$a$: Equivalent widths listed in parentheses are for 7 stars 
for which stellar \ovi\ contamination is likely (see \S~3.2). The line profiles for these objects 
are shown in Fig.~3. Measurements for these stars are not used to infer information about \ovi\ in the LISM. 
$b:$ We assume a distance of 100 pc. 
}
\end{deluxetable}

\begin{deluxetable}{lccc}
\tablecolumns{4}
\tablewidth{0pc} 
\tabletypesize{\scriptsize}
\tablecaption{Detection of \ovi\ $\lambda$1038 \label{t2a}} 
\tablehead{\colhead{WD name}   & \colhead{$\bar{v}_a$}    &   \colhead{$\log N_a$} &   \colhead{$W_\lambda$}\\
\colhead{}    & \colhead{(\km)}  &   \colhead{(cm$^{-2}$)}&   \colhead{(m\AA)}}
\startdata
WD\,0004$+$330  &$-8.9 \pm 6.3		 $	&$	12.75 \pm 0.12         	 $	&$3.5 \pm 1.1$        \\
WD\,0027$-$636  &$+14.7 \pm 5.4    $	   	&$   13.52 \pm 0.07   		    $	&$19.5\pm 3.7$        \\
WD\,0715$-$703  &$-15.2\pm 8.7   	$	&$    13.30 \pm 0.15		  $	&$9.6 \pm 3.8$        \\
WD\,1254$+$223  &$+16.4 \pm 5.4	        $	&$   13.12 \pm 0.09    	       $	&$8.1 \pm 1.7$        \\
WD\,1314$+$293  &$+12.4 \pm 9.5  $		&$     12.91 \pm 0.13 		    $	&$5.1 \pm 1.5$        \\
WD\,1528$+$487  &$-26.0   \pm	  13.2	$	&$	13.19   \pm   0.10	$  	&$10.1\pm 2.3$        \\
WD\,2000$-$561  &$-29.1 \pm 7.4		   $	&$   13.68 \pm 0.10		   $	&$28.0\pm 7.1$        \\
WD\,2321$-$549  &$+4.7  \pm 6.1    $	   	&$  13.54 \pm 0.12		   $	&$20.3\pm 5.1$ 
\enddata
\end{deluxetable}

\begin{deluxetable}{lcccc}
\tablecolumns{5}
\tablewidth{0pc} 
\tabletypesize{\scriptsize}
\tablecaption{Average Velocities of LISM \ovi, \cii, and \oi$^a$ \label{t3}} 
\tablehead{\colhead{WD name}      &   \colhead{$\bar{v}_a({\mbox \ovi})$}&   \colhead{$\bar{v}_a({\mbox \cii})$} &   \colhead{$\bar{v}_a({\mbox \oi})$} &   \colhead{$\bar{v}_a({\mbox \ovi})-\bar{v}_a({\mbox \cii})$}  \\
\colhead{}    &  \colhead{(\km)}  &  \colhead{(\km)}  &  \colhead{(\km)} }
\startdata
WD\,0004$+$330 & $  -3.8   \pm  3.6 $	 & $  +0.1 \pm 0.2  $  & $  -1.0 \pm 0.5  $   	   & $  -3.9   \pm  3.6 $    \\
WD\,0455$-$282 & $  -23.6  \pm   4.6$	 & $ +14.0 \pm 0.6  $  & $ +16.9 \pm 2.5  $  	   & $  -37.6  \pm   4.6$   \\
WD\,0603$-$483 & $+25.6   \pm	 7.1$	 & $  +15.3\pm 1.4  $  & $  +16.6\pm 2.2  $  	   & $+10.3   \pm    7.2$   \\
WD\,0715$-$703 & $  -6.7   \pm   3.8$	 & $  -9.0 \pm 0.4  $  & $  -8.1 \pm 0.4  $  	   & $  +2.3   \pm   3.8$   \\
WD\,0809$-$728 & $-3.7    \pm	 7.1$	 & $  -4.1 \pm 1.0  $  & $  -12.0\pm 1.3  $$^b$	   & $+0.4    \pm    7.2$   \\
WD\,0830$-$535 & $+10.5   \pm	 7.7$	 & $  +9.2 \pm 0.8  $  & $ +12.7 \pm 2.1  $  	   & $+1.3    \pm    7.7$   \\
WD\,0937$+$505 & $+23.1   \pm	 6.4$	 & $ -5.1  \pm 1.5  $  & $ -8.8  \pm 1.2  $$^b$	   & $+28.2   \pm    6.4$   \\
WD\,1017$-$138 & $  +21.1  \pm  13.5$	 & $ -7.5  \pm 0.9  $  & $ -13.5 \pm 1.3  $  	   & $  +28.6  \pm  13.5$   \\
WD\,1100$+$716 & $+4.1    \pm	 3.5$	 & $ -14.8 \pm 0.6  $  & $ -20.7 \pm 1.1  $$^b$	   & $+18.9   \pm    3.6$   \\
WD\,1211$+$332 & $+4.0    \pm	 5.5$	 & $ -17.9 \pm 0.4  $  & $ -17.1 \pm 0.5  $	   & $+21.9   \pm    5.5$   \\
WD\,1234$+$481 & $  -14.7  \pm  3.8 $	 & $ -28.9 \pm 0.6  $  & $ -26.5 \pm 0.7  $  	   & $  +14.2  \pm  3.8 $   \\
WD\,1254$+$223 & $  +19.4  \pm  3.1 $	 & $ -5.0  \pm 0.7  $  & $ -3.6  \pm 1.6  $  	   & $  +24.4  \pm  3.1 $   \\
WD\,1314$+$293 & $  +13.1   \pm 3.7 $	 & $ -6.8 \pm 0.6   $  & $ -6.4  \pm 1.4  $  	   & $  +19.9	\pm 3.7 $   \\
WD\,1528$+$487 & $  -72.1  \pm  5.0 $	 & $ -57.5 \pm 1.0  $  & \nodata$^c$	     	   & $  -14.6  \pm  5.0 $   \\
	       & $  -13.7  \pm  4.9 $	 & $ -21.6 \pm 0.3  $  & $ -21.1 \pm 0.6  $  	   & $  +7.9   \pm  4.9 $   \\
WD\,1603$+$432 & $-27.6   \pm	 9.4$	 & $ -26.2 \pm 1.2  $  & $ -30.1 \pm 1.6  $  	   & $-1.4    \pm    9.5$   \\
WD\,1631$+$781 & $  -16.4  \pm  5.1 $	 & $ -11.8 \pm 0.3  $  & $ -12.1 \pm 0.5  $  	   & $  -4.6   \pm  5.1 $   \\
WD\,1634$-$573 & $  -21.2 \pm   3.7 $	 & $ -19.6 \pm 0.3  $  & $ -19.9 \pm 0.4  $  	   & $  -1.6   \pm  3.7 $   \\
WD\,1636$+$351 & $  -12.5  \pm 5.0  $	 & $ -33.4 \pm 0.6  $  & $ -33.2 \pm 1.0  $  	   & $  +20.9  \pm 5.0  $   \\
WD\,1648$+$407 & $-15.0   \pm	 6.5$	 & $ -27.2\pm 2.2   $  & $ -33.9 \pm 1.8  $$^b$	   & $+12.2   \pm    6.9$   \\
WD\,1800$+$685 & $  -14.3  \pm  2.9 $	 & $ -15.9 \pm 0.5  $  & $ -16.3 \pm 0.4  $  	   & $  +1.6   \pm  2.9 $   \\
WD\,1950$-$432 & $-4.9    \pm	 7.3$	 & $ -7.5  \pm 0.7  $  & $ -8.9  \pm 0.9  $  	   & $+2.6    \pm    7.3$   \\
WD\,2004$-$605 & $  -23.2  \pm  6.4 $	 & $ -28.1 \pm 0.5  $  & $ -29.0 \pm 0.6  $  	   & $  +4.9   \pm  6.4 $   \\
WD\,2116$+$736 & $+3.5    \pm	 7.9$	 & $ -18.0 \pm 0.5  $  & $ -18.9 \pm 0.4  $  	   & $+21.5  \pm    7.9$   \\
WD\,2124$-$224 & $ -1.9   \pm  10.6 $	 & $ -14.8 \pm 0.7  $  & $ -16.3 \pm 0.6  $  	   & $ +12.9  \pm  10.6 $   
\enddata
\tablecomments{$a$: Velocities are listed for 24 WDs from Table~3 with $\ge 2\sigma$ \ovi\ 
detections and no evidence for stellar contamination. $b$: \oi\ is complicated by the 
airglow line. $c$: A high velocity component is present in \cii, but not 
in \oi. }
\end{deluxetable}

\begin{deluxetable}{lcc}
\tablecolumns{3}
\tablewidth{0pc} 
\tabletypesize{\scriptsize}
\tablecaption{Values of $n$(\ovi) in the LISM$^a$  \label{t5}} 
\tablehead{\colhead{Method} &   \colhead{$n({\mbox \ovi})$}    &  \colhead{$n({\mbox \ovi}) \pm \sigma$}  \\
\colhead{}   &  \colhead{median}  &  \colhead{average}\\
\colhead{}   &  \colhead{($10^{-8}$ cm$^{-3}$)}  &  \colhead{($10^{-8}$ cm$^{-3}$)} }
\startdata
Median and average determined from data for 39 stars where		&  3.7 & $ 4.4 \pm 2.8^b  $   \\ 
the 2$\sigma$ upper limits for 14 stars are treated as detections.	&    &    \\
& & \\
Median and average determined from data for 24 stars where  		& 4.3 & $ 4.6 \pm 3.2^b  $   \\
\ovi\ has been detected with $\ge 2\sigma$ significance.		&    &    \\
& & \\
Estimate of the average from data for 39 stars where the		&\nodata& $ 3.6 \pm 0.5^c   $   \\
contributions for the 2$\sigma$ upper limits are evaluated 		&    &    \\
using the Kaplan-Meier analysis method.    		     	     	&    &    
\enddata
\tablecomments{$a$: Medians and averages are derived from the 39 measurements and limits
in Table~\ref{t2} for WDs with little or no stellar contamination. 
The high velocity component of \ovi\ toward WD\,1528$+$487 is not included.
$b$: The listed error is the dispersion about the average value listed. 
$c$: The listed error is the Kaplan-Meier estimator error on the derived average value 
(see \S~4.1).
}
\end{deluxetable}

\begin{deluxetable}{lccl}
\tablecolumns{4}
\tablewidth{0pc} 
\tabletypesize{\scriptsize}
\tablecaption{LISM \ovi\ and \oi\ Column Densities$^a$ \label{t4}} 
\tablehead{\colhead{WD name} &  \colhead{$d$}      &  \colhead{$\log N_a$(\ovi)}&   \colhead{$\log N$(\oi)}  \\
\colhead{}     &  \colhead{(pc)} &  \colhead{(cm$^{-2}$)}  &  \colhead{(cm$^{-2}$)} }
\startdata
WD\,0004$+$330     &	   97		    &  $  12.79  \pm 0.09	     $    & $  16.35 \pm 0.15		    $	   \\
WD\,0455$-$282     &	   102  	    &  $  13.42  \pm   0.07 $		  & $  14.91 :  	    $	    \\
WD\,0549$+$158     &	   49		    &  $ < 12.62	     $  	  & $  14.27 \pm\,^{0.07}_{0.09}  $    \\
WD\,0715$-$703     &	   94		    &  $  13.23  \pm   0.07   $ 	  & $  15.90 \pm\,^{0.10}_{0.09}  $    \\
WD\,1017$-$138     &	   90		    &  $  13.29   \pm  0.11 $		  & $  15.95 \pm\,^{0.45}_{0.22}  $    \\
WD\,1211$+$332     &	   115		    &  $  12.38\pm\,^{0.17}_{0.28} $	  & $  15.74 \pm\ 0.05  $   		 \\
WD\,1234$+$481     &	   129  	    &  $  12.91   \pm	0.08 $  	   & $  15.63 \pm\,^{0.08}_{0.07}  $	\\
WD\,1254$+$223     &	   67		    &  $  13.10   \pm	0.05 $  	   & $  14.25 \pm\,^{0.06}_{0.05}  $	\\
WD\,1314$+$293     &	   68 		    &  $  12.94   \pm	  0.06 $	   & $  14.51 \pm 0.03  	 $    \\
WD\,1528$+$487     &	   140  	    &  $  13.27    \pm   0.06  $	  & $  15.80 \pm 0.08		    $	 \\
WD\,1615$-$154     &	   55		    &  $ < 12.94	     $  	  & $  15.78 \pm\,^{0.10}_{0.09}  $    \\
WD\,1631$+$781     &	   67		    &  $  12.52 \pm\,^{0.12}_{0.17}$	  & $  15.90 \pm 0.09		    $	 \\
WD\,1634$-$573     &       37 		    &  $     13.04 \pm 0.06	    $	   & $  15.51 \pm 0.03       $    \\
WD\,1636$+$351     &	   109  	    &  $12.95\pm\,^{0.12}_{0.20}$	  & $  15.71 \pm\,^{0.24}_{0.13}  $    \\
WD\,1800$+$685     &	   159  	    &  $12.96\pm 0.07		$	  & $  16.12 \pm\,^{0.14}_{0.12}  $    \\
WD\,1844$-$223     &	   62		    &  $ < 12.84	     $  	& $  15.97 \pm 0.08		  $    \\
WD\,2004$-$605     &	   58		    &  $  13.00  \pm	0.10 $  	& $  15.65 \pm 0.08		  $    \\
WD\,2124$-$224     &	   224		    &  $    13.07   \pm   0.11  $	& $  15.94 \pm 0.03		  $    \\
WD\,2309$+$105     &	   79		    &  $ < 12.53	    $		& $  15.67 \pm 0.04		  $    
\enddata
\tablecomments{$a$:
The \ovi\ column densities are from the AOD values listed in Table~\ref{t2} for WDs with 
no evidence of stellar contamination. The limits are 2$\sigma$.
The \oi\ column densities are from 
Lehner et al. (2003) and references therein, except toward WD\,1254$+$223 which is from 
Oliveira et al. (2005). See Table~1 for distance references. }
\end{deluxetable}

\clearpage

\begin{figure}[tbp]
\epsscale{1}
\plotone{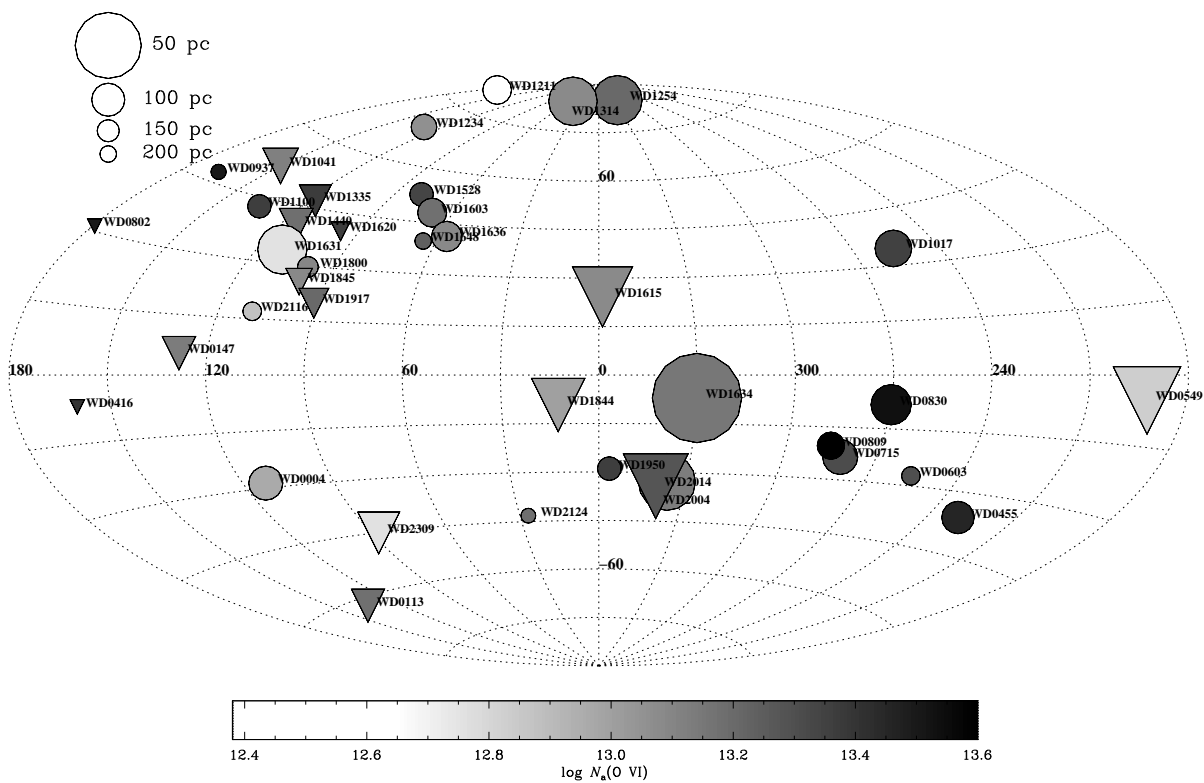}
\caption{WD locations in  Galactic coordinates ($l,b$) for the 39 stars from Table~\ref{t1} not 
likely affected by stellar blending (a distance of 100 pc is assumed for the purpose of this plot 
for WD\,0113$+$002). 
The size of each symbol is inversely proportional to  the distance of the line of sight, 
and the shading of the symbol indicates the total column density of \ovi. Note that {\em circles} are \ion{O}{6} 
detections, while the {\em triangles} are 2$\sigma$ upper limits for  \ion{O}{6}.
\label{ovimap}}
\end{figure}

\begin{figure}[tbp]
\epsscale{1}
\plotone{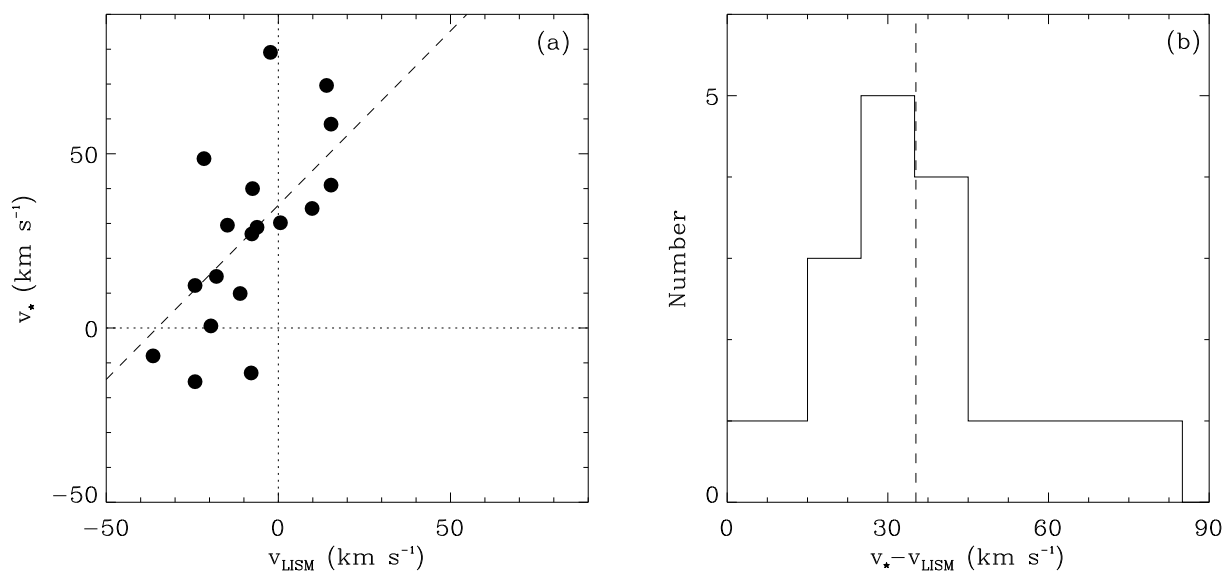}
\caption{In (a) the WD stellar heliocentric velocity, $v_\star$, is compared to the 
LISM velocity for the measurements listed in Table~1. In (b) the number
histogram of $v_\star - v_{\rm LISM}$ is displayed. The stellar velocity
is on average redshifted by 35 \km\  relative to the gas in the LISM. Most
of this shift is probably from the gravitational redshift of the WDs. 
The dashed line in (a) and (b) shows the average value of $v_\star - v_{\rm LISM}$.
\label{stellar}}
\end{figure}

\begin{figure}[tbp]
\epsscale{1}
\plotone{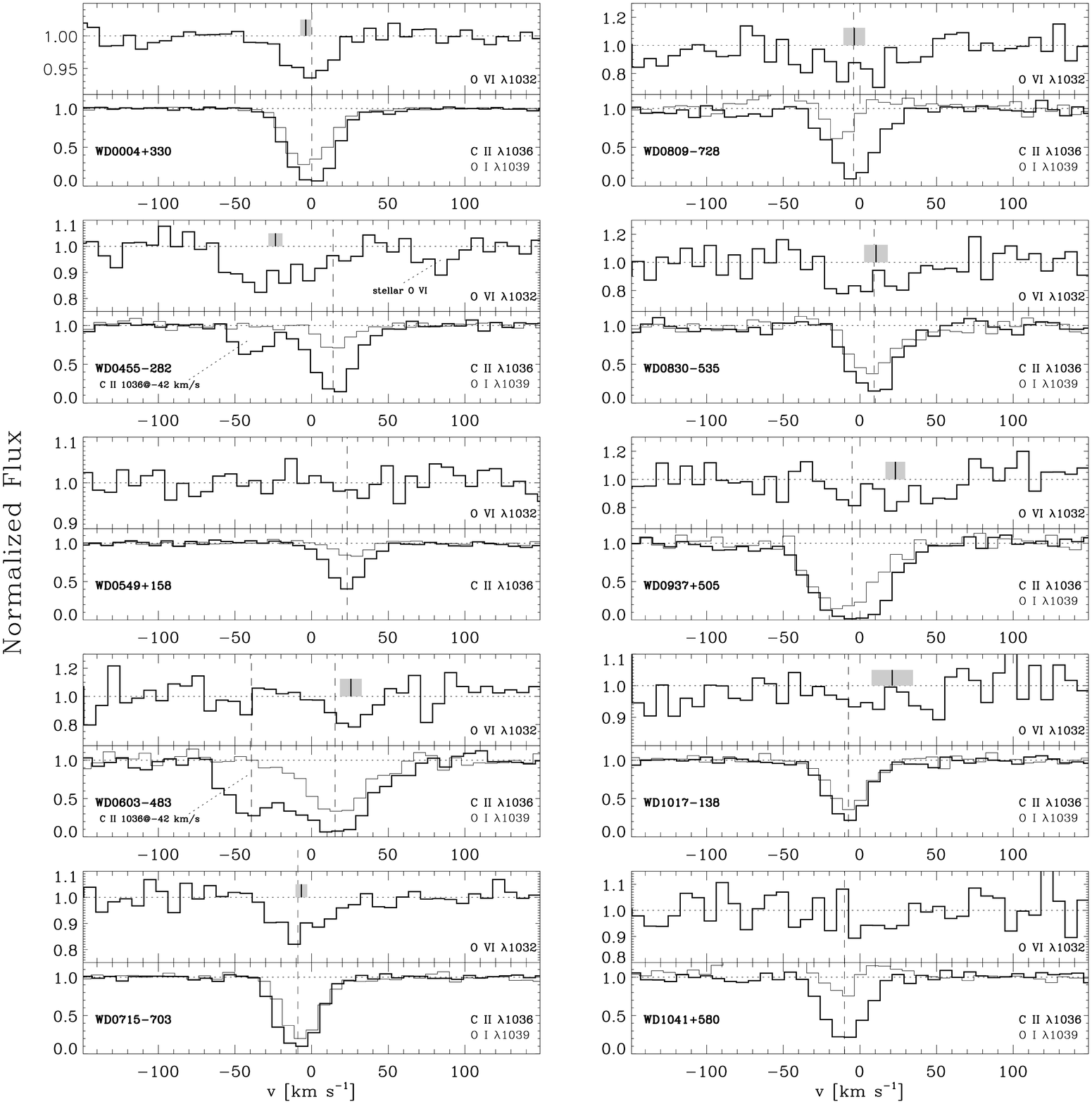}
\caption{Normalized profiles versus the heliocentric velocity of LISM  \ovi, \cii, and \oi\
absorption. The  {\fuse}\/ velocity was corrected
using the LISM values in Table~\ref{t1}, except for the cases where we do not have accurate absolute velocities. 
The vertical dotted line is the \cii\  average velocity. The tick mark above the \ovi\ profile is the \ovi\ AOD velocity with
the gray area showing its error. We only display the 2$\sigma$ upper limits from  Table~\ref{t1} when $\log N($\ovi$)< 13.00$.
\label{plotnew}}
\end{figure}

\begin{figure}[tbp]
\epsscale{1}
\plotone{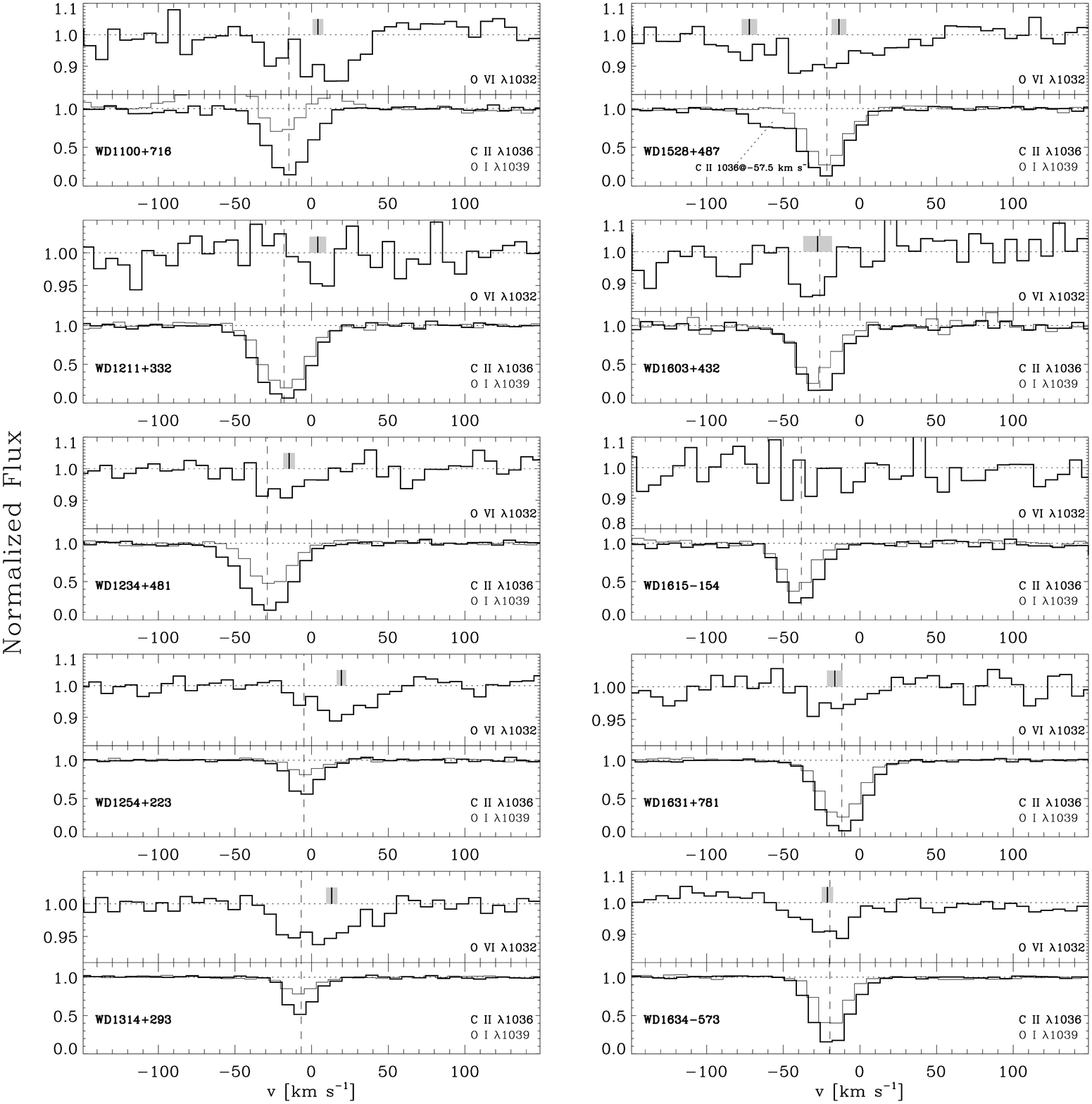}
\figurenum{\ref{plotnew}}
\caption{continued.}
\end{figure}

\begin{figure}[tbp]
\epsscale{1}
\plotone{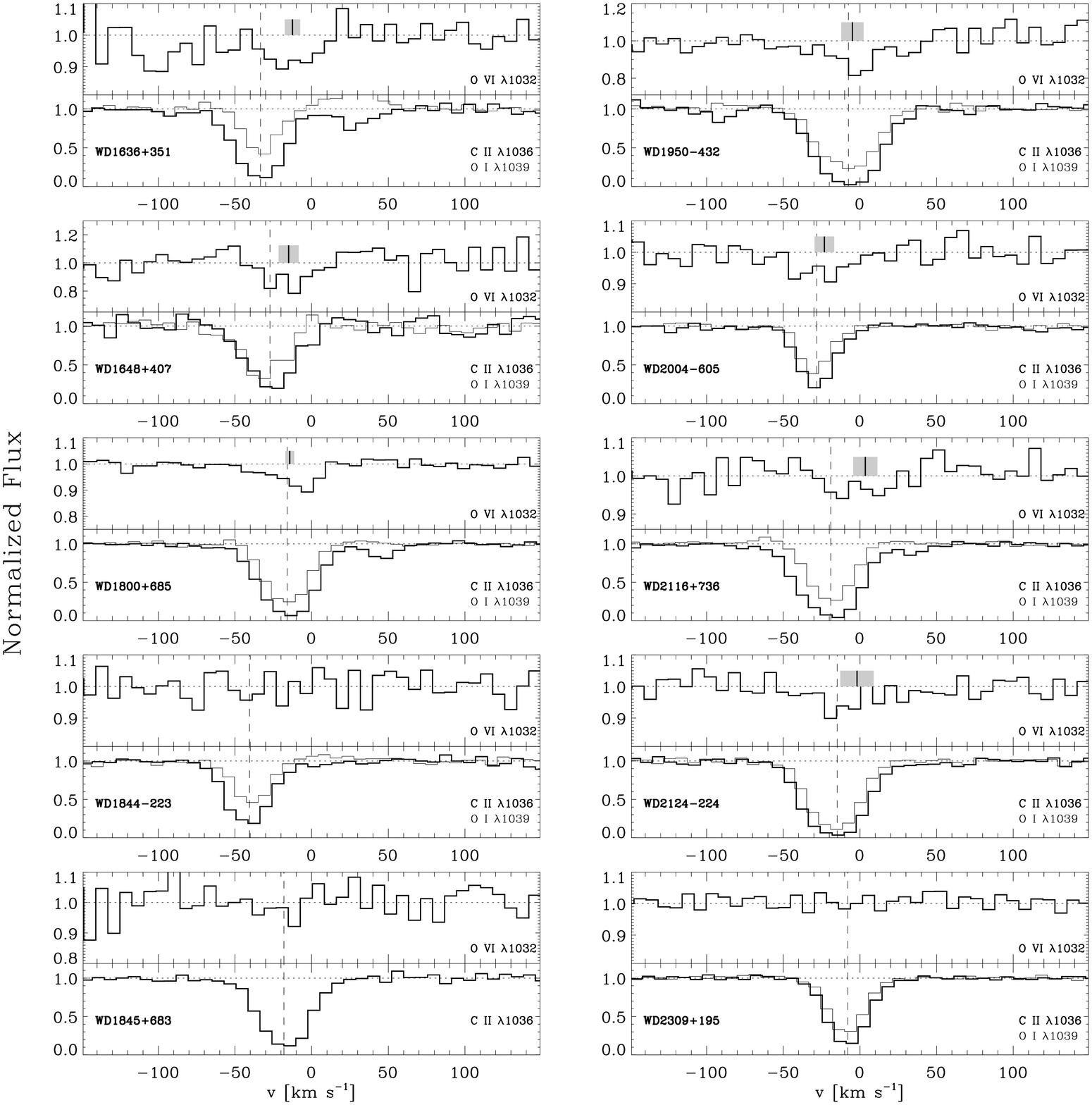}
\figurenum{\ref{plotnew}}
\caption{continued.}
\end{figure}

\begin{figure}[tbp]
\epsscale{1}
\plotone{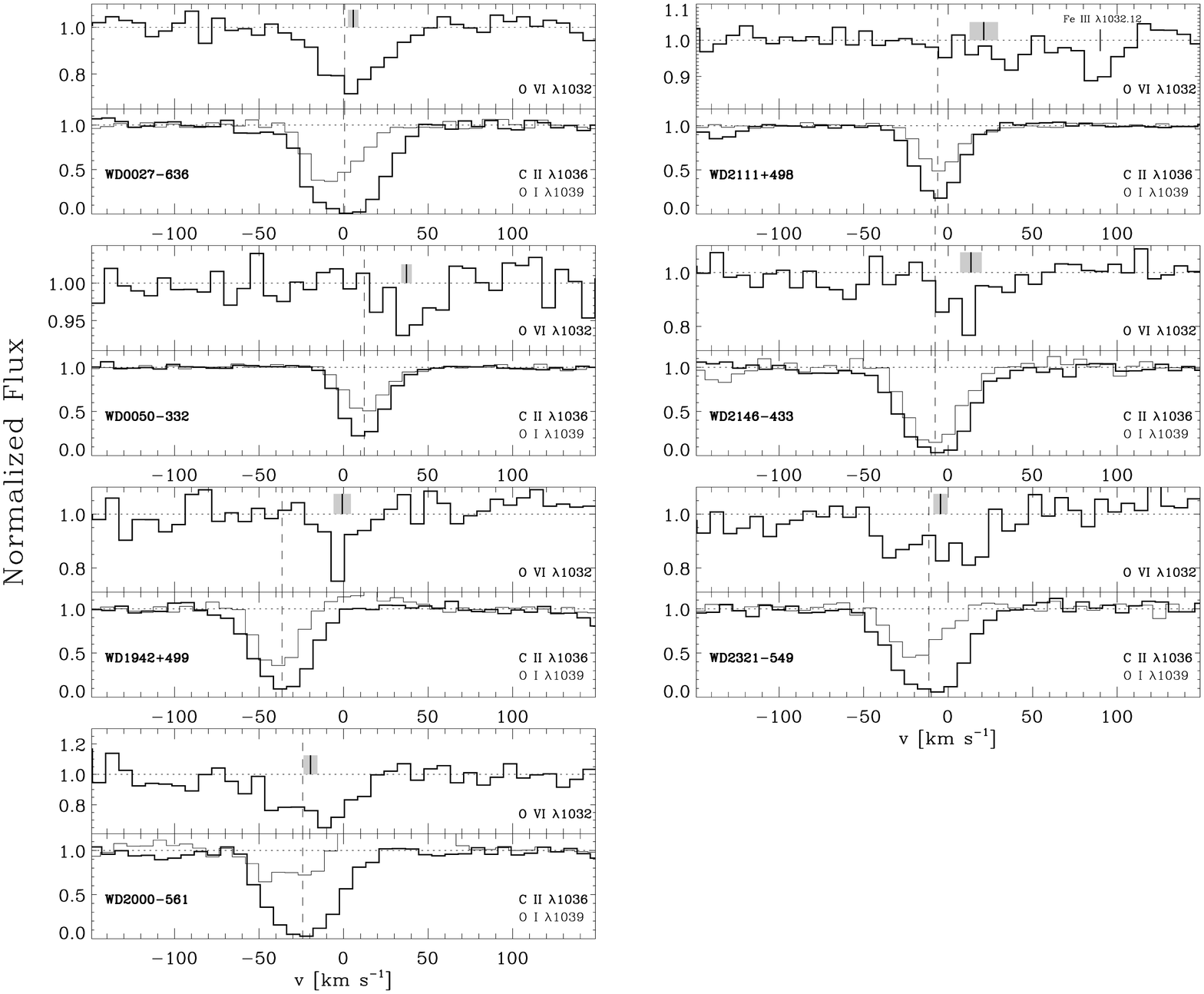}
\caption{Normalized profiles for \ovi, \cii, and \oi\ versus the heliocentric 
velocity. Same as in Fig~\ref{plotnew} except for  those WDs where much of
the observed \ovi\ absorption may be stellar rather LISM (see \S~3.2).
\label{plotnew1}}
\end{figure}

\begin{figure}[tbp]
\epsscale{1}
\plotone{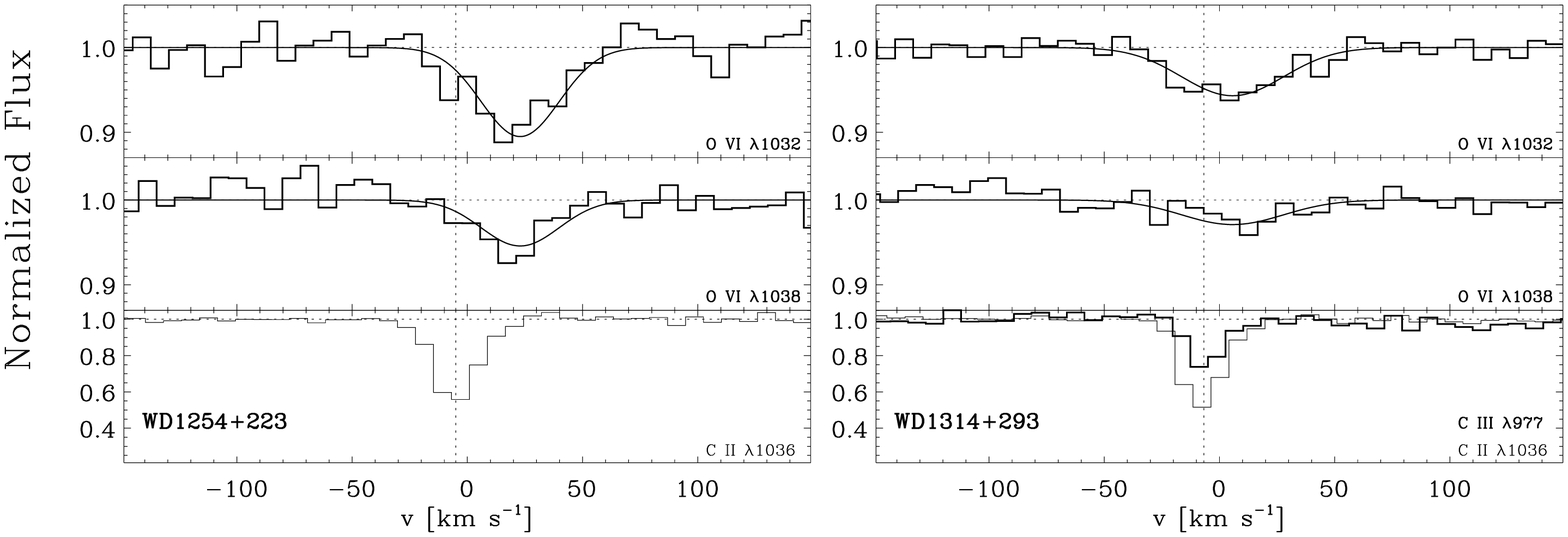}
\caption{Normalized flux versus heliocentric velocity and one component fit to the \ovi\ absorption 
lines toward WD\,1254+223 and WD\,1314+293. \cii\ absorption is displayed for both stars. For WD\,1314+293
\ciii\ is also displayed.
\label{model}}
\end{figure}

\begin{figure}[tbp]
\epsscale{1}
\plotone{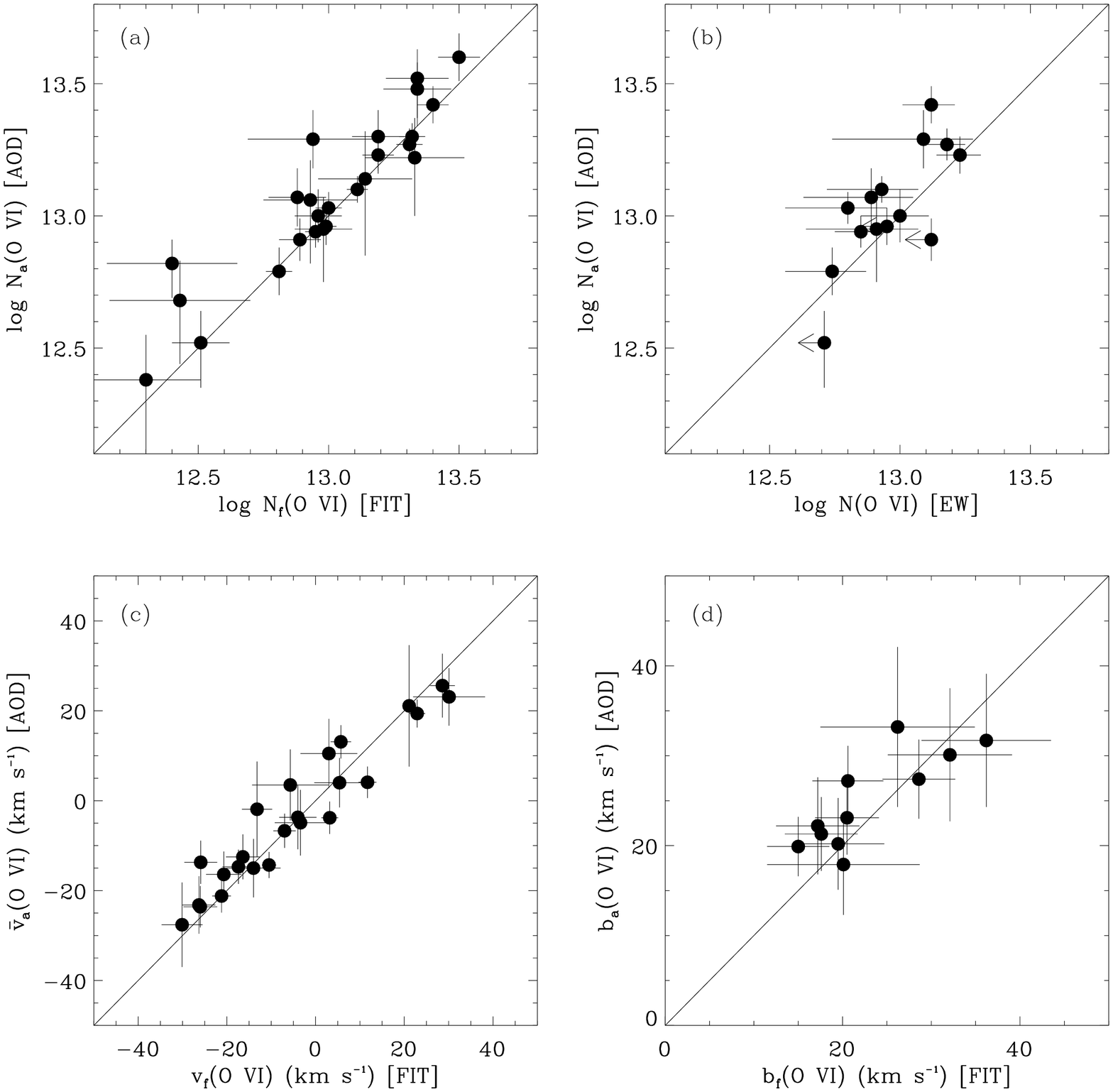}
\caption{Comparison plots: (a) AOD and FIT column densities; (b)
AOD and equivalent width (Oegerle et al. 2004, data with arrows for 2$\sigma$ upper limits) 
column densities; (c )AOD and FIT velocities; (d) AOD and FIT $b$-parameters.
\label{comp}}
\end{figure}

\begin{figure}[tbp]
\epsscale{1}
\plotone{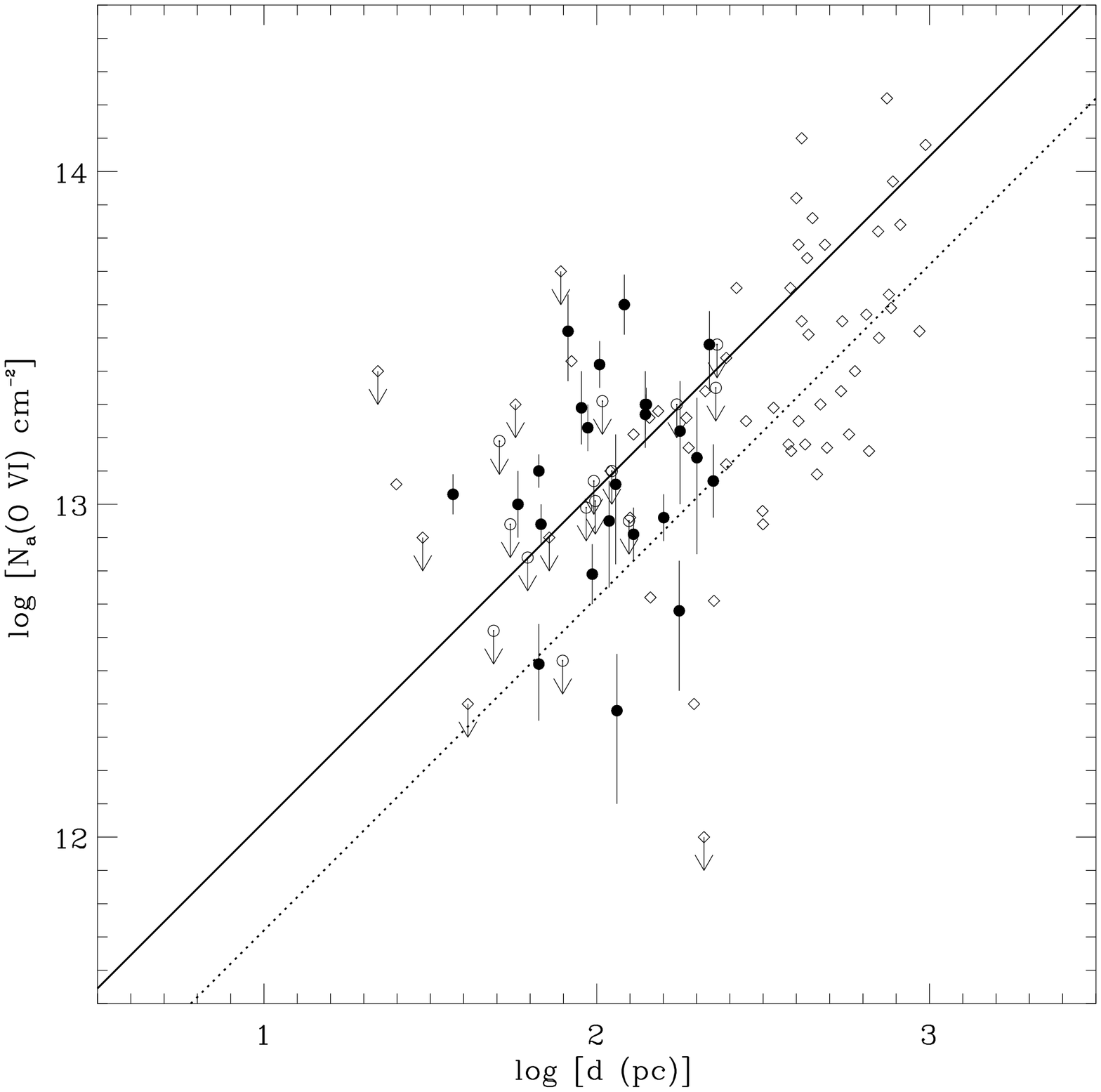}
\caption{The logarithmic column density of \ovi\ is plotted against the logarithmic distance to the star.
Filled circles are the LISM $\ge 2\sigma$ detections from Table~\ref{t2}. Open circles with 
arrows are 2$\sigma$ upper limits. Detections and  2$\sigma$ upper limits from Jenkins (1978a)
for stars with $\log d < 3$ are displayed with the open diamonds, without and with arrows. 
The solid line corresponds to $n($\ovi$)=3.6 \times 10^{-8}$ cm$^{-3}$;
the dotted line corresponds to $n($\ovi$)=1.7 \times 10^{-8}$ cm$^{-3}$.
\label{ovidist}}
\end{figure}

\begin{figure}[tbp]
\epsscale{1}
\plotone{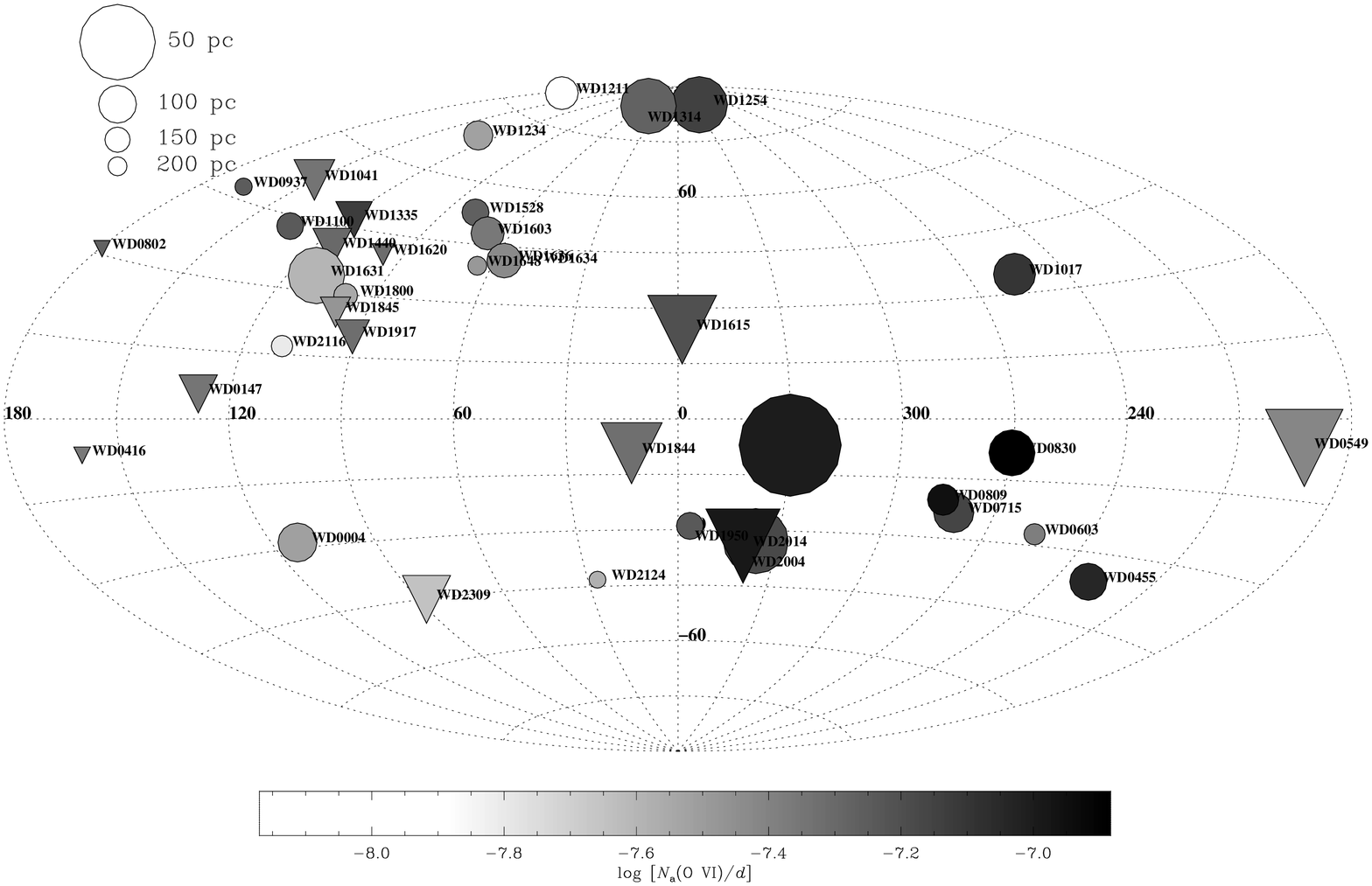}
\caption{WD locations in Galactic coordinates ($l,b$) for the 39 stars from Table~\ref{t1} not 
likely affected by stellar blending problems. Same symbol code as Fig.~\ref{ovimap} except
the grey scale indicates the average line of sight density $N_a($\ovi$)/d$ ({\em circles}) or
density limit ({\em triangles}).
\label{novimap}}
\end{figure}

\begin{figure}[tbp]
\epsscale{1}
\plotone{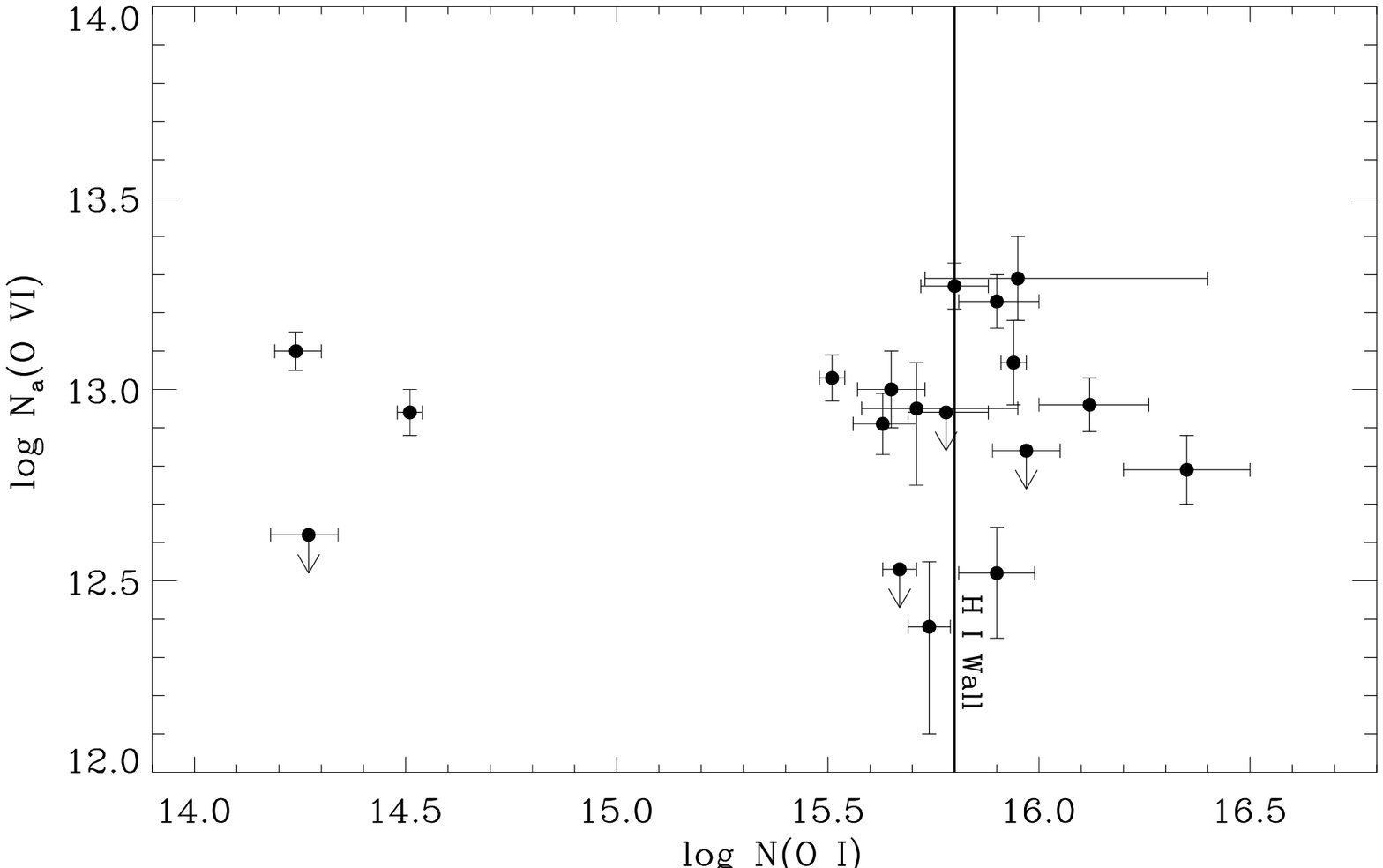}
\caption{The logarithmic column density of \ovi\ (AOD) is plotted against the logarithmic column density of \oi\
from Lehner et al. (2003). The Local Bubble 
data points are delimited by the vertical line ($\log N($\oi$)\la 15.80$ dex). 
We did not plot the data for WD\,0455$-$282 because $N$(\oi) is uncertain. Data with arrows 
are 2$\sigma$ upper limits. 
\label{ovicomp}}
\end{figure}

\begin{figure}[tbp]
\epsscale{0.5}
\plotone{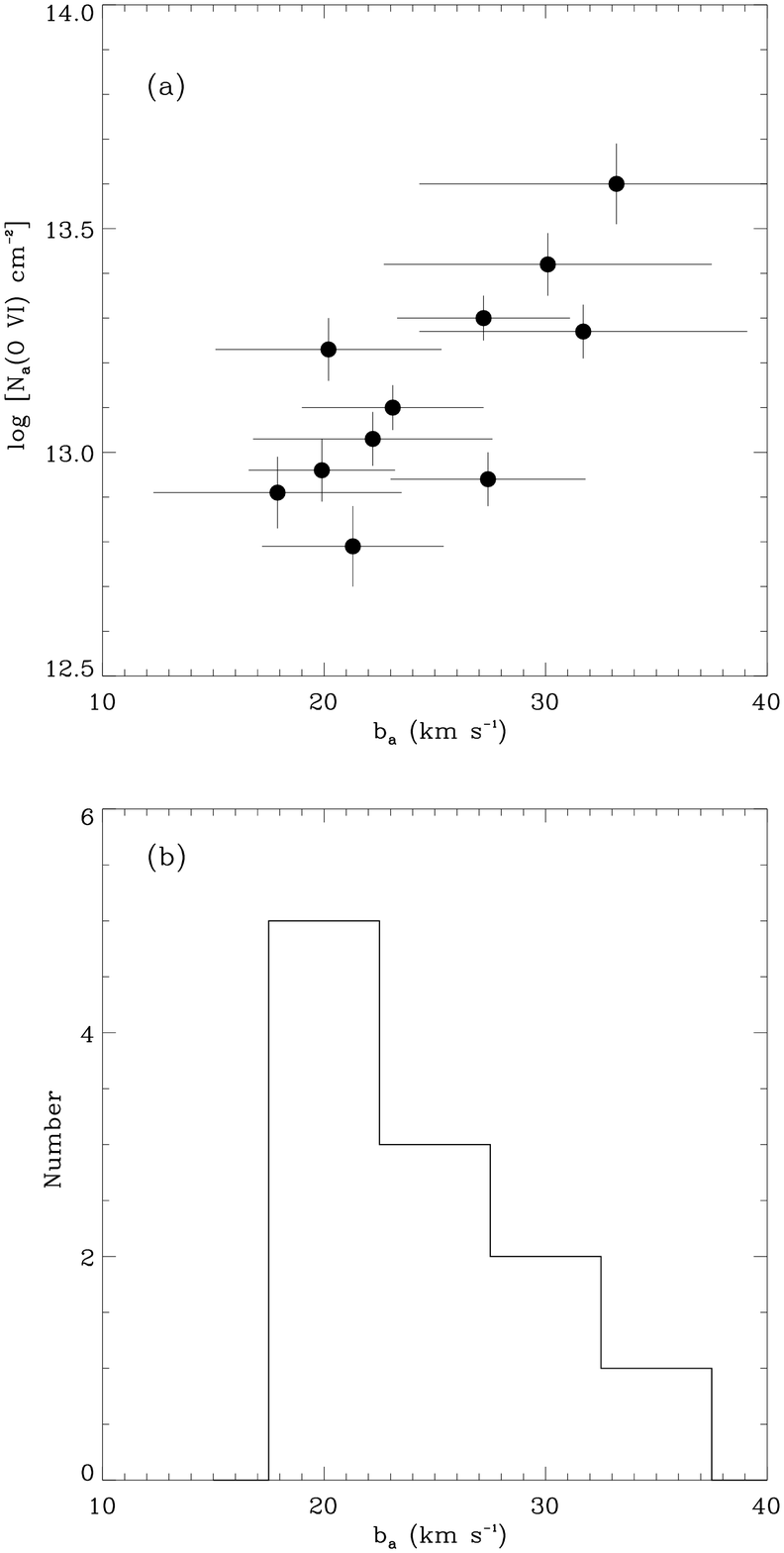}
\caption{(a) The logarithmic column density of \ovi\ is plotted against the Doppler parameter.  (b) Histogram of the Doppler parameter.
\label{fig12}}
\end{figure}

\begin{figure}[tbp]
\epsscale{1}
\plotone{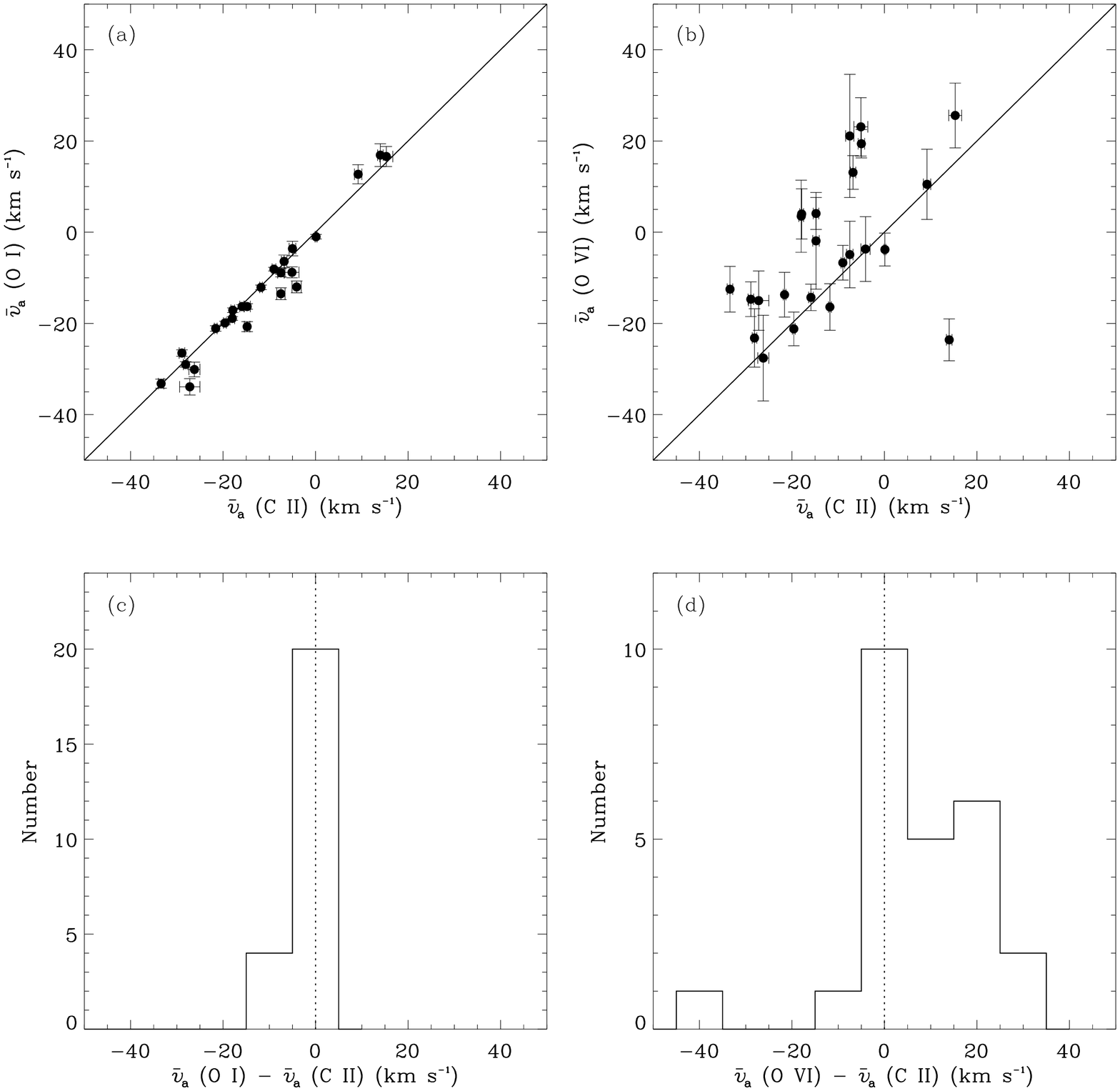}
\caption{Average  velocities of \ovi\ and \oi\ absorptions are plotted against the 
average  velocity of \cii\ absorption in (a) and (b). Note that we did not include \oi\ if 
followed by a $b$ in Table~4.  Histograms of  $\bar{v}_a($\oi$)-\bar{v}_a($\cii) and $\bar{v}_a($\ovi$)-\bar{v}_a($\cii)
are shown in (c) and (d)
\label{velcen}}
\end{figure}

\begin{figure}[tbp]
\epsscale{1}
\plotone{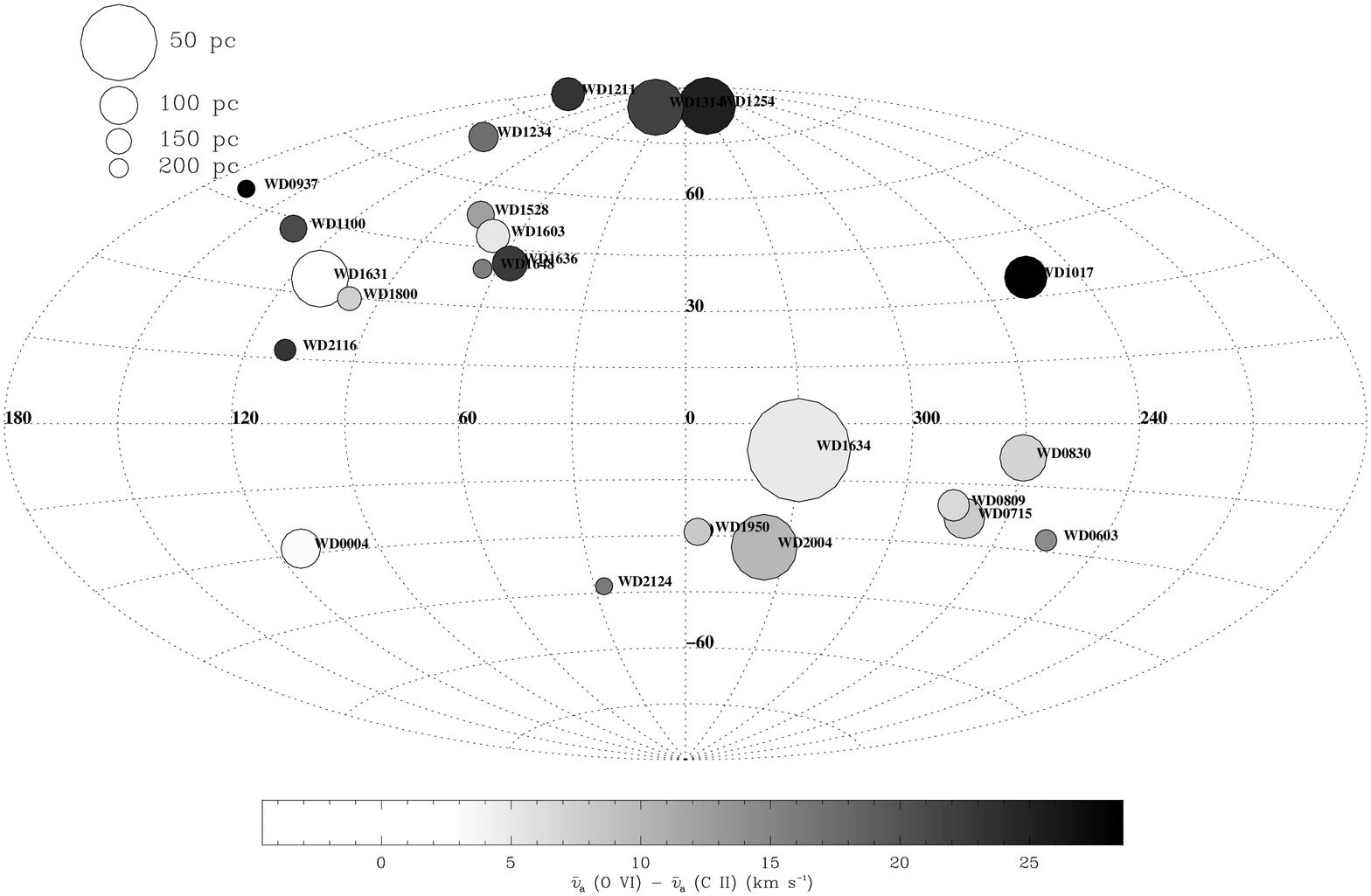}
\caption{ WD locations in Galactic coordinates ($l, b$) of the 24 stars 
from Table~\ref{t3} except WD\,0455$-$282 with $\ge 2 \sigma$ \ovi\ detections and not likely affected by 
stellar contamination. Same symbol code as Fig.~\ref{ovimap} except
the grey scale indicates $\Delta v = \bar{v}_a($\ovi$)-\bar{v}_a($\cii). 
Note that the WDs with the  larger values of $\Delta v$
lie north of the Galactic plane.
\label{deltav}}
\end{figure}

\end{document}